\documentclass[preprint,amsmath,amssymb,aps,pra,nofootinbib,longbibliography,superscriptaddress]{revtex4-1}
\usepackage{graphicx}
\usepackage{dcolumn}
\usepackage{bm}
\usepackage{amsthm,mathrsfs}
\usepackage{mathtools,bbm}
\usepackage{physics}
\usepackage{subfig}
\usepackage{array}
\usepackage{multirow}
\usepackage{tikz}
\usepackage{appendix}

\usetikzlibrary{positioning}
\usetikzlibrary{matrix}
\usetikzlibrary{arrows,decorations.pathmorphing}

\newcolumntype{E}{>{\centering\arraybackslash} m{1.0cm}}

\usepackage{hyperref}
\usepackage{xcolor}
\hypersetup{
    colorlinks,
    linkcolor={blue!70!black},
    citecolor={blue!70!black},
    urlcolor={blue!70!black}
}

\begin{document}

\title{Large molasses-like cooling forces for molecules using polychromatic optical fields: A theoretical description}

\author{Konrad Wenz}
\email{k.wenz@columbia.edu}
\affiliation{
 Department of Physics, Columbia University, 538 West 120th Street, New York, NY 10027
}
\author{Ivan Kozyryev}%
\affiliation{
 Department of Physics, Columbia University, 538 West 120th Street, New York, NY 10027
}
\author{Rees L. McNally}%
\affiliation{
 Department of Physics, Columbia University, 538 West 120th Street, New York, NY 10027
}
\author{Leland Aldridge}%
\affiliation{
 Department of Physics, Gonzaga University, 502 East Boone Avenue, Spokane, WA 99201
}
\author{Tanya Zelevinsky}%
\affiliation{
 Department of Physics, Columbia University, 538 West 120th Street, New York, NY 10027
}

\date{\today}

\begin{abstract}
Recent theoretical investigations have indicated that rapid optical cycling should be feasible in complex polyatomic molecules with diverse constituents, geometries and symmetries. However, as a composite molecular mass grows, so does the required number of photon scattering events necessary to decelerate and confine molecular beams using laser light. Utilizing coherent momentum exchange between light fields and molecules can suppress spontaneous emission and significantly reduce experimental complexity for slowing and trapping. Working with BaH as a test species, we have identified a robust, experimentally viable configuration to achieve large molasses-like cooling forces for molecules using polychromatic optical fields addressing both $X-A$ and $X-B$  electronic transitions, simultaneously. Using numerical solutions of the time-dependent density matrix as well as Monte Carlo simulations, we demonstrate that creation of Suppressed Emission Rate (SupER) molasses with large capture velocities ($\sim 40$ m/s) is generically feasible for polyatomic molecules of increasing complexity that have an optical cycling center. Proposed SupER molasses are anticipated to not only extend quantum control to novel molecular species with abundant vibrational decay channels, but also significantly increase trapped densities for previously laser-cooled diatomic and triatomic species. 

\end{abstract}

\pacs{Atomic Physics}%
\maketitle
\newpage
\section{Introduction}

\subsection{Direct Molecular Laser Cooling}

Optical control over atomic spatial degrees of freedom is one of the cornerstones of modern atomic physics \cite{chu1998nobel,phillips1998nobel} and quantum technologies \cite{weiss2017quantum,bongs2019taking}. In recent years, laser cooling and trapping methods have been successfully extended to a handful of molecular species \cite{McCarron2018laser,tarbutt2018laser}. Yet despite more than a decade of active research efforts, only three diatomic species (SrF \cite{norrgard2016submillikelvin}, CaF \cite{truppe2017molecules,anderegg2018dipole} and YO \cite{ding2020sub}) have been trapped in three dimensions (3D) at microkelvin temperatures. While one-dimensional (1D) laser cooling of cryogenic molecular beams of diatomic BaH \cite{mcnally2020} and YbF \cite{lim2018laser}, triatomic SrOH \cite{kozyryev2017sisyphus}, YbOH \cite{augenbraun2020laser} and CaOH \cite{baum2020CaOH},  and even hexatomic CaOCH\textsubscript{3} \cite{mitra2020direct} molecules has been achieved, the number of scattered photons demonstrated ($\sim100-1,000$) is still at least an order of magnitude below what is needed to achieve radiative slowing and 3D trapping. Therefore, the question of general prospects of utilizing laser slowing, 3D cooling and trapping for molecular species with new internal structures (e.g. BaH) or increased vibrational complexity (e.g. triatomics) remains largely unanswered\footnote{Alternative methods for molecular cooling and trapping that do not rely on repeated photon scattering have also been demonstrated \cite{prehn2016optoelectrical,reens2017controlling,segev2019collisions,momose2017magnetic}. We refer interested readers to review articles that compare and contrast different methods and how they can address various fields of scientific research \cite{McCarron2018laser, tarbutt2018laser,carr2009cold,bohn2017cold,balakrishnan2016perspective}.}. 

Traditional Doppler slowing and cooling relies on a repeated process of directional photon absorption (resulting in $\hbar k$ momentum transfer) followed by spontaneous emission to the initial set of states for the cycle to repeat \cite{metcalf2003review}. While in many atoms, the use of specific angular momentum configurations for the ground and excited states together with the appropriate laser polarization can lead to an effective ``two-level'' system, the absence of strict vibrational selection rules for molecular electronic decays necessitates novel approaches to molecular laser cooling \cite{McCarron2018laser}. The probability of decay into a given vibrational level is described by the square of the overlap integral between the excited (double prime) and ground (single prime) vibrational wavefunctions $\mathcal{F}_{v''v'}$, also known as a Frank-Condon Factor (FCF) for that transition \cite{tarbutt2018laser}. For certain diatomic species with small off-diagonal FCFs (i.e. $v''\neq v'$), one or two additional lasers can be used to repump molecules from excited vibrational levels $v''>0$ back to the ground vibrational state $v''=0$, enabling scattering of $\gtrsim 10^4$ photons \cite{barry2012laser,hemmerling2016laser,yeo2015rotational} needed to slow molecular beams to below the capture velocity of a 3D molecular magneto-optical trap with $v_{\rm{cap}}\approx5-10$ m/s \cite{williams2017characteristics,steinecker2019sub}. However, even for light triatomic species with relatively diagonal FCFs like CaOH ($\mathcal{F}_{00} = 0.954$), eight additional repumping lasers are needed to scatter $\sim 10^4$ photons \cite{baum2020CaOH,baum2020thesis,baum2020CaOH10k}, thus, presenting a significant technical challenge for extending Doppler slowing and trapping methods to heavier (e.g. YbOH) or more complex (e.g. CaOCH\textsubscript{3}) molecules. Towards this end, various alternative techniques have been developed for efficient momentum transfer from the laser light to atoms or molecules, while minimizing spontaneous emissions \cite{metcalf2017colloquium}. To date, the emphasis has been on developing novel experimental methods to achieve molecular slowing to $v_{\rm{cap}}$ with a small number of spontaneously emitted photons \cite{long2019suppressed,jayich2014continuous,Chieda2011}, thus reducing the number of required repumping lasers. Here we present a novel cooling scheme that uses multifrequency light to rapidly dampen molecular motion in a wide range of velocity classes $v_{\rm{eff}}\gg v_{\rm{cap}}$, while minimizing the number of spontaneous decay cycles. The proposed Suppressed Emission Rate (SupER) molasses could be either combined with coherent slowing techniques or used with previously magneto-optically trapped species to capture and cool molecules with $v > v_{\rm{cap,MOT}}$. 

\subsection{Coherent Optical Forces}

Widely utilized optical slowing and cooling methods for atomic gases usually use a single optical frequency to address a specific ``two-level'' transition. Such radiative methods are characterized by the maximum force $F_{\rm{rad}}=\hbar k\Gamma_{\rm{sp}}/2$ affecting velocity classes within $\Delta v_{\rm{rad}}\approx \Gamma_{\rm{sp}}/k$, with the intrinsic spontaneous decay rate $\Gamma_{\rm{sp}}$ limiting the maximum force as well as the capture range \cite{metcalf2003review}. While a conservative dipole force arising from the gradient of the light shift can lead to strong confining forces, its utility in cooling atomic or molecular motion is severely limited since it averages out to zero over a spatial scale larger than light wavelength $\lambda$ \cite{metcalf2017colloquium}. However, already more than thirty years ago it has been theorized that the dipole force can be ``rectified'' to maintain a constant sign over position scales much larger than $\lambda$ by adding a second light field to spatially modulate the atomic energy levels and, therefore, the sign of the detuning for the initial dipole force laser field \cite{kazantsev1987rectification,kazantsev1989rectification}. Shortly afterwards, Grimm and co-workers have conclusively demonstrated the effect of the rectified dipole force (RDF) on a sodium atomic beam achieving $F_{\rm{RDF}}\approx 4F_{\rm{rad}}$ \cite{grimm1990observation}, a factor of 2.5 lower than the initial prediction due to the presence of transverse atomic velocities larger than $v_{\perp}\sim\Gamma_{\rm{sp}}/k$ \cite{ovchinnikov1993rectified}. In order to remedy the issue of a small velocity capture range of the RDF, other methods for generating large coherent optical forces have been proposed that realize coherent control of light - atom momentum exchange by tailoring the inversion of the atomic populations \cite{cashen2003optical}. Only recently, however, have the effects of such coherent optical methods been conclusively demonstrated in molecules using triatomic SrOH \cite{Kozyryev2018} and diatomic CaF \cite{Gould2018CaFdeflection}. 

\par The use of counterpropagating amplitude-modulated light waves, leading to a stimulated light pressure on atoms, has been proposed as a viable method to achieve large force magnitudes $F\gg F_{\rm{rad}}$ over a wide velocity range $v_{\perp}\gg\Gamma_{\rm{sp}}/k$ \cite{voitsekhovich1989observation, Yatsenko1991}. While the magnitude of the stimulated bichromatic force (BCF) can be explained using an intuitive resonant optical $\pi$-pulse interpretation \cite{Yatsenko1991,Soding1997}, an understanding of the large velocity capture range requires a doubly-dressed atom picture \cite{Yatsenko2004}. In the simplest case, a two-level system interacts with collinearly superimposed bichromatic standing waves with equal intensities $I_{\rm{BCF}}$ and symmetrically detuned by $\pm \lvert\delta \gg \Gamma_{\rm{sp}}\rvert$ from atomic resonance. By imposing a relative phase offset $\chi$ between the counterpropagating laser fields, the directionality of the force can be controlled by fixing the relative timing between the resulting beat pulse trains (Fig. \ref{fig:bcf_schematic}) \cite{Soding1997}. Choosing $I_{\rm{BCF}}$ and $\delta$ such that the Rabi frequency integrated over a single beat pulse area satisfies $\Omega t_{\pi}\approx\pi$, efficient transfer of atomic population between the ground and excited states can be achieved at a rate of $\delta/\pi\gg\Gamma_{\rm{sp}}$ \cite{metcalf2003review}. Since each directional $\pi$ pulse transfers $\hbar k$ momentum to the atom, the order of magnitude for the bichromatic force\footnote{As shown in App. \ref{sec:Variance-estimation}, the expression presented here is in fact exact for a two-level system: $F_{\mathrm{BCF,2-level}}=\hbar k\delta/\pi$.} is $F_{\rm{BCF}}\sim \hbar k\delta/\pi\gg F_{\rm{rad}}$. Even though a spontaneous emission rate can be significantly suppressed by reducing the excited state fraction with a properly designed pulse sequence \cite{Galica2013, long2019suppressed}, any spontaneously emitted photons will lead to quantum state decoherence and potential reversal of the momentum transfer direction. For an asymmetric $\chi$ phase choice required for a directional momentum transfer this leads to only order of unity reduction in the net magnitude of photon transfers averaged over a time greater than $1/\Gamma_{\rm{sp}}$ \cite{Soding1997}.  However, such processes can have important consequences for the limiting temperature of the ensemble when coherent optical forces are employed for cooling of lab-frame velocity. Careful understanding of limiting temperatures in stimulated transfer cooling methods has proven challenging \cite{Partlow2004,norcia2018narrow}, and here we develop a novel method of doing so using a continuous-time Markov Chain model detailed in Apps. \ref{sec:Variance-estimation} and \ref{sec:App-B}. 

\par When considering BCF and other stimulated light forces like RDF, it is important to properly account for the atom's or molecule's finite velocity that could significantly affect the magnitude of the experimentally achievable decelerations (seen in the initial RDF experiments, for example \cite{ovchinnikov1993rectified}). In order to create force profiles accurately depicting velocity dependence necessitates solving Liouville-von Neumann equations for density matrix evolution in the rotating wave, fixed-velocity approximations, followed by obtaining the force averaged over the ensemble $\langle F\rangle=-\mathrm{tr}\,(\rho \grad H)$, where $\rho$ is the density matrix \cite{metcalf2017colloquium,Yatsenko2004}. However, intuitively, the velocity capture range for the bichromatic force can be interpreted as arising from the relative dephasing between consecutive beat notes. Once the Doppler shift $kv$ becomes comparable to the Rabi frequency $\Omega \sim \delta$, the $\pi$ pulses no longer lead to efficient population transfer, thus limiting the affected velocity range to $\triangle v \sim \delta/k\gg \Gamma_{\rm{sp}}/k$ \cite{chieda2012bichromatic} under conditions of large BCF detuning $\delta\gg\Gamma_{\rm{sp}}$.  

\subsection{\label{sec:Cooling-properties} Cooling Properties of the Bichromatic Force}

\par
\begin{figure}[!h]
\centering
\includegraphics[scale=0.8]{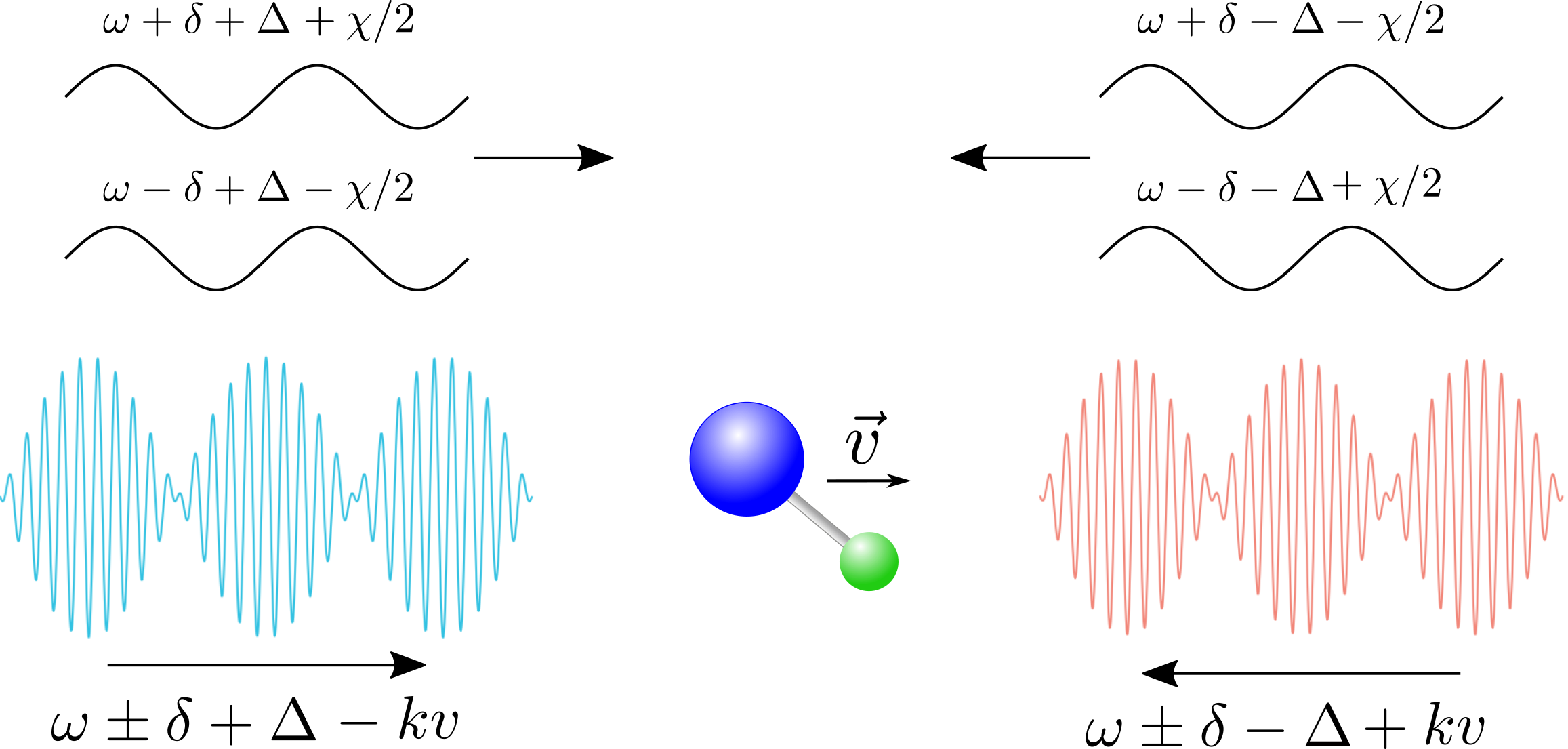}
\caption{\small Interaction of a moving two-level system with a bichromatic light field. Pulse trains are created by a pair of counter-propagating two-color ($\pm \delta$) beams offset from each other by a phase $\chi$. To center the force profile around a non-zero velocity, pulses on one side have to be red-detuned, while pulses approaching from the other side have to be blue-detuned.}
\label{fig:bcf_schematic}
\end{figure}

\begin{figure}[!h]
\centering
\includegraphics[width=0.7\textwidth]{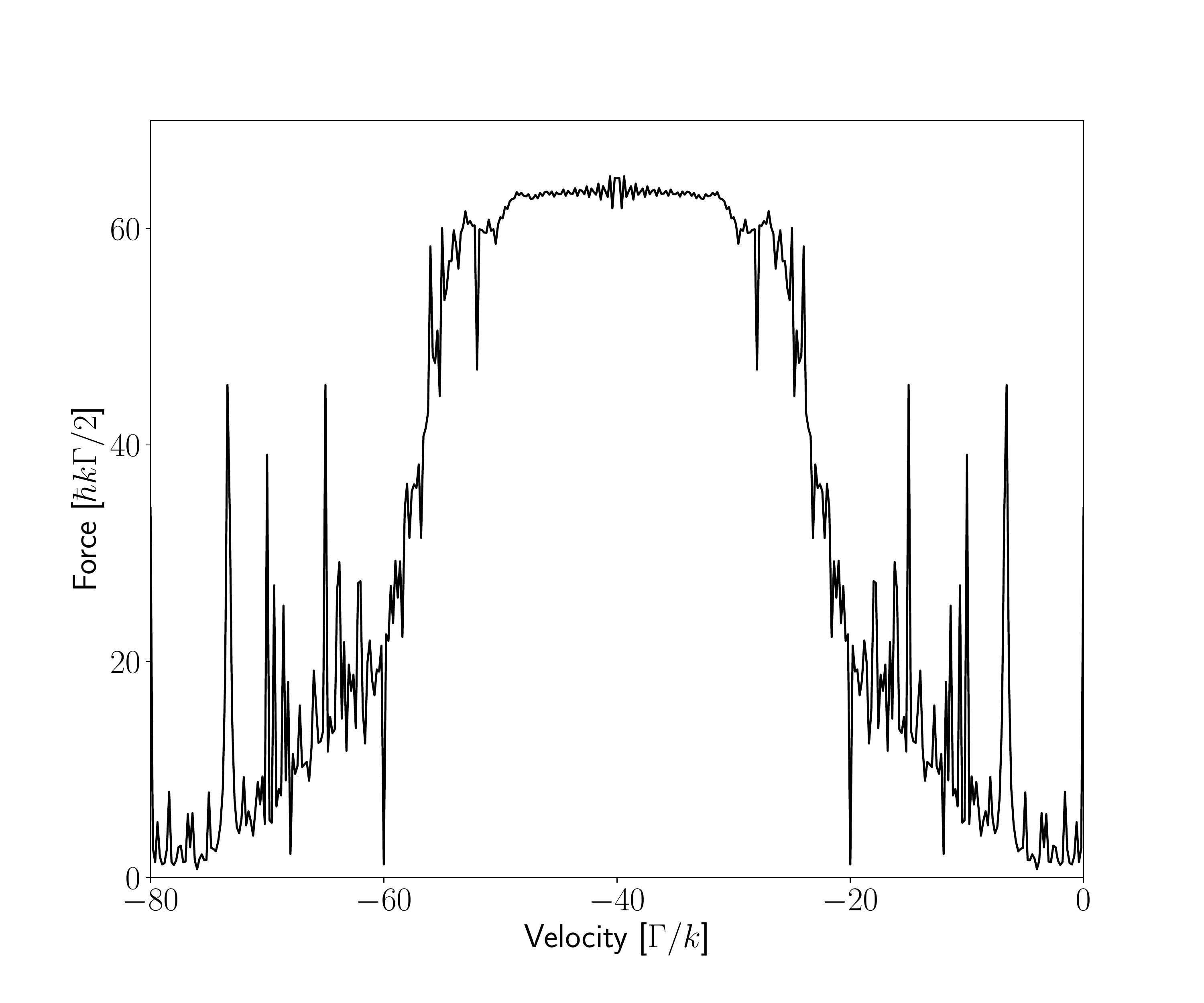}
\caption{\small A shifted bichromatic force profile for $\delta=100\,\Gamma$, $\Omega=\sqrt{3/2}\delta$, $\chi=45^{\circ}$ with opposing beams detuned by $\Delta=-40\,\Gamma$. The profile is centered at $v=\Delta/k=-40\,\Gamma/k$ and was obtained by numerically solving Liouville-von Neumann equations for density matrix evolution in the rotating wave, fixed-velocity approximations. The maximum force is equal to $\hbar k \delta/\pi\approx 63\, \hbar k \Gamma/2$ and the width of the profile can be estimated to be equal to $\delta/2k=50\,\Gamma/k$. Sharp vertical spikes are Doppleron resonances arising when integer numbers of red-detuned absorptions and blue-detuned emissions of photons (or vice versa) are resonant with the transition \cite{minogin1979resonant}. This will occur when $\left(\delta + kv\right)/\left(\delta - kv\right)$ is rational. However, in previous measurements of the bichromatic force profiles \cite{williams1999measurement,williams2000bichromatic}, these narrow Doppleron resonances were not observed and are not expected to have any significant effect on real physical systems.}
\label{fig:bcf_shifted}
\end{figure}

 Since the bichromatic force does not vanish for atoms at rest and involves mostly coherent state transfers, it may seem surprising that rapid cooling of atomic beams has been achieved using BCF configurations\footnote{In fact, transverse cooling of the metastable helium beam using the bichromatic force has been demonstrated in the regime of $\lesssim 1.5$ emitted photons \cite{corder2015JOSAB,corder2015PRL}.}. However, the sharp edges of the BCF profiles can be used for compressing velocity distribution and achieving cooling of atomic motion. The frequencies of counter-propagating dual-frequency beams can be offset by opposite amounts, creating a situation where the atom or molecule undergoes efficient $\pi$-pulse transfers from $\omega_0\pm\delta$ beat notes only at non-zero velocities (Fig. \ref{fig:bcf_schematic}). The Doppler shift experienced by an atom or molecule is $\pm kv$, so by letting $\Delta=kv_0$ we can center the force profile around a chosen non-zero velocity $v_0$. By shifting the bichromatic force profile to be centered around a non-zero velocity, efficient longitudinal cooling of atomic beams has been demonstrated \cite{Soding1997,chieda2012bichromatic}. In Fig. \ref{fig:bcf_shifted}, we show an example of such a shifted force profile centered at $v_0=-40\,\Gamma/k$ obtained for a relative shift of $\Delta=-40\,\Gamma$. Our parameters were chosen by optimizing the peak force at the profile's center.

\newpage 

\par The use of large stimulated optical forces for achieving 3D cooling and kelvin-deep trapping of atoms and molecules was one of the primary motivations for the initial extensive development of such methods \cite{voitsekhovich1989observation,kazantsev1987rectification}. However, despite thirty years of research, the application of coherent stimulated forces to zero-velocity cooling (i.e. compression of velocity distribution towards zero lab-frame velocity) and confinement of atomic and molecular samples has been limited \cite{jayich2014continuous,ooi2003laser,ooi2010laser}. Partlow and co-workers have performed a landmark experiment on a helium beam to use spatially separated, shifted bichromatic force profiles with opposite phase $\chi$ and frequency shift $\Delta$ \cite{Partlow2004}. Such a 1D collimation scheme required two sequential interaction regions acting on atoms with positive and negative initial velocities, respectively, and thus leading to experimental results emulating the effects of optical molasses. However, as pointed out by Partlow and co-workers \cite{Partlow2004}, the underlying physical process was not resulting in a true damping force for velocities of interest and led to a different physical behavior than optical molasses cooling.

\par Here, we use the inherent multilevel structure of molecular radicals that limits cooling efficiency of traditional Doppler molasses to propose a novel 1D laser cooling scheme with significantly higher velocity range and damping coefficient. By addressing two separate, yet radiatively coupled, two-level systems with polychromatic optical fields, we discover that it is possible to achieve a large velocity damping force with Suppressed Emission Rate (SupER). Furthermore, in the experimentally accessible regime, we demonstrate the feasibility of damping molecular motion to millikelvin temperatures on microsecond timescales. We show how SupER molasses force profiles can be created in a simple 4-level system, develop a new mathematical model for estimating the final temperature, and perform Monte Carlo simulations of the cooling dynamics to confirm the analytical estimates. Throughout the paper we use the barium monohydride (BaH) molecule as a test species for our time-dependent density matrix calculations, but also suggest a more general level scheme common to many diatomic and polyatomic radicals that could be utilized to create large 1D molasses-like forces. Therefore, we identify a way to use the internal complexity of molecular systems to enable their efficient quantum control for a wide variety of proposed applications \cite{carr2009cold}. Our work makes an important step towards experimental realization of kelvin-deep macroscopic ($r\gg \lambda$) optical traps for molecules proposed more than thirty years ago \cite{kazantsev1987rectification,voitsekhovich1989observation}.

\section{Suppressed emission rate molasses}

\par To realize cooling force profiles arising from a rapid coherent momentum exchange between light fields and molecules, we identify an appropriate multi-level system that would allow us to combine two asymmetric (shifted) force-velocity profiles (e.g. shown in Fig. \ref{fig:bcf_shifted}) without significant reduction in force vs velocity characteristics. For ease of theoretical tractability and computational simplicity, our analysis if performed in one dimension (1D). While 1D/2D cooling methods are already of significant interest for molecular cooling experiment, we believe that extension of our observations to the full 3D case should be also possible, yet beyond the scope of the present work. 

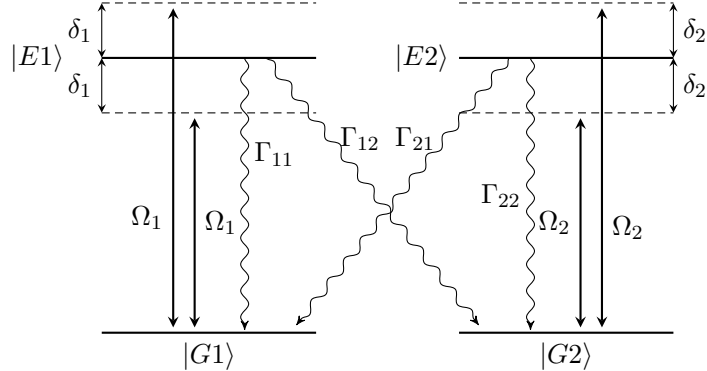
\begin{figure}[!h]
\centering
\begin{tikzpicture}[
scale=0.95, every node/.style={transform shape},
level/.style={thick},
energy/.style={thin,<->,shorten >=0.1pt,shorten <=0.1pt,>=stealth},
energyZ/.style={ultra thin,<->,shorten >=0.1pt,shorten <=0.1pt,>=stealth,line width=0.02},
virtual/.style={thin, densely dashed},
virtual_d/.style={thin, dotted},
radiative/.style={black,->,>=stealth',shorten >=1pt,decorate,decoration={snake,amplitude=1.5}},
trans/.style={thick,<->,shorten >=2pt,shorten <=2pt,>=stealth},
]

\draw[level] (-10cm,-5em) -- (-7cm,-5em) node[midway,below] {$\ket{G1}$};
\draw[level] (-7cm,5em) -- (-10cm,5em) node[left] {$\ket{E1}\ \ \ $};
\draw[level] (-5cm,-5em) -- (-2cm,-5em) node[midway,below] {$\ket{G2}$};
\draw[level] (-2cm,5em) -- (-5cm,5em) node[left] {$\ket{E2}$};

\draw[virtual] (-10cm,7em) -- (-7cm,7em);
\draw[virtual] (-10cm,3em) -- (-7cm,3em);

\draw[trans] (-8.7cm,-5em) -- (-8.7cm, 3em) node[midway,right] {$\Omega_1$};
\draw[trans] (-9cm,-5em) -- (-9cm, 7em) node[pos=0.35,left] {$\Omega_1$};

\draw[energy] (-10cm,3em) -- (-10cm,5em) node[midway,left] {$\delta_1$};
\draw[energy] (-10cm,5em) -- (-10cm,7em) node[midway,left] {$\delta_1$};

\draw[radiative] (-8cm,5em) -- (-8cm,-5em) node[pos=0.35,right] {$\Gamma_{11}$};
\draw[radiative] (-7.7cm,5em) -- (-4.7cm,-4.8em) node[pos=0.3,right] {$\Gamma_{12}$};

\draw[virtual] (-5cm,7em) -- (-2cm,7em);
\draw[virtual] (-5cm,3em) -- (-2cm,3em);

\draw[trans] (-3.3cm,-5em) -- (-3.3cm, 3em) node[midway,left] {$\Omega_2$};
\draw[trans] (-3cm,-5em) -- (-3cm,7em) node[pos=0.32,right] {$\Omega_2$};

\draw[energy] (-2cm,5em) -- (-2cm,7em) node[midway,right] {$\delta_2$};
\draw[energy] (-2cm,3em) -- (-2cm,5em) node[midway,right] {$\delta_2$};

\draw[radiative] (-4cm,5em) -- (-4cm,-5em) node[pos=0.5,left] {$\Gamma_{22}$};
\draw[radiative] (-4.3cm,5em) -- (-7.3cm,-4.8em) node[pos=0.3,left] {$\Gamma_{21}$};

\end{tikzpicture}
\caption{\small Diagram of BCF in two coupled 2-level systems that represent the structures necessary for realizing large molasses-like force profiles described in the text.} \label{fig:BCF_BaH_diagram}
\end{figure}

As initially pointed out by Partlow and co-workers \cite{Partlow2004}, it is impossible to simultaneously apply both force profiles on one single 2-level system without drastically perturbing individual force profiles. In order to circumvent this limitation, instead we consider two almost independent 2-level systems that are coupled only by spontaneous decay as shown in Fig. \ref{fig:BCF_BaH_diagram}. As described in Sec. \ref{sec:real-molecules} such a system can be generically realized in diatomic and polyatomic radicals with optical cycling properties. For initial studies we consider an idealized 4-level system with  $\Gamma_{11}+\Gamma_{12}=\Gamma_{22}+\Gamma_{21}\equiv\Gamma$ with $\Gamma_{11}=\Gamma_{22}$ and $\Gamma_{12}=\Gamma_{21}$, as well as $\lambda_1=\lambda_2\equiv 2\pi/k$. While this assumption facilitates some initial theoretical calculations and demonstrates general aspects of coherent momentum exchange leading to suppressed emission rate molasses force profiles, we relax the simplifying assumptions in the next step when working with realistic systems.

\par The simplest realization of SupER molasses in the toy model would have Rabi rate and phase difference set to the optimum conditions identified for stimulated bichromatic force configuration: $\Omega_1=\Omega_2=\sqrt{3/2}\,\delta$ and $|\chi|=45^{\circ}$ \cite{metcalf2017colloquium}, where Rabi rates are those of \emph{every} component shown in Fig. \ref{fig:bcf_schematic}. In order to obtain two asymmetric profiles those 2-level systems have to have opposite signs of the detunings and phase differences: $\Delta_1=-\Delta_2$ and $\chi_1=-\chi_2$. For simplicity, we assume for now that $\delta_1=\delta_2\equiv\delta$, with $\delta=100\,\Gamma$ used in calculations. While in the toy model $\delta$ has to only be much larger than $\Gamma$, in real systems we need it to be larger than naturally occurring energy splittings, such as hyperfine splitting, which quite often are on the order of couple $\Gamma$. Our specific choice, while arbitrary, should be applicable in many situations and be realizable using off-the-shelf acousto-optic modulators (AOMs).
\par
\begin{figure}[!h]
\centering
\includegraphics[width=0.7\textwidth]{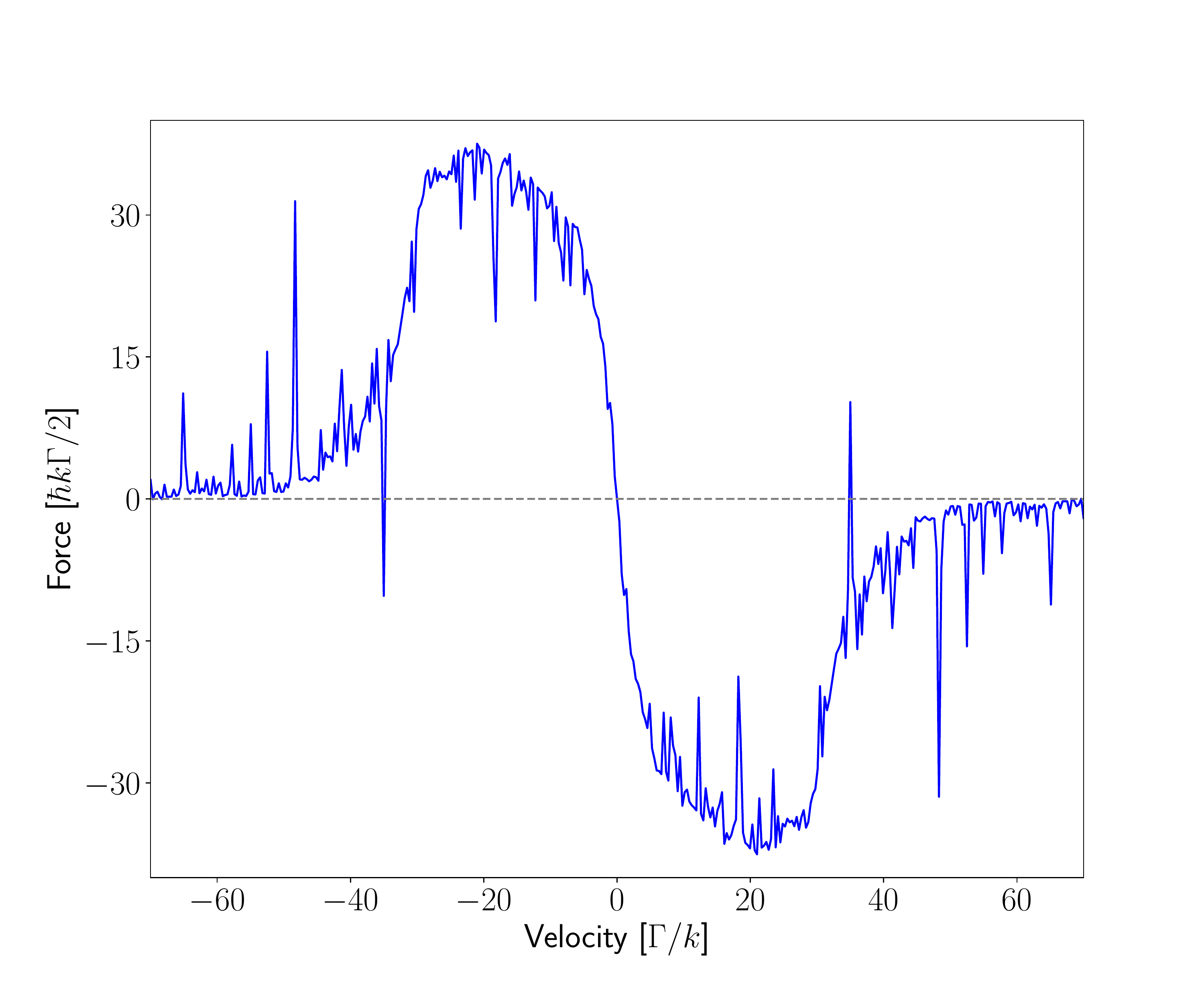}
\caption{\small BCF molasses force profile in a symmetrized BaH toy model. Profile was obtained for $\Delta_{\pm}=\mp 15\,\Gamma$, $\delta=100\,\Gamma$, $\Omega_0=\sqrt{3/2}\delta$ and $\chi_{\pm}=\pm 45^{\circ}$. }
\label{fig:full_bcf_mol}
\end{figure}

 We found the most optimal profile for $\Delta_1=-15\,\Gamma$ and depict it in Fig. \ref{fig:full_bcf_mol}. We should note that in such symmetrized system given no additional selection rules the light from both 2-level subsystems would couple to the other subsystem. In a real system, like that in BaH, light frequencies are vastly different and don't couple to both transitions simultaneously, yet still enable realization of SupER molasses as discussed below.

\par
\begin{figure}[!h]
\centering
\includegraphics[width=0.7\textwidth]{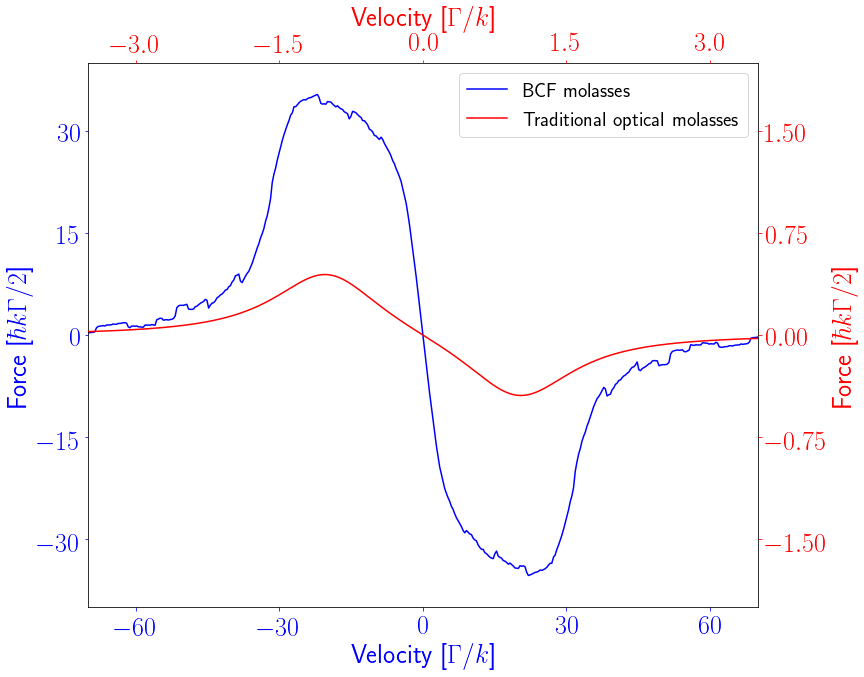}
\caption{\small Comparison of BCF molasses force profile (blue graph and axes) in a symmetrized 4-level toy model and a regular optical molasses force profile (red graph and axes). BCF profile was obtained for $\Delta_{\pm}=\mp 15\,\Gamma$, $\delta=100\,\Gamma$, $\Omega_0=\sqrt{3/2}\delta$ and $\chi_{\pm}=\pm 45^{\circ}$ and was smoothed using moving average filter, while the radiative force profile was drawn for $I=I_{\mathrm{sat}}$ and detunings $\delta_{\pm}=\pm 1\,\Gamma$. Both scales differ by a factor of 20. While in the case of radiative forces, the slope is about 0.44 $\hbar k^2/2$ for given parameters, in the BCF force profile it is close to 7 $\hbar k^2/2$, a factor of 16 higher.}
\label{fig:BCF_rad_mol}
\end{figure}

 As shown in Fig. \ref{fig:full_bcf_mol}, the resulting force profile shows remarkably strong forces and high capture velocities - the force peaks at around 35 $\hbar k\Gamma/2$ at velocity of about $\pm 20\,\Gamma/k$, while at the same time the slope around zero velocity, representing velocity damping coefficient, is quite linear and steep. We can benchmark this force vs velocity curve against normal optical molasses realized with radiative forces. The comparison is shown in Fig. \ref{fig:BCF_rad_mol} after smoothing the BCF-induced force profile with a moving average filter in order to smooth out sharp spikes arising from multiphoton reasonances. Qualitatively, the obtained force profile perfectly resembles the radiative Doppler molasses, but on a much bigger scale: the slope near zero velocity is 16 times higher compared to radiative force in Fig. \ref{fig:BCF_rad_mol}, peak forces are substantially bigger, and so are capture velocities.

\begin{figure}[!h]
\centering
\includegraphics[width=0.7\textwidth]{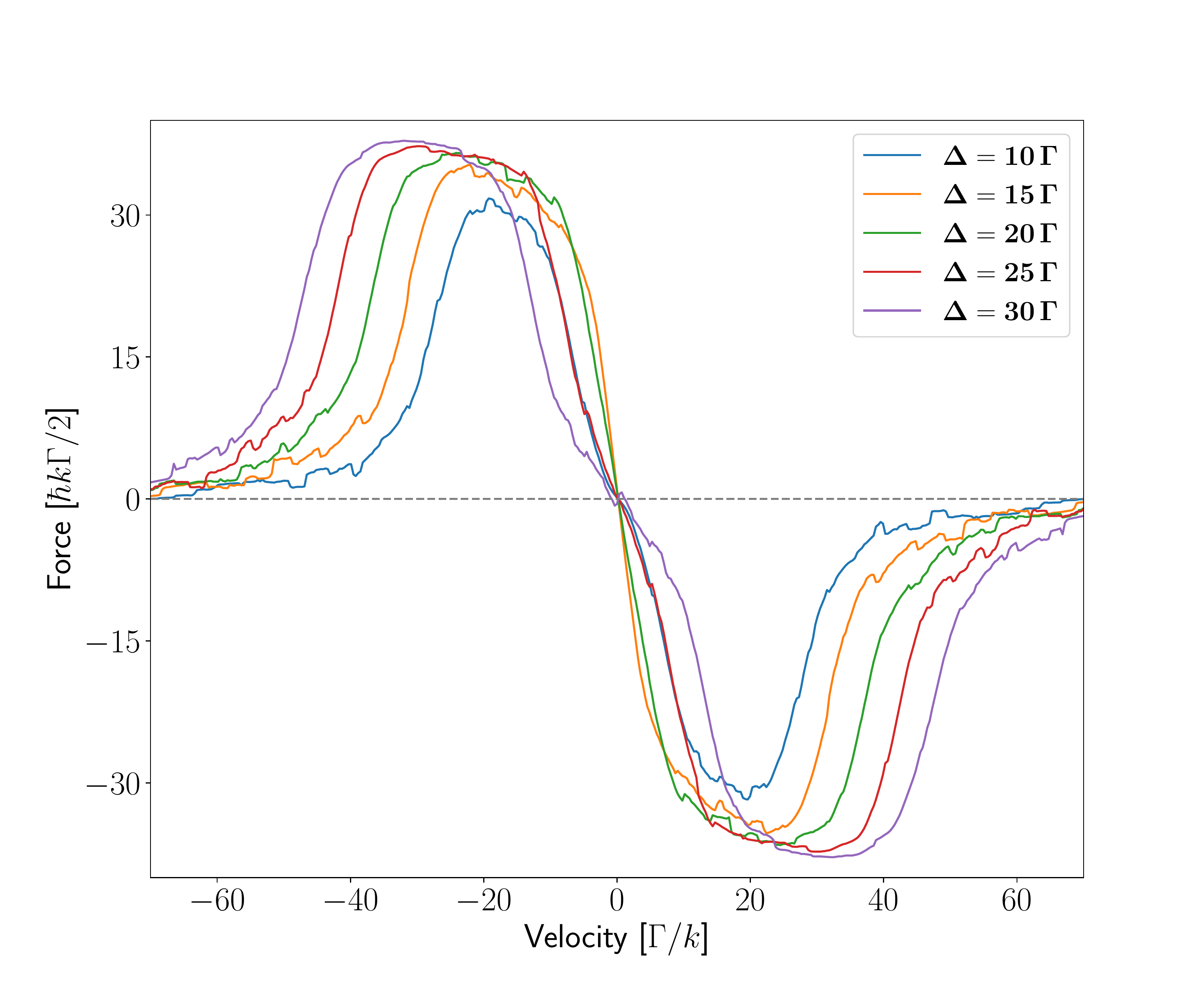}
\caption{\small BCF molasses force profiles in a symmetrized BaH toy model after applying moving average filter. Profiles were obtained for $\delta=100\,\Gamma$, $\Omega_0=\sqrt{3/2}\delta$ and $\chi_{\pm}=\pm 45^{\circ}$.}
\label{fig:BCF_mol_comp}
\end{figure}

\par
\begin{figure}[!h]
\hspace*{-2.5cm}
\begin{tabular}{c c}
\subfloat[\small $\Delta=\pm 15\,\Gamma$.]{{
\includegraphics[scale=0.25]{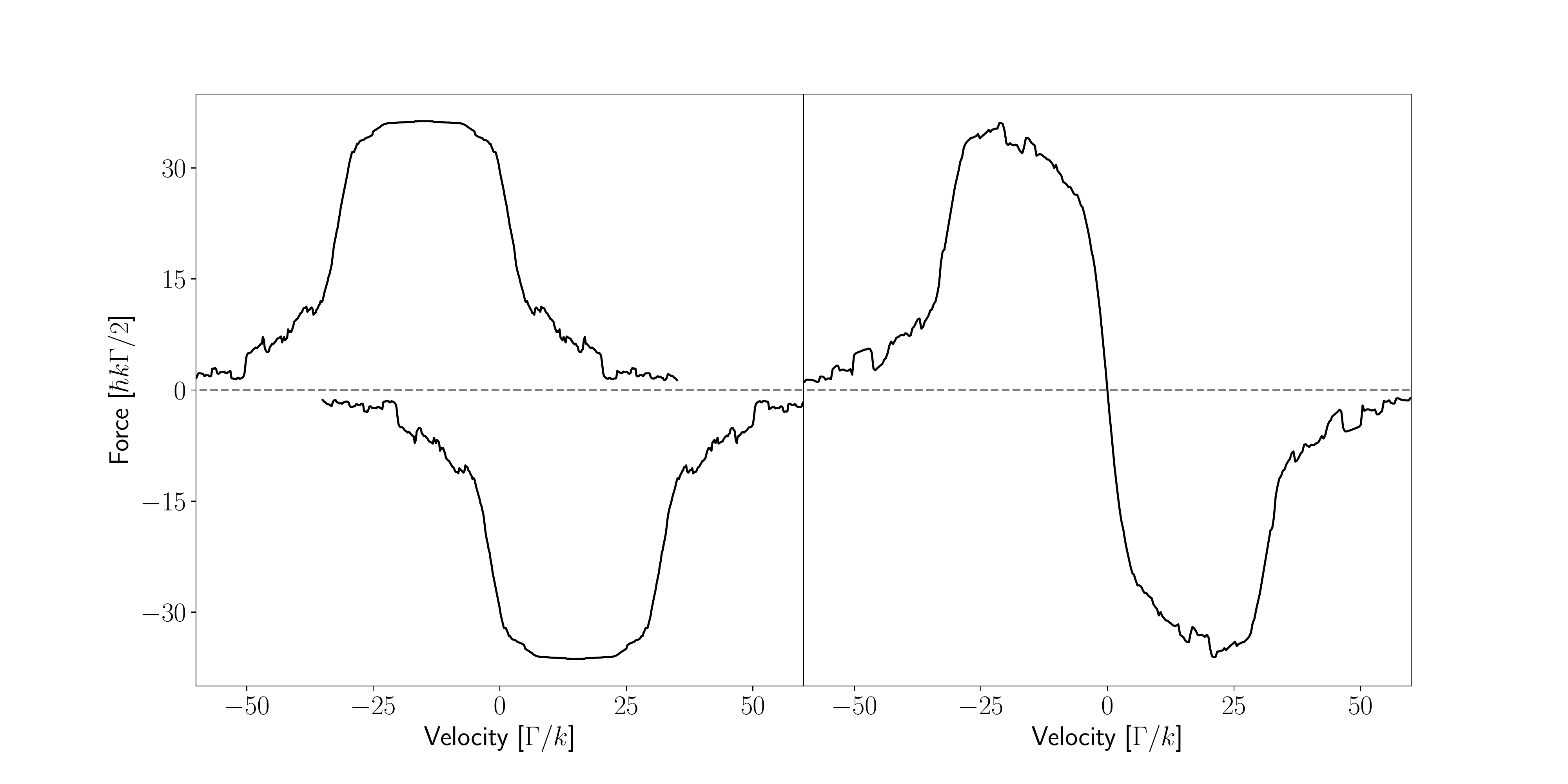}
}}
&
\subfloat[\small $\Delta=\pm 20\,\Gamma$.]{{
\includegraphics[scale=0.25]{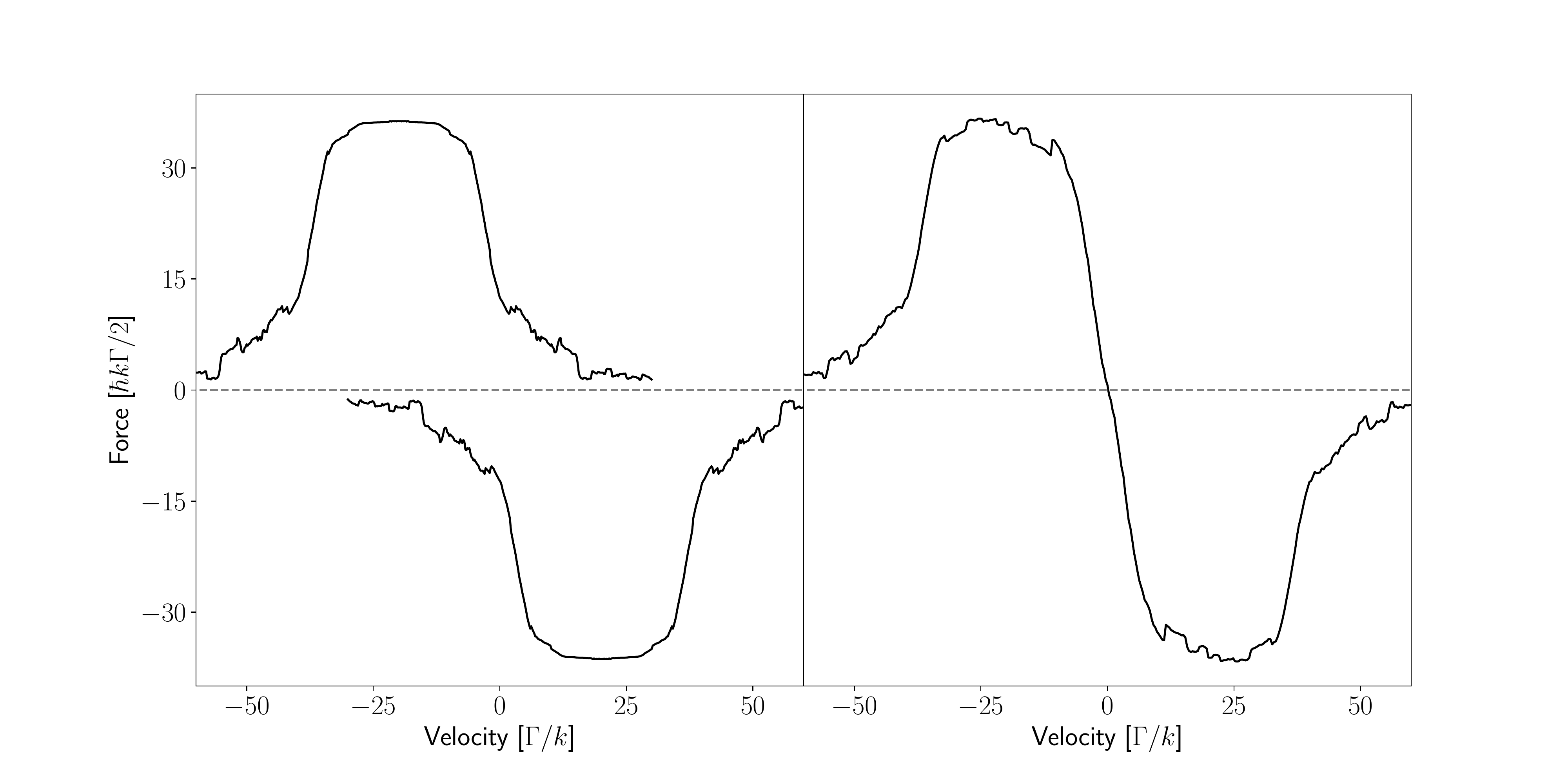}
}}
\\
\multicolumn{2}{c}{
\subfloat[\small$\Delta=\pm 30\,\Gamma$.]{{
\includegraphics[scale=0.25]{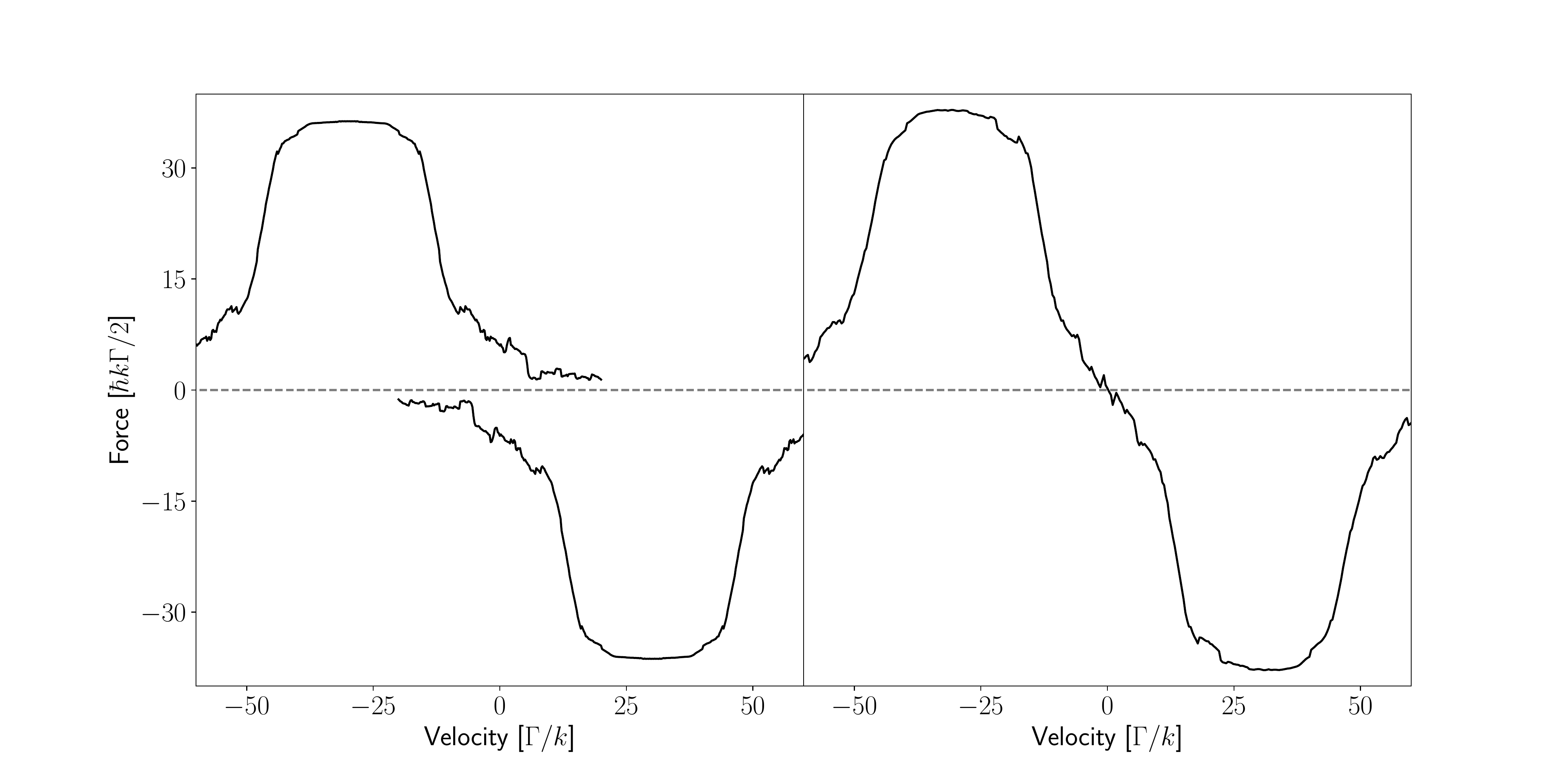}
}}
}
\end{tabular}
\caption{\small Two opposite BCF profiles (left panels) compared with force profiles obtained from a full calculation performed for the 4-level system (right panels). Results are smoothed with a moving average filter in order to highlight the overall features. All profiles were obtained for $\delta=100\,\Gamma$, $\Omega_0=\sqrt{3/2}\delta$ and $\chi_{\pm}=\pm 45^{\circ}$. }
\label{fig:bcf_vs_molasses}

\end{figure}

 Rapid adjustment of the shape and magnitude of the cooling profile is easily achievable experimentally. Capture velocity and the damping force are modified when changing profile-shifting detuning $\Delta$ (see Fig. \ref{fig:bcf_schematic}), which can be controlled by an AOM, by changing its RF drive frequency. The magnitude of the force depends on the choice of BCF detuning $\delta$ and the Rabi rate $\Omega$ for each two-level system shown in Fig. \ref{fig:BCF_BaH_diagram}. In Fig. \ref{fig:BCF_mol_comp} we present comparison of smoothed force profiles for $\Delta$ ranging from 10 $\Gamma$ to 30 $\Gamma$. These resulting force-velocity profiles are effectively created by a sum of two opposing, shifted and re-scaled 2-level BCF profiles, as shown in Fig. \ref{fig:bcf_vs_molasses}. The scaling effect appears there, because in our 4-level model the atom or molecule spends on average less time in either of the 2-level subsystem interacting with their own respective bichromatic fields than in a simple situation of a BCF-driven 2-level system. In fact, using the $\pi$-pulse approach (described in detail in Appendix \ref{sec:App-B}), we can predict that for our toy model such factor will be exactly equal to 4/7, and so the peak forces expected for any detuning $\delta$ are $F_{\mathrm{BCF,mol}}=4/7\ \hbar k\delta/\pi$. However, this scaling factor has some limitations. In case of fields with more than 2-colors, the direct solution of 4-level system yields forces higher than one would obtain by re-scaling a 2-level system solution, thus reaffirming the necessity to perform a full calculation with all levels and laser frequencies as done here rather than using scaled analytical results from isolated two-level systems.

\par
\begin{figure}[!h]
\begin{minipage}[c][10cm][t]{0.49\textwidth}
  \vspace*{\fill}
  \centering
  \includegraphics[width=0.95\textwidth]{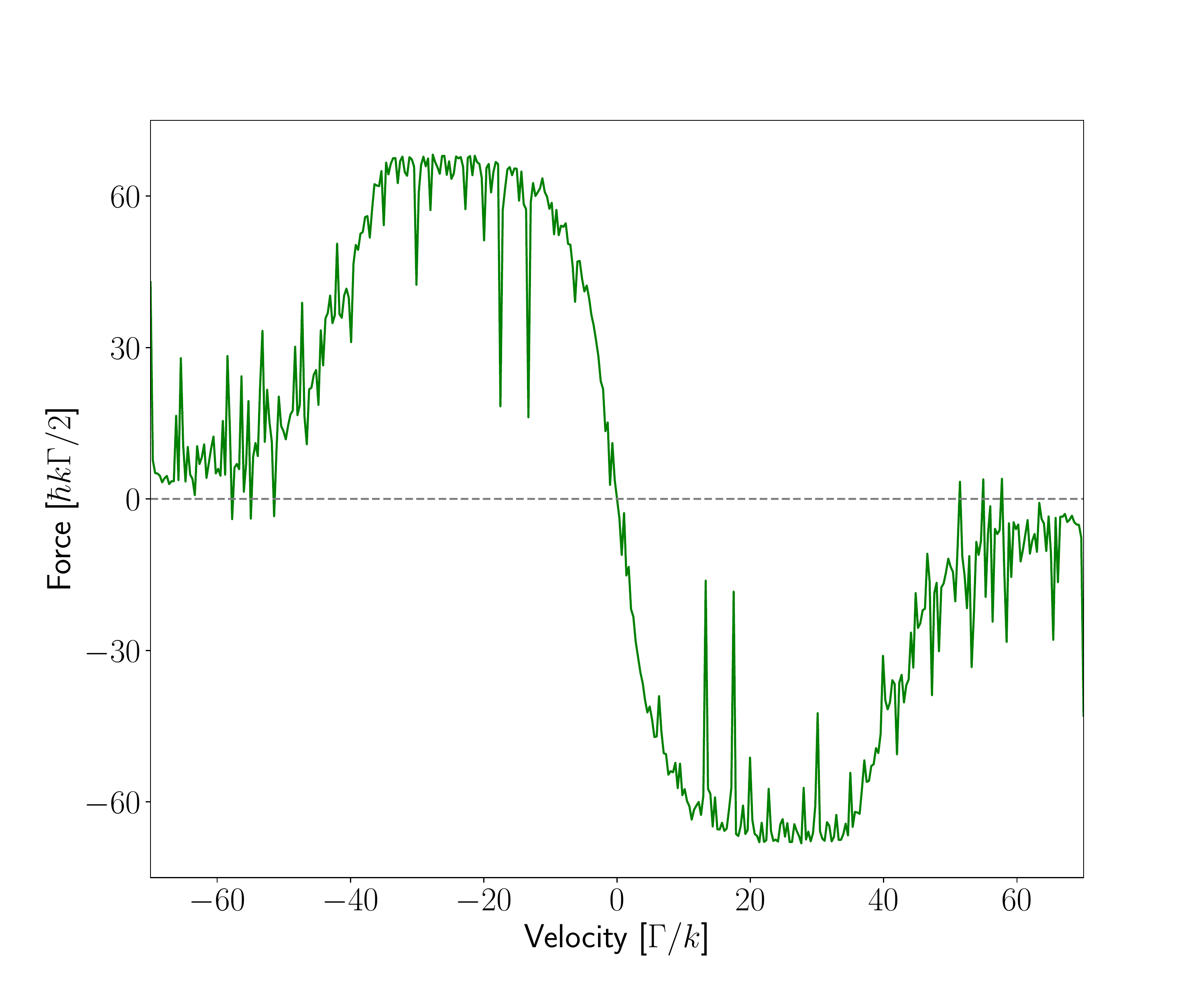}
\caption{\small 4-color molasses force profile in a symmetrized BaH toy model. Profile was obtained for $\Delta_{\pm}=\mp 15\,\Gamma$, $\delta=100\,\Gamma$, $\Omega_0=1.16\,\delta$ and $\chi_{\pm}=\pm 25^{\circ}$.}
\label{fig:TCF_mol_20}
\end{minipage}
\begin{minipage}[c][10cm][t]{0.49\textwidth}
  \vspace*{\fill}
  \centering
  \includegraphics[width=0.95\textwidth]{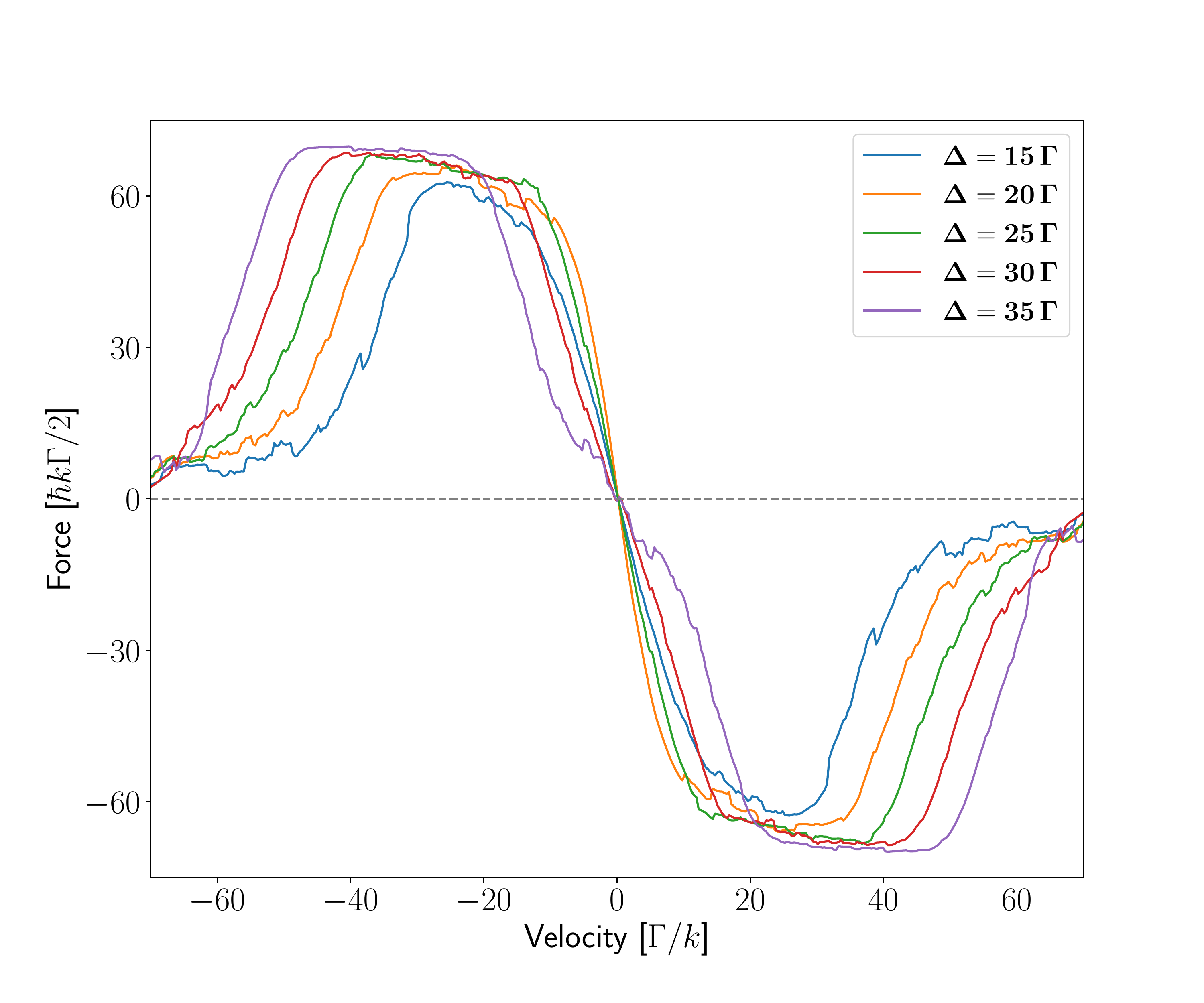}
\caption{\small 4-color molasses force profiles in a symmetrized BaH toy model after applying moving average filter. Profiles were obtained for $\delta=100\,\Gamma$, $\Omega_0=1.16\, \delta$ and $\chi_{\pm}=\pm 25^{\circ}$.}
\label{fig:TCF_mol_comp}
\end{minipage}
\end{figure}

 The force profile can also be obtained for any $2n$-color forces, though it is not immediately obvious what the benefits of adding additional frequencies are. If we assume that we always operate with a certain power per frequency component, in a simple 2-level system moving from 2- to 4-color laser fields increases the force and velocity range quite substantially while decreasing the time spent in the excited state \cite{Galica2013}. Indeed, as shown in Fig. \ref{fig:TCF_mol_20}, the 4-color profile does not look that much different from the BCF profile. However, the peak force is much larger and remains such for higher velocities, which could result in higher capture velocity. In Fig. \ref{fig:TCF_mol_comp} we compare 4-color force profiles for various shifts $\Delta$. We found the highest peak forces for $\abs{\chi}=25^{\circ}$ and $\Omega=1.16\,\delta$, where $\Omega$ is rate of every component.

\par
\begin{figure}[!h]
\begin{minipage}[c][11cm][t]{0.49\textwidth}
  \vspace*{0.4cm}
  \centering
\includegraphics[width=0.95\textwidth]{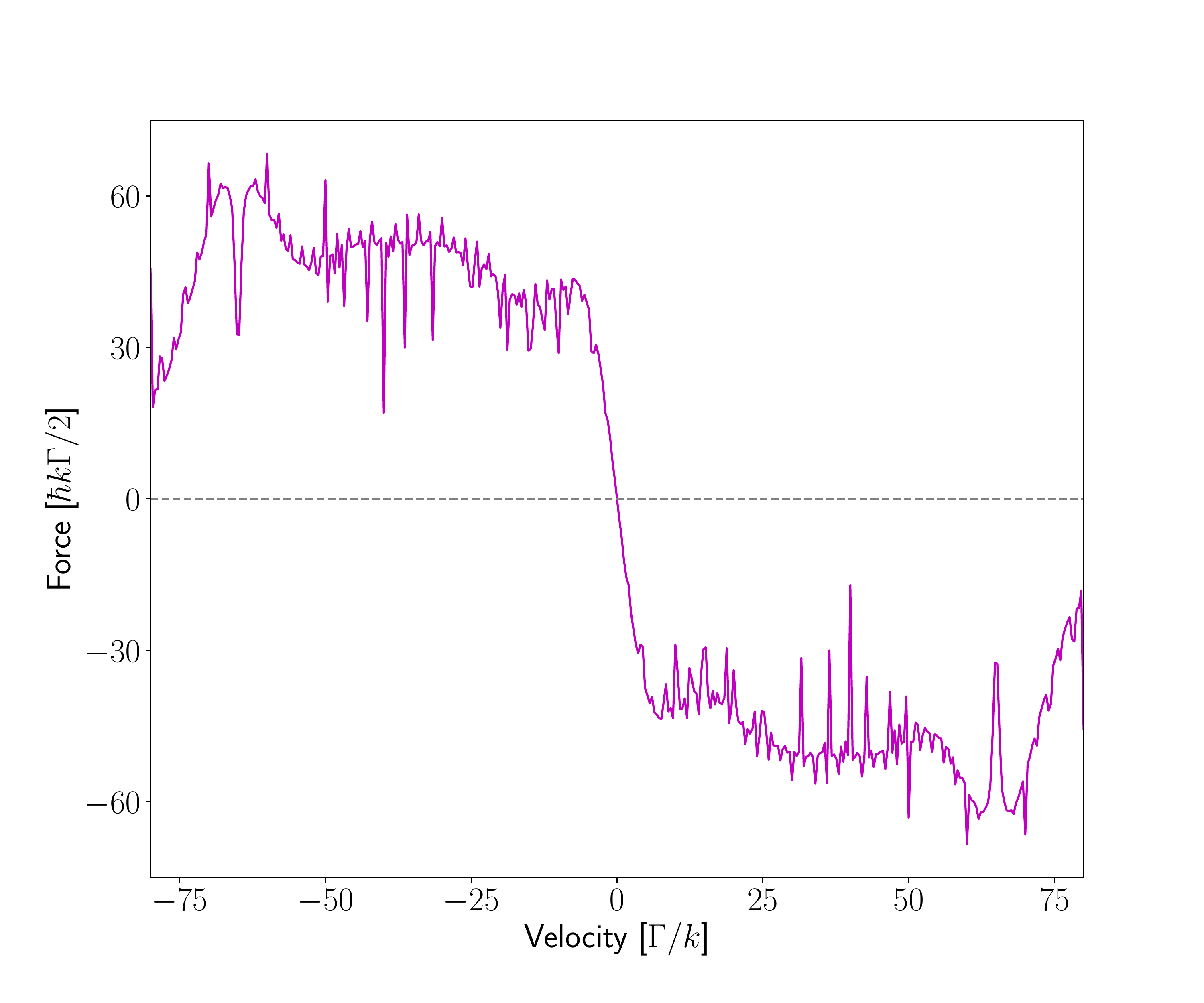}
\caption{\small Wide 4-color molasses force profile in a symmetrized BaH toy model. Profile was obtained for $\Delta_{\pm}=\mp 35\,\Gamma$, $\delta=100\,\Gamma$, $\Omega_0=\,\delta$ and $\chi_{\pm}=\pm 30^{\circ}$. }
\label{fig:TCF_mol_wide}
\end{minipage}
\begin{minipage}[c][11cm][t]{0.49\textwidth}
  \vspace*{\fill}
  \centering
\includegraphics[width=0.95\textwidth]{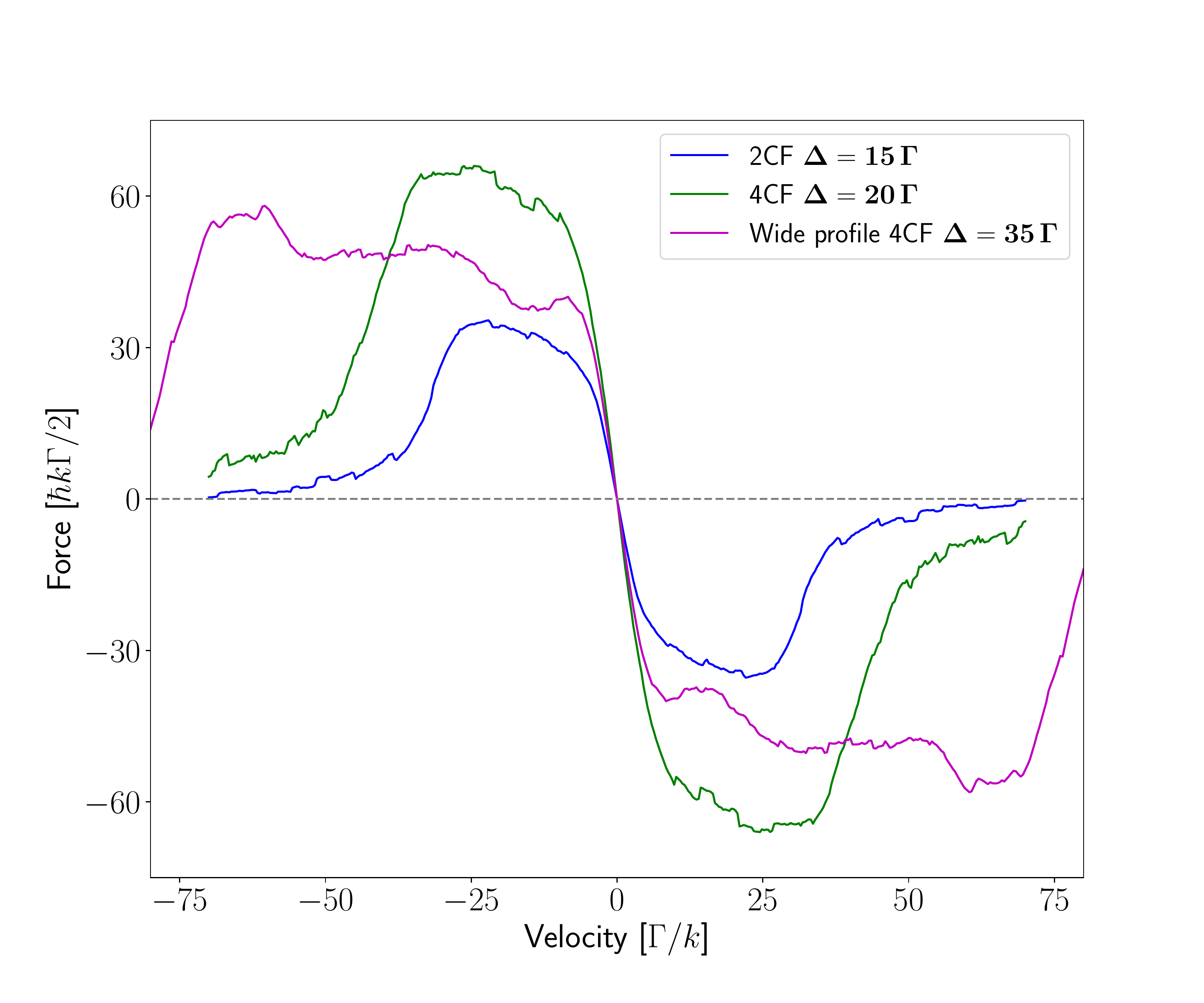}
\caption{\small Comparison of molasses force profiles in a symmetrized 4-level system. By adding colors and appropriately changing the parameters, profiles with different characteristics can be created.}
\label{fig:3_mol_comp}
\end{minipage}
\end{figure}

 The 4-color force profiles can be quite strongly adjusted by appropriately changing the phase $\chi$, which experimentally is controlled by a relative length of the optical delay line between the counter-propagating multi-frequency laser beams of the same color (see Fig. \ref{fig:bcf_schematic}), and Rabi rate $\Omega$. In Fig. \ref{fig:TCF_mol_wide} we show a much wider profile with capture velocities as high as $60\,\Gamma/k$ (which for molecule like BaH is equivalent to $\approx\ $72 m/s) and forces of the order of $50\, \hbar k\Gamma/2$ that was obtained for $\abs{\chi}=30^{\circ}$ and $\Omega=\delta$. Finally, we present comparison of 2-color molasses profile, narrow 4-color force profile and wide 4-color force profile in Fig. \ref{fig:3_mol_comp}. All these SupER molasses have high capture velocities, high peak forces and steep slopes ranging from 7 $\hbar k^2/2$ for the BCF to 9.5 $\hbar k^2/2$ for the narrow 4-color profile, enabling rapid damping of molecular motion.

\newpage

\section{Temperature in super molasses}

\par Having developed a mathematical approach to the $\pi$-pulse model of the polychromatic force dynamics using continuous-time Markov chains (CTMC) (App. \ref{sec:Variance-estimation}), we are now able to estimate the final temperature of atoms or molecules under the influence of SupER molasses. The diffusion coefficient, which determines the limiting temperature, should not only include terms related to spontaneous emission from the excited states, but also the term specific to this model, which is related to the uncertainty in ``position'' in the excitation-stimulated emission cycle. This term is effectively the variance in momentum transfer from the optical field to the atom or molecule. In our model, the system quickly relaxes to a steady state that's achieved at long times $t\gg \frac{1}{\Gamma_{12}+\Gamma_{21}}$ and the variance term can be written as (App. \ref{sec:App-B}):
\begin{displaymath}
\mathrm{Var}\,p_{PCF}=\hbar^2k^2\frac{\delta^2}{\pi^2\Gamma}\sigma^2(\varepsilon)\,t,
\end{displaymath}

\noindent where $\sigma^2(\varepsilon)$ is dependent on fraction of time $\varepsilon$ spent in the excited state per cycle (it is \emph{not} the ensemble average excited state population $\rho_{ee}$ appearing in the density matrix and it is discussed in Appendix \ref{sec:Variance-estimation} as well as in Ref. \cite{Chieda2012thesis}). Apart from the contribution from PCF, total variance of the momentum transfer will have a contribution arising from the spontaneous decay as well:
\begin{displaymath}
\mathrm{Var}\,p=\hbar^2k^2\frac{\delta^2}{\pi^2\Gamma}\sigma^2(\varepsilon)t+\frac{1}{3}\hbar^2k^2\rho_{ee}\Gamma\, t,
\end{displaymath}

\noindent with $\rho_{ee}$ being the time-averaged excited state population (in case of the 4-level system we have shown, $\rho_{ee}=\rho_{E1}+\rho_{E2}$). We should also note that the variance shown here is variance of momentum transfer in one dimension and is the reason behind the $1/3$ factor in the second term on the right hand side in the equation above. 
\par Obtained result shows a quadratic dependence on $\hbar k \delta$ which is consistent with results experimentally shown in Ref. \cite{Partlow2004} and first estimated in Ref. \cite{Dalibard1985} for dipole forces. In case of the bichromatic force used in the helium collimation experiment, authors obtained diffusion coefficient of $\approx  (\hbar k \delta)^2/2\Gamma$ \cite{Partlow2004}, while the value we obtain from our model is $\approx 0.6 (\hbar k \delta)^2/\Gamma$. 
\par From the calculated variance we can obtain the diffusion coefficient $D$:
\begin{equation}
D\equiv\frac{1}{2}\frac{d}{dt}p^2(t)=\frac{1}{2}\left(\hbar^2k^2\frac{\delta^2}{\pi^2\Gamma}\sigma^2(\varepsilon)+\frac{1}{3}\hbar^2k^2\rho_{ee}\Gamma\right), \label{eqn:diffusin_coefficient}
\end{equation}

\noindent which is then related to the limiting temperature $T_L$. The PCF molasses force around $v=0$ is linear and can simply be written as $F=-\beta v$, where $-\beta$ is the slope. Using the definition of the diffusion coefficient and by assuming mass of the atom or molecule is $M$, at equilibrium one can write:
\begin{displaymath}
Mv^2=\frac{D}{\beta}.
\end{displaymath}

\noindent By associating the limiting temperature $T_L$ with kinetic energy, i.e. $Mv^2=k_B T_L$ and using the definition of Doppler temperature $T_D=\hbar \Gamma/2k_B$, we get:
\begin{equation}
T_L=2T_D\frac{\hbar k^2}{2\beta}\left(\frac{1}{3}\rho_{ee}+\frac{\delta^2}{\pi^2\Gamma^2}\sigma^2(\varepsilon)\right). \label{eqn:limit_temp}
\end{equation}

\noindent In the equation above the parameter $\beta$ is  measured in natural units for this problem - $(\hbar k\Gamma/2)/(\Gamma/k)=\hbar k^2/2$. Because $\sigma^2(\varepsilon)$ is on the order of 10, in Eq.(\ref{eqn:limit_temp}) the spontaneous emission term is negligible given detuning $\delta$ typical for this problem. Slope $\beta$ that appears in this equation can be estimated (following \cite{Partlow2004}) to be $\beta\sim \delta/4\pi\Gamma\ \hbar k^2/2$ for presented profile, which leads to $T_L\sim 8\,\sigma^2(\varepsilon)\,\delta/\pi\,\ \ T_D$. For example, for BCF molasses presented here $\sigma^2(\varepsilon_{BCF})\approx 20$ and $\delta=100\,\Gamma$, so $T_L\approx 5.09\times 10^3\ T_D$, which for BaH is about $137\,$mK. For a more realistic and asymmetric system $\sigma^2(\varepsilon_{BCF})\approx 21$ and so, given the slope of similar magnitude, the final temperature would be almost exactly the same. The linear dependence of $T_L$ on the detuning shows that it might be beneficial to keep $\delta$ as small as possible, while still keeping $\delta\gg\Gamma$ condition fulfilled. 
\par While the estimated limiting temperature appears high compared to the Doppler limit, it can actually be made lower. The temperature at equilibrium in such system is not only determined by the slope of the profile, but also by how strongly the polychromatic forces act on the atom or molecule around zero velocity. In the derivation of the formula for variance of momentum transfer (App. \ref{sec:App-B}) we assumed that in every state considered the force acting on an atom or molecule is $2\hbar k\delta/\pi$. However, around zero velocity that does not need to be the case. Assuming that the force around $v=0$ is equal to $F_0$, we can simply substitute $\hbar k\delta/\pi\sim F_0$. If we removed all the constants appearing naturally in both $\beta$ and $F_0$, we'd simply obtain:
\begin{equation}
T_L\sim T_D\frac{F_0^2}{\beta}\sigma^2(\varepsilon), \label{eqn:limit_temp2}
\end{equation}

\noindent clearly showing that the final temperature mainly depends on the ratio $F^2_0/\beta$. This shows us that effectively, depending on how the basic 2-level system profiles (Fig. \ref{fig:bcf_vs_molasses}) are aligned to create a full molasses-like forces, we obtain different values of $\beta$ and $F_0$. For small shifts $\Delta$ both force and slope are high as shown in Fig. \ref{fig:bcf_vs_molasses}(a). If we make our detuning too high, like in Fig. \ref{fig:bcf_vs_molasses}(c), the force around zero velocity becomes small, but because the profiles are far from the center, the slope of the effective profile is very small as well. 
\par In between there should exist an optimal configuration, where slopes of single PCF profiles are each other's continuation. There, the slope should remain high, while the force $F_0$ should be relatively small and Fig. \ref{fig:bcf_vs_molasses}(b) depicts such situation. In this configuration the smallest temperature ought to be achieved. Given that width of BCF profiles is approximately $\delta/2k$, we should expect that the most optimal cooling forces will appear around $\Delta_{\pm}=\mp \delta/4k$. To confirm these estimates, We analyze the effective forces, damping coefficients and limiting temperature $T_L$ using Monte Carlo simulations of an exact realization of the 4-state model described in App. \ref{sec:App-B}.

\section{Monte Carlo Simulations \label{sec:MC-simulations}}

\par To prove that the estimated temperature follows the model we have created we present results of a Monte Carlo simulation of an ideal 4-level toy model for a 2-color light field. In it we assumed that the system consists of four states (described in App. \ref{sec:App-B}). While in the $\pi$-pulse model we assumed that the force in every state is $2\hbar k\delta/\pi$, here we assume it is velocity-dependent and follows a typical BCF force profile, re-scaled in such a way that at maximum it is equal to the mentioned $2\hbar k\delta/\pi$ value. Using such values results in effective force profiles seen in figures provided before. It is important to note that the temperature model and obtained formulas presented in App. \ref{sec:Variance-estimation} and \ref{sec:App-B} work in the regime where the interaction can actually be described in the framework of a $\pi$-pulse model. This occurs for specific parameters (such as $\chi=45^{\circ}$ for 2-color fields) and for velocities, for which the force is at its maximum. The model might not hold at the edges of force-velocity profiles ($|v-v_0|>\delta/2k$), which is the regime we expect to be in in the cooling process ($v\sim 0$). Therefore, the proportionality constant in Eq. \ref{eqn:limit_temp2} could be difficult to predict, given that $\sigma^2(\varepsilon)$ stems from the near-perfect $\pi$-pulse behavior. Additionally, the slope $\beta$ that appears in the formula might not be the effective slope that can be read directly from provided effective force-velocity profiles.

\par Occurrence of transitions between different states in the simulations was assumed to follow Poissonian statistic with rates given in the appendix \ref{sec:Variance-estimation}. We have also included recoils from spontaneous emission events even though they play a very limited role. The force profiles used were re-scaled versions of those seen in Fig. \ref{fig:bcf_vs_molasses}(a) and Fig. \ref{fig:bcf_vs_molasses}(b), which should show a typical (the former) and a perfect (the latter) configuration for the molasses. In the simulation these profiles were separate and, like in mentioned figures, already smoothed with a moving average filter. We have also assumed that the molecule undergoing the cooling process is BaH with $\Gamma= 2\pi\times 1.15$ MHz, $M=139$ u and $k=2\pi/1060.7867$ nm. Experimentally, to obtain these profiles for our test species the lasers would have to be detuned from the resonance by $\delta\approx 115$ MHz, and, assuming beams with uniform power density and diameter of 5 mm, have a total power of $\sim 4.4$ W.

\par The effective damping coefficients (slope) that we expect for $\Delta=\pm 15\,\Gamma$ are $\beta\approx 4.9\, \hbar k^2/2$ and $\beta\approx 3.1\, \hbar k^2/2$ for $\Delta=\pm 20\,\Gamma$, while the force around $v=0$ is expected to be $F_0\approx 103.4\, \hbar k \Gamma/2$ and $F_0\approx 42.4\, \hbar k \Gamma/2$ respectively. If the developed model holds in both described regimes, we expect that the limiting temperature will be (in the natural units):
\begin{equation}
T_L= T_D\frac{F_0^2}{\beta}\frac{\sigma^2(\varepsilon)}{8}, \label{eqn:limit_temp3}
\end{equation}

\noindent which should lead to $T_L\approx 147$ mK for $\Delta=\pm 15\,\Gamma$ and $T_L\approx 39$ mK for $\Delta=\pm 20\,\Gamma$. Eq. \ref{eqn:limit_temp3} can also be written using the notation used in the appendix as:
\begin{equation}
T_L= T_D\frac{F_{\mathrm{IV}}^2}{\beta}\frac{\sigma_{\mathrm{IV}}^2(\varepsilon)}{2\mu^2_{\mathrm{IV}(\varepsilon)}}, \label{eqn:limit_temp4}
\end{equation}

\noindent where $F_{\mathrm{IV}}$ is already the effective and re-scaled force around $v=0$ that is obtained in a 4-state model (App. \ref{sec:App-B}) and that can be read directly from the presented force profiles, $\sigma_{\mathrm{IV}}^2(\varepsilon)=\sigma^2(\varepsilon)/2$ and is equal to $10.06$ for BCF in a symmetric system, and $\mu_{\mathrm{IV}}(\varepsilon)$ is the re-scaling factor equal to $4/7$ for BCF in such system.

\par
\begin{figure}[!h]
\begin{minipage}[c][14cm][t]{0.49\textwidth}
  \vspace*{\fill}
  \centering
  \includegraphics[scale=0.22]{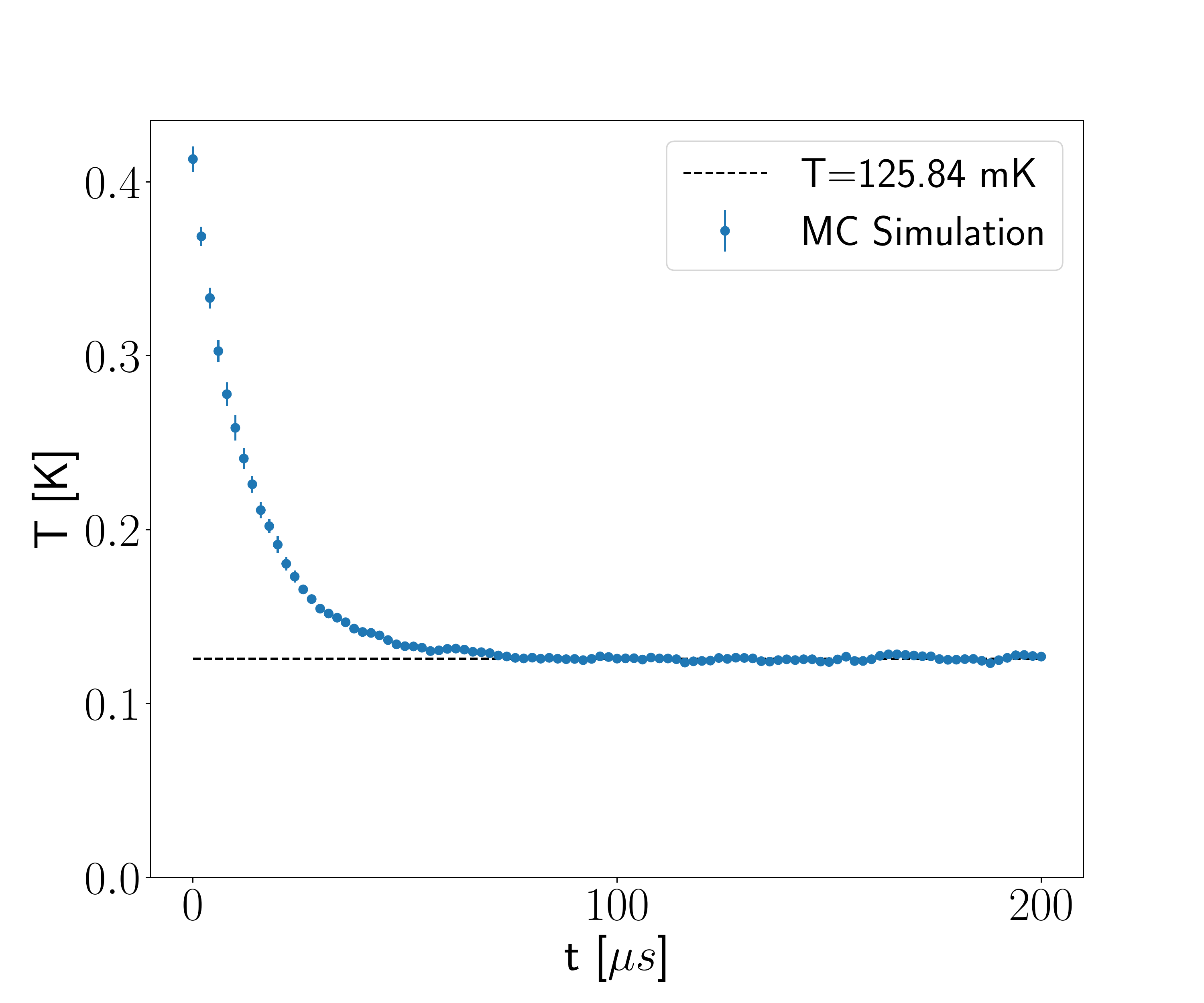}
  \label{fig:highTdecay}\par
  \includegraphics[scale=0.22]{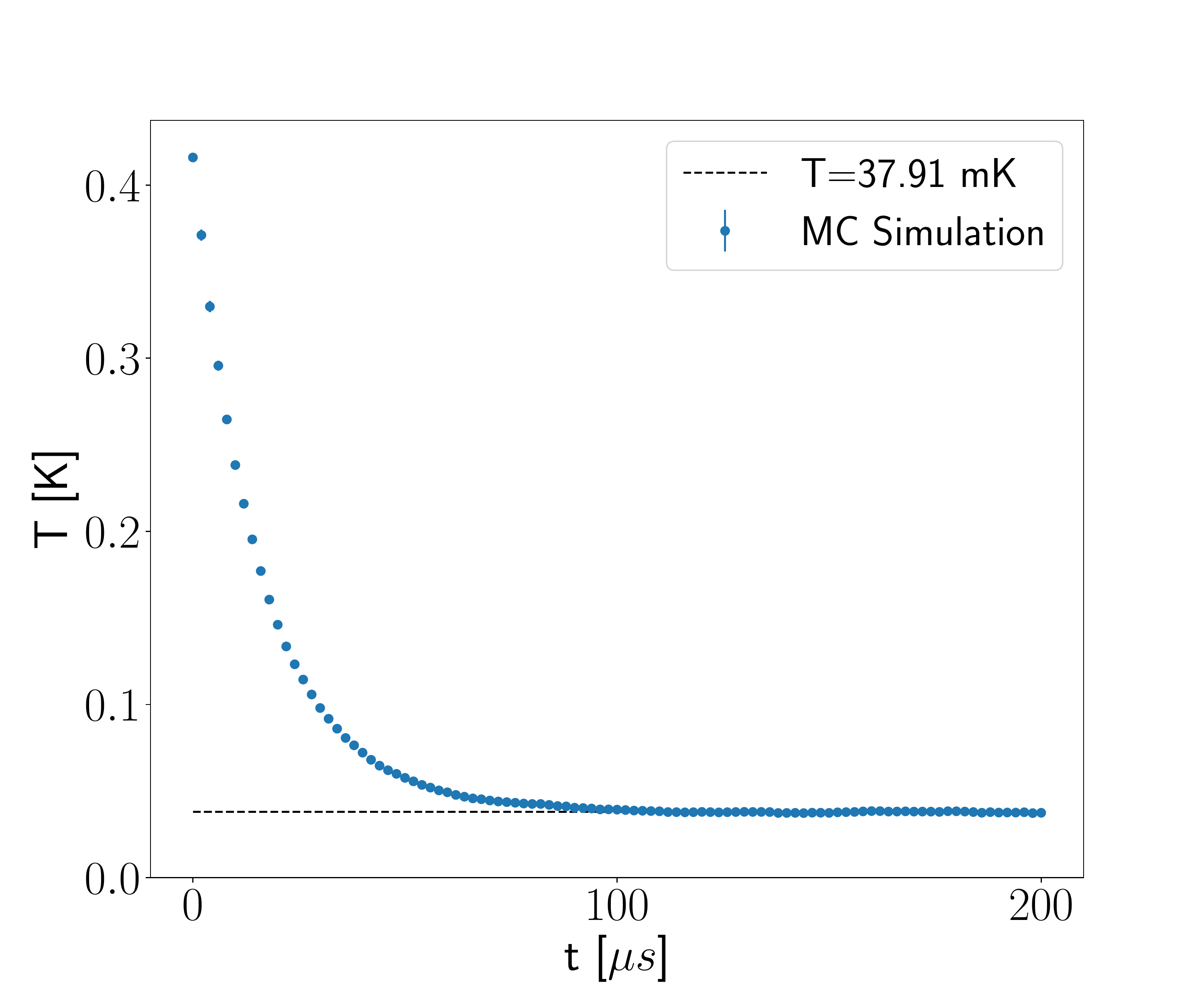}
  \caption{\small Temperature decay for $\Delta=\pm 15\,\Gamma$ molasses reaching $T_L=125.9$ mK (top) and for $\Delta=\pm 20\,\Gamma$ molasses reaching $T_L=37.9$ mK (bottom).}
  \label{fig:lowTdecay}
\end{minipage}
\begin{minipage}[c][14cm][t]{0.49\textwidth}
  \vspace*{\fill}
  \centering
  \includegraphics[width=0.974\textwidth]{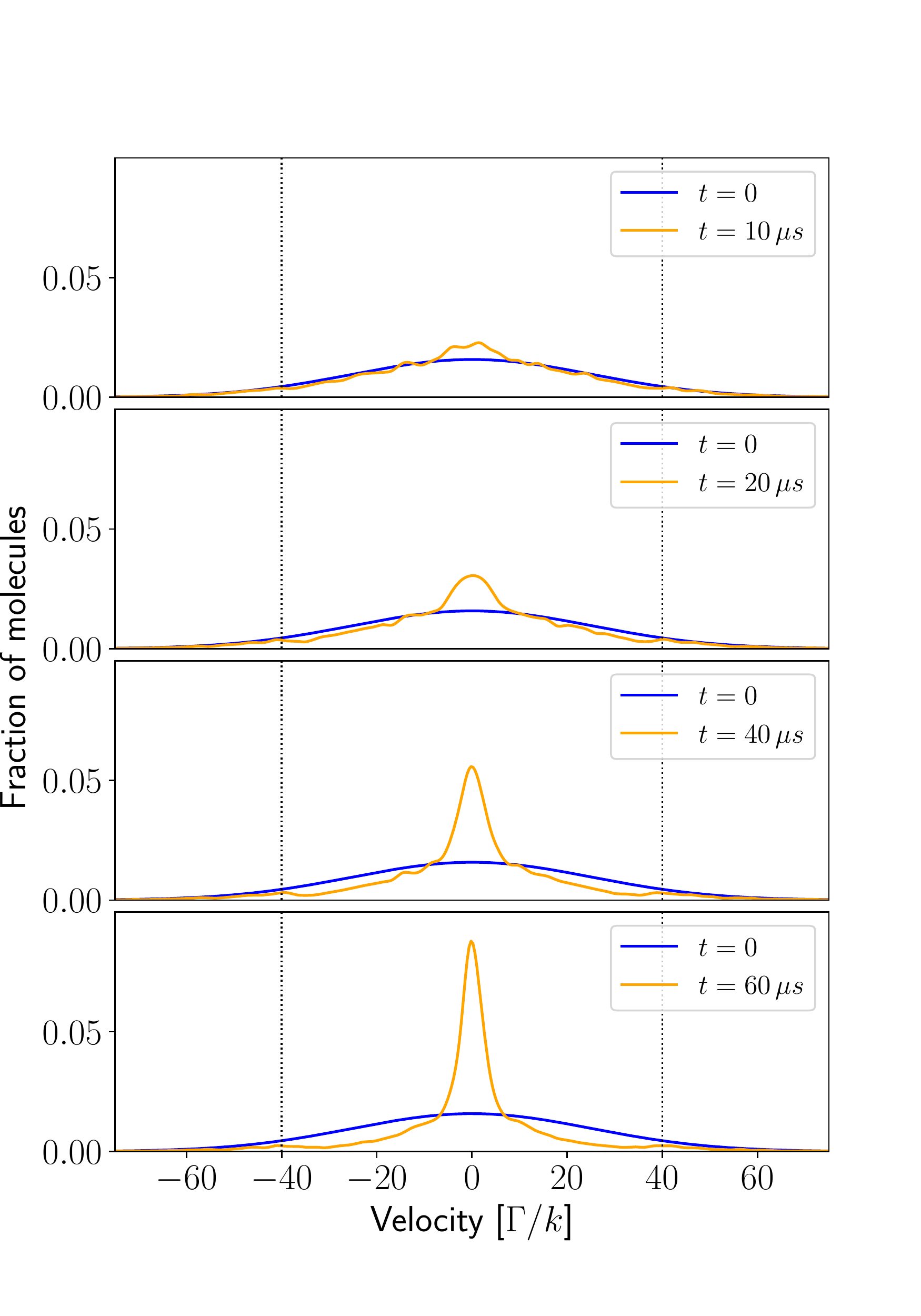}
  \caption{\small Velocity profiles after $10$, $20$, $40$ and $60\,\mu s$ showing changes experienced in $\Delta=\pm 20\,\Gamma$ BCF molasses. Dotted lines mark approximate capture velocity of $40\,\Gamma/k$.}
  \label{fig:capture_velocity}
\end{minipage}
\end{figure}

 In the simulation we started with molecules distributed with $\sigma_v=5\,$m/s , which for BaH corresponds to temperature of approximately 0.4 K. The simulation was run in steps of 5 ns for a total time of $200\,\mu$s, which was more than enough to reach the final limiting temperature for both considered SupER molasses configurations. Figure \ref{fig:lowTdecay} shows molecular temperature at different times averaged over multiple simulations for $\Delta=\pm 15\,\Gamma$ molasses (top frame) and for $\Delta=\pm 20\,\Gamma$ molasses (bottom frame). The limiting temperatures obtained were $125.84$ mK and $37.91$ mK respectively, relatively close to the CTMC model estimates. The model was also proven to be correct by the steady state populations which quickly relaxed to predicted levels of $\eta_{C_1}=\eta_{C_2}=11/28$ and $\eta_{W_1}=\eta_{W_2}=3/28$ (App. \ref{sec:App-B}). Assuming that the temperature follows an exponential decay curve, i.e. $T(t)=T_L+(T_0-T_L)\exp(-t/\tau_{\beta})$, where $\tau_{\beta}=M/2\beta$ is the characteristic decay time, we can find the actual effective damping rates. For $\Delta=\pm 15\,\Gamma$ we obtained $\tau_{\beta}\approx 13.38\,\mu s$ giving $\beta\approx 4.66\,\hbar k^2/2$, and $\tau_{\beta}\approx 16.27\,\mu s$ leading to $\beta\approx 3.83\,\hbar k^2/2$ for $\Delta=\pm 20\,\Gamma$. Both results are close to provided estimates.

\par To investigate the capture velocity we have also performed simulations for $\Delta=\pm 20\,\Gamma$ molasses with molecules starting with $\sigma_v=30\,$m/s corresponding to $T_0\approx 14.7$ K for BaH. Simulation was performed to reach final time of $200\,\mu s$, which was enough to show the approximate capture velocity. Figure \ref{fig:capture_velocity} shows comparison between velocity distributions at $t=0$ and at times of $t=10,\,20,\,40,\,60\,\mu s$. We can see that the molecules with $\abs{v}\lesssim 40\,\Gamma/k$ ($\sim 47\,$m/s for BaH) accumulate around $v=0$ showing that we can consider this to be the effective capture velocity of the SupER molasses for our parameters.  In general, we could expect the capture velocity to be equal to the typical width of the force vs velocity profile of $\sim \delta/2k$. Note that the simulated cooling time of SupER molasses for BaH agrees with the characteristic BCF timescale of $\tau_{\rm{BCF}}=\pi/\left(4\omega_r\right)\approx100\,\mu$s \cite{metcalf2017colloquium} where $\omega_r\equiv\hbar k^2/\left(2M\right)\approx2\pi\times 1.3\times10^3$ s$^{-1}$ is the recoil frequency for BaH. The value of $\tau_{\rm{BCF}}$ represents the timescale over which a molecules is accelerated across the full velocity range $\sim\delta/k$ and gives an approximate upper bound on the cooling time in the molasses configuration.

\section{Application to Real Molecules \label{sec:real-molecules}}

With the theoretical results established for an ideal 4-level system (the minimum number of states required for the proposed scheme), we relax the simplifying assumptions in order to apply our cooling method to real molecular systems. We develop a general scheme to obtain rotational closure (while satisfying the requirements dictated by the 4 level toy model presented above), at the same time accounting for spin-rotation, fine and nuclear hyperfine structure for molecules with an unpaired electron spin. As we show below, additional internal substructure present in molecular radicals enables generic experimental realization of the SupER molasses cooling scheme.

\begin{figure}[h]
\centering
\includegraphics[width=1 \textwidth]{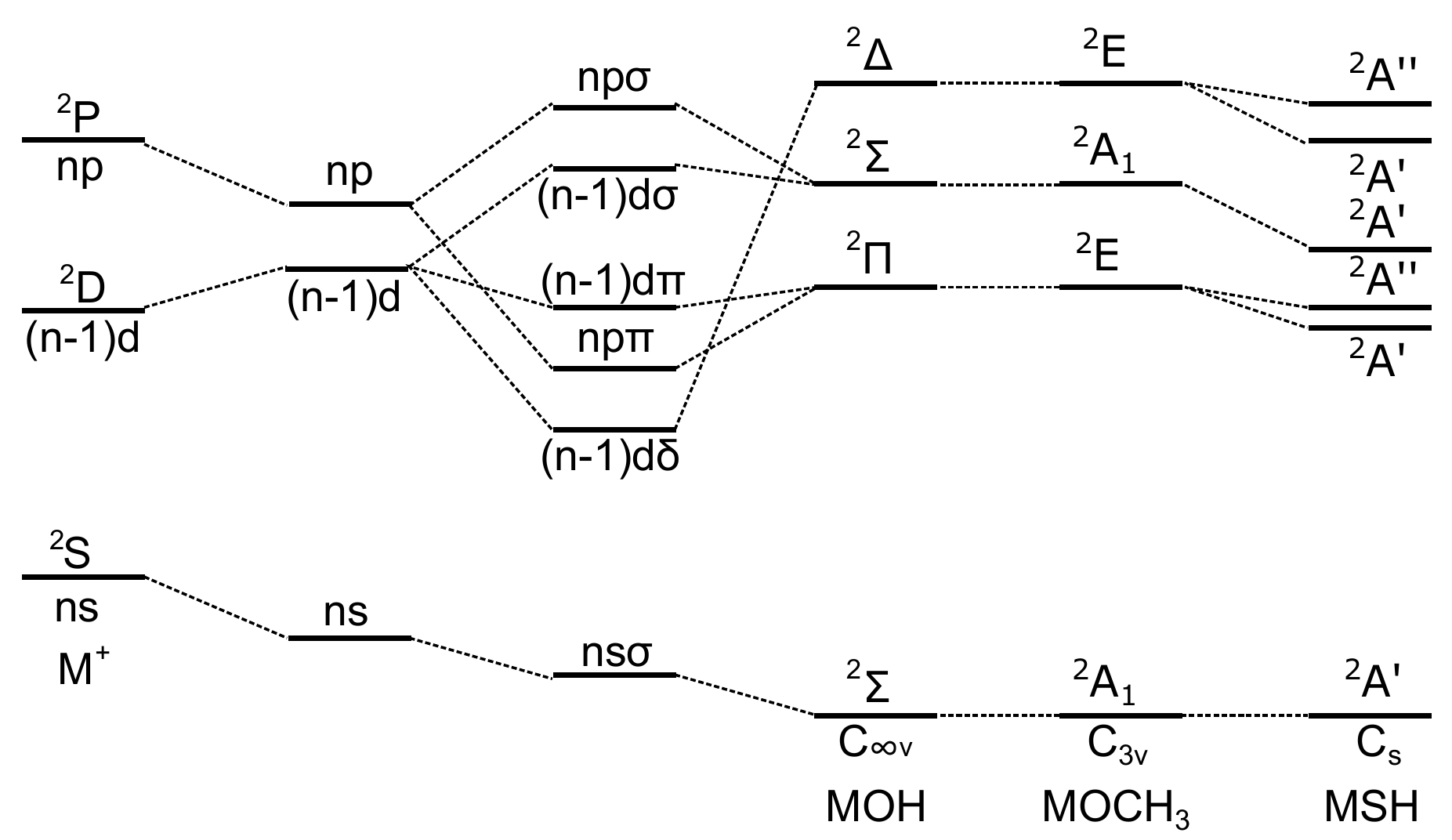}
\caption{\small Energy correlation diagram for alkaline earth metal (M) atoms bonded to ionic ligands (L). As can be seen from the diagram, multiple closely-lying electronic states generically arise for ML molecules with diverse constituents and symmetries. Using the lowest two non-degenerate excited electronic states (e.g. $^2\Pi$ and $^2\Sigma$) for molecules of $C_{\infty v}$ or $^2A'$ and $^2A''$ for $C_S$ molecules will allow application of the molasses methods described here. While we indicated selective ligands in the diagram, other possible constituents include M = Ca, Sr, Ba and L = F, OH, CCH, OCH\textsubscript{3}, CH\textsubscript{3}, SH. For a comprehensive list of ML monovalent derivativies of the alkaline earth metals that have an electronic structure of the type presented here refer to Refs. \cite{Bernath1991,Bernath1997}. }
\label{fig:ML-electronic}
\end{figure}

\subsection{Rotational Level Schemes}

\begin{figure}[!h]
\centering

\begin{tikzpicture}[
scale=0.8, every node/.style={transform shape},
level/.style={thick},
energy/.style={thin,<->,shorten >=1pt,shorten <=1pt,>=stealth},
virtual/.style={densely dashed},
elong/.style={thin, dotted},
radiative/.style={black,->,>=stealth',shorten >=1pt,decorate,decoration={snake,amplitude=1.5}},
trans/.style={thick,<->,shorten >=2pt,shorten <=2pt,>=stealth},
]

\node at (-9cm,-10em) {$X^2\Sigma^+$};
\node at (-9cm,-11.5em) {$N=0,\,J=\frac{1}{2}$};
\node at (-9cm,-13em) {$\mathcal{P}=+1$};

\draw[level] (-4cm,-11em) -- (-7cm,-11em) node[above,pos=0.95] {$F=1$};
\draw[level] (-4cm,-13em) -- (-7cm,-13em) node[below,pos=0.95] {$F=0$};

\draw[elong] (-5.5cm,-12em) -- (0.5cm,-12em);

\draw[energy] (-6.7cm,-13em) -- (-6.7cm,-11em) node[midway,right] {$\Delta^{X\Sigma}_{F_{1/2}}$};

\node at (-9cm,12em) {$B^2\Sigma^+$};
\node at (-9cm,10.5em) {$N=1,\,J=\frac{1}{2}$};
\node at (-9cm,9em) {$\mathcal{P}=-1$};
\node at (-9cm,7.5em) {$\tau_{B\Sigma},\,\Gamma_{B\Sigma}$};

\draw[level] (-4cm,11em) -- (-7cm,11em) node[above,pos=0.95] {$F=1$};
\draw[level] (-4cm,8em) -- (-7cm,8em) node[below,pos=0.95] {$F=0$};

\draw[elong] (-5.5cm,9.5em) -- (0.5cm,9.5em);

\draw[virtual] (-5cm,6em) -- (-2.8cm,6em);
\draw[virtual] (-5cm,13em) -- (-2.8cm,13em);

\draw[energy] (-6.7cm,8em) -- (-6.7cm,11em) node[midway,right] {$\Delta^{B\Sigma}_{F_{1/2}}$};

\draw[radiative] (-5.8cm,7.5em) -- (-8.5cm,1em) node[midway,left] {$1-\mathcal{F}^{B\Sigma}_{00}$};
\draw[radiative] (-5.2cm,9.5em) -- (-5.2cm,-11.9em) node[midway,left] {$1/3\,$};

\draw[trans] (-4.8cm,-12em) -- (-4.8cm,6em);
\draw[trans] (-4.3cm,-12em) -- (-4.3cm,13em) node[pos=0.45,right] {$\Omega_{B\Sigma_{1/2}}$};

\draw[energy] (-3cm,6em) -- (-3cm,9.5em) node[midway,right] {$\delta_{B\Sigma_{1/2}}$};
\draw[energy] (-3cm,9.5em) -- (-3cm,13em) node[midway,right] {$\delta_{B\Sigma_{1/2}}$};

\node at (9.5cm,-5em) {$X^2\Sigma^+$};
\node at (9.5cm,-6.5em) {$N=2$};
\node at (9.5cm,-8em) {$\mathcal{P}=+1$};
\node at (8cm,-10em) {$J=\frac{3}{2}$};
\node at (8cm,-4em) {$J=\frac{5}{2}$};

\draw[level] (4cm,-3em) -- (7cm,-3em) node[above,pos=1.05] {$F=3$};
\draw[level] (4cm,-5em) -- (7cm,-5em) node[below,pos=1.05] {$F=2$};
\draw[level] (4cm,-9em) -- (7cm,-9em) node[above,pos=1.05] {$F=2$};
\draw[level] (4cm,-11em) -- (7cm,-11em) node[below,pos=1.05] {$F=1$};

\draw[elong] (5.3cm,-7em) -- (-0.5cm,-7em);
\draw[elong] (5.6cm,-10em) -- (4cm,-10em);
\draw[elong] (5.6cm,-4em) -- (4cm,-4em);

\draw[energy] (0cm,-12em) -- (0cm,-7em) node[midway,right] {$\Delta_N$};
\draw[energy] (5.9cm,-11em) -- (5.9cm,-9em) node[midway,right] {$\Delta^{X\Sigma}_{F_{3/2}}$};
\draw[energy] (5.9cm,-5em) -- (5.9cm,-3em) node[midway,right] {$\Delta^{X\Sigma}_{F_{5/2}}$};
\draw[energy] (5.5cm,-10em) -- (5.5cm,-4em) node[midway,right] {$\Delta^{X\Sigma}_J$};

\node at (9cm,15em) {$B^2\Sigma^+$};
\node at (9cm,13.5em) {$N=1,\,J=\frac{3}{2}$};
\node at (9cm,12em) {$\mathcal{P}=-1$};
\node at (9cm,10.5em) {$\tau_{B\Sigma},\,\Gamma_{B\Sigma}$};

\draw[level] (4cm,14em) -- (7cm,14em) node[above,pos=0.95] {$F=2$};
\draw[level] (4cm,12em) -- (7cm,12em) node[below,pos=0.95] {$F=1$};

\draw[elong] (-0.5cm,13em) -- (5.5cm,13em);

\draw[virtual] (2.8cm,10.5em) -- (5cm,10.5em);
\draw[virtual] (2.8cm,15.5em) -- (5cm,15.5em);

\draw[energy] (0cm,9.5em) -- (0cm,13em) node[midway,right] {$\Delta^{B\Sigma}_J$};
\draw[energy] (6.2cm,12em) -- (6.2cm,14em) node[midway,right] {$\Delta^{B\Sigma}_{F_{3/2}}$};

\draw[radiative] (5.8cm,11.5em) -- (8.5cm,5em) node[midway,right] {$1-\mathcal{F}^{B\Sigma}_{00}$};
\draw[radiative] (5.2cm,13em) -- (5.2cm,-6.9em) node[midway,right] {$\,2/3$};

\draw[trans] (4.8cm,-10em) -- (4.8cm,10.5em);
\draw[trans] (4.3cm,-10em) -- (4.3cm,15.5em) node[pos=0.45,left] {$\Omega_{B\Sigma_{3/2}}$};

\draw[energy] (3cm,10.5em) -- (3cm,13em) node[midway,left] {$\delta_{B\Sigma_{3/2}}$};
\draw[energy] (3cm,13em) -- (3cm,15.5em) node[midway,left] {$\delta_{B\Sigma_{3/2}}$};

\draw[trans] (5.8cm,-2.5em) -- (8.5cm,3em) node[pos=0.5,right] {$\ \Omega_{\mathrm{aux}}$};
\draw[radiative] (4.7cm,13em) -- (-3.9cm,-11.9em) node[pos=0.4,above] {$1/3 \ \ $};
\draw[radiative] (-4.7cm,9.5em) -- (3.9cm,-6.9em) node[pos=0.5,above] {$\ 2/3$};

\end{tikzpicture}

\caption{\small Possible realization of polychromatic molasses-like forces in a real molecular level structure using one electronic state with a large spin-rotation splitting $\Delta^{B\Sigma}_{J}$.} \label{fig:sc1}
\end{figure}
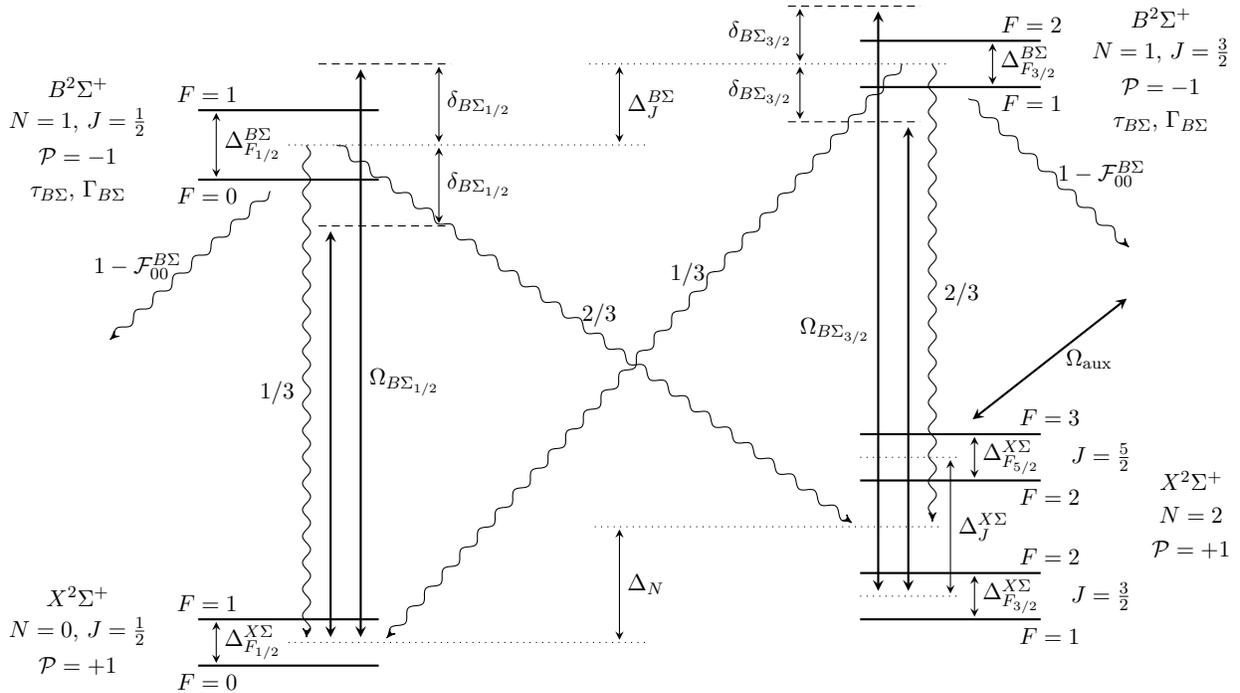

As detailed above, in order to realize large cooling forces primarily due to coherent momentum exchange between multi-frequency laser beams and molecules we must work with two separate excited states that each decay into both ground states. One way this can be accomplished for molecular radicals is when the $\lvert G1\rangle$ and $\lvert G2\rangle$ states in Fig. \ref{fig:BCF_BaH_diagram} are the rotational ground $(N''=0)$ and second $(N''=2)$ excited levels in the ground vibrational manifold, while the excited states consist of the ground vibrational, first rotational $(N'=1)$ level of an excited state manifold with two different sufficiently separated sub-manifolds (e.g. spin-rotation components). Angular momentum selection rules dictate that these excited states will decay to both ground states we have selected. This scheme is shown, using as general of notation as possible, in Fig. \ref{fig:sc1}. This scheme provides for the general requirements outlined by the toy 4-level model, but there are additional nuances introduced by potential hyperfine and spin-rotation splittings in various states.

\par Figure \ref{fig:ML-electronic} demonstrates energy correlation diagrams for molecular radicals of different structural symmetries. In the presented scheme, the $B^2\Sigma^+$ electronic state (second excited electronic state arising from the mixing between the $\left(n-1\right)d\sigma$ and $np\sigma$ orbitals), which exists in all diatomic and polyatomic molecules that have been considered for optical cycling applications thus far \cite{dirosa2004laser,augenbraun2020laser,kozyryev2016MOR,klos2020prospects,Ivanov2019,Augenbraun2020ARM}, is chosen as the excited state. Angular momentum selection rules dictate that only the included hyperfine and rotational states take part in the cycle, although there are some basic criteria for this scheme to work. Firstly, because we would like to use different $J'$ manifolds as $\lvert E1\rangle$ and $\lvert E2\rangle$ levels in Fig. \ref{fig:BCF_BaH_diagram} of our 2-level subsystems, the spin-rotation splitting in the excited electronic state $\Delta_J^{B\Sigma}$ has to be much larger than either of the PCF detunings $\delta_{B\Sigma_{1/2}}$ or $\delta_{B\Sigma_{3/2}}$. Otherwise, the $N''=0$ state might couple to $J'=3/2$ or $N''=2$ to $J'=1/2$, reducing the velocity damping coefficients. Similarly, $\Delta_N$, and so the rotational constant $B_{\mathrm{rot}}$ as well, must be much larger than either of mentioned detunings - for typically investigated detunings of $\sim 10^2\,\Gamma$, a rotational constant order of magnitude bigger ($B_{\mathrm{rot}}\gtrsim 10^3\,\Gamma$) should be sufficient. Secondly, for the same reason we have to make sure that both transitions are separated in frequency by more than the PCF detunings, which means that rotational splitting $\Delta_N$ and spin-rotation splitting $\Delta_J^{B\Sigma}$ have to be quite different. 

\par While the branching ratios in this scheme are not balanced, polychromatic forces acting on both subsystems can be balanced by an appropriate choice of PCF detunings and Rabi rates (App. \ref{sec:App-B}). However, the FCF for the $X^2\Sigma^+(\nu''=0)\leftrightarrow B^2\Sigma^+(\nu'=0)$ transition have to be high enough to allow multiple scattering events to occur. Albeit, they don't have to be perfect - one of the reasons polychromatic fields are extremely promising in the context of molecules is that they generate high forces while suppressing spontaneous emission \cite{Chieda2011}. FCFs of $\mathcal{F}^{B\Sigma}_{00}\gtrsim 0.95$ should be sufficiently high to allow the forces to create observable effects. Additionally, a repump laser might be added and population recycled, as was shown in multiple diatomic and polyatomic systems \cite{McCarron2018laser}.

\par Finally, hyperfine splittings and potential creation of dark states in the $N''=2$ manifold have to be discussed. Ideally, we would like the hyperfine splittings in a sub-level to be smaller than the PCF detunings, e.g. $\Delta_{F_{1/2}}^{X\Sigma}\approx\Delta_{F_{1/2}}^{B\Sigma}\ll\delta_{B\Sigma_{1/2}}$ in the $N''=0\leftrightarrow J'=1/2$ coupling. In the case of $N''=2\leftrightarrow J'=3/2$ transitions we also have the ground state spin-rotation splitting $\Delta^{X\Sigma}_J$ to take into account. If it is smaller than the PCF detunings, then all hyperfine states will be coupled resulting in creation of 12 dark states. If the splitting is larger than the detunings, we can couple only the $J=3/2$ manifold in the ground state without creating any dark states, but with population accumulating in the $J=5/2$ manifold. In both cases we could use an auxiliary transition, driven by $\Omega_{\mathrm{aux}}$, to a different excited state, like $A^2\Pi_{1/2}$, which would bring the population back into the cycle. Alternatively, dark state remixing method could be used such as adding a magnetic field, polarization switching or microwave-induced AC Stark shift.

\par
\begin{figure}[!h]
\centering

\begin{tikzpicture}[
scale=0.8, every node/.style={transform shape},
level/.style={thick},
energy/.style={thin,<->,shorten >=1pt,shorten <=1pt,>=stealth},
virtual/.style={densely dashed},
elong/.style={thin, dotted},
radiative/.style={black,->,>=stealth',shorten >=1pt,decorate,decoration={snake,amplitude=1.5}},
trans/.style={thick,<->,shorten >=2pt,shorten <=2pt,>=stealth},
]

\node at (-9cm,-10em) {$X^2\Sigma^+$};
\node at (-9cm,-11.5em) {$N=0,\,J=\frac{1}{2}$};
\node at (-9cm,-13em) {$\mathcal{P}=+1$};

\draw[level] (-4cm,-11em) -- (-7cm,-11em) node[above,pos=0.95] {$F=1$};
\draw[level] (-4cm,-13em) -- (-7cm,-13em) node[below,pos=0.95] {$F=0$};

\draw[elong] (-5.5cm,-12em) -- (0.5cm,-12em);

\draw[energy] (-6.7cm,-13em) -- (-6.7cm,-11em) node[midway,right] {$\Delta^{X\Sigma}_{F_{1/2}}$};

\node at (-9cm,12em) {$A^2\Pi_{1/2}$};
\node at (-9cm,10.5em) {$R=0,\,J=\frac{1}{2}$};
\node at (-9cm,9em) {$\mathcal{P}=-1$};
\node at (-9cm,7.5em) {$\tau_{A\Pi},\,\Gamma_{A\Pi}$};

\draw[level] (-4cm,11em) -- (-7cm,11em) node[above,pos=0.95] {$F=1$};
\draw[level] (-4cm,8em) -- (-7cm,8em) node[below,pos=0.95] {$F=0$};

\draw[elong] (-5.5cm,9.5em) -- (0.5cm,9.5em);

\draw[virtual] (-5cm,6em) -- (-2.8cm,6em);
\draw[virtual] (-5cm,13em) -- (-2.8cm,13em);

\draw[energy] (-6.7cm,8em) -- (-6.7cm,11em) node[midway,right] {$\Delta^{A\Pi_{1/2}}_{F}$};

\draw[radiative] (-5.8cm,7.5em) -- (-8.5cm,1em) node[midway,left] {$1-\mathcal{F}^{A\Pi_{1/2}}_{00}$};
\draw[radiative] (-5.2cm,9.5em) -- (-5.2cm,-11.9em) node[midway,left] {$2/3\,$};

\draw[trans] (-4.8cm,-12em) -- (-4.8cm,6em);
\draw[trans] (-4.3cm,-12em) -- (-4.3cm,13em) node[pos=0.45,right] {$\Omega_{A\Pi_{1/2}}$};

\draw[energy] (-3cm,6em) -- (-3cm,9.5em) node[midway,right] {$\delta_{A\Pi_{1/2}}$};
\draw[energy] (-3cm,9.5em) -- (-3cm,13em) node[midway,right] {$\delta_{A\Pi_{1/2}}$};

\node at (9.5cm,-5em) {$X^2\Sigma^+$};
\node at (9.5cm,-6.5em) {$N=2$};
\node at (9.5cm,-8em) {$\mathcal{P}=+1$};
\node at (8cm,-10em) {$J=\frac{3}{2}$};
\node at (8cm,-4em) {$J=\frac{5}{2}$};

\draw[level] (4cm,-3em) -- (7cm,-3em) node[above,pos=1.05] {$F=3$};
\draw[level] (4cm,-5em) -- (7cm,-5em) node[below,pos=1.05] {$F=2$};
\draw[level] (4cm,-9em) -- (7cm,-9em) node[above,pos=1.05] {$F=2$};
\draw[level] (4cm,-11em) -- (7cm,-11em) node[below,pos=1.05] {$F=1$};

\draw[elong] (5.3cm,-7em) -- (-0.5cm,-7em);
\draw[elong] (5.6cm,-10em) -- (4cm,-10em);
\draw[elong] (5.6cm,-4em) -- (4cm,-4em);

\draw[energy] (0cm,-12em) -- (0cm,-7em) node[midway,right] {$\Delta_N$};
\draw[energy] (5.9cm,-11em) -- (5.9cm,-9em) node[midway,right] {$\Delta^{X\Sigma}_{F_{3/2}}$};
\draw[energy] (5.9cm,-5em) -- (5.9cm,-3em) node[midway,right] {$\Delta^{X\Sigma}_{F_{5/2}}$};
\draw[energy] (5.5cm,-10em) -- (5.5cm,-4em) node[midway,right] {$\Delta_J$};

\node at (9cm,15em) {$A^2\Pi_{3/2}$};
\node at (9cm,13.5em) {$R=0,\,J=\frac{3}{2}$};
\node at (9cm,12em) {$\mathcal{P}=-1$};
\node at (9cm,10.5em) {$\tau_{A\Pi},\,\Gamma_{A\Pi}$};

\draw[level] (4cm,14em) -- (7cm,14em) node[above,pos=0.95] {$F=2$};
\draw[level] (4cm,12em) -- (7cm,12em) node[below,pos=0.95] {$F=1$};

\draw[elong] (-0.5cm,13em) -- (5.5cm,13em);

\draw[virtual] (2.8cm,10.5em) -- (5cm,10.5em);
\draw[virtual] (2.8cm,15.5em) -- (5cm,15.5em);

\draw[energy] (0cm,9.5em) -- (0cm,13em) node[midway,right] {$\Delta_{\Omega}$};
\draw[energy] (6.2cm,12em) -- (6.2cm,14em) node[midway,right] {$\Delta^{A\Pi_{3/2}}_{F}$};

\draw[radiative] (5.8cm,11.5em) -- (8.5cm,5em) node[midway,right] {$1-\mathcal{F}^{A\Pi_{3/2}}_{00}$};
\draw[radiative] (5.2cm,13em) -- (5.2cm,-6.9em) node[midway,right] {$\,1/2$};

\draw[trans] (4.8cm,-10em) -- (4.8cm,10.5em);
\draw[trans] (4.3cm,-10em) -- (4.3cm,15.5em) node[pos=0.45,left] {$\Omega_{A\Pi_{3/2}}$};

\draw[energy] (3cm,10.5em) -- (3cm,13em) node[midway,left] {$\delta_{A\Pi_{3/2}}$};
\draw[energy] (3cm,13em) -- (3cm,15.5em) node[midway,left] {$\delta_{A\Pi_{3/2}}$};

\draw[trans] (5.8cm,-2.5em) -- (8.5cm,3em) node[pos=0.5,right] {$\ \Omega_{\mathrm{aux}}$};
\draw[radiative] (4.7cm,13em) -- (-3.9cm,-11.9em) node[pos=0.4,above] {$1/2 \ \ $};
\draw[radiative] (-4.7cm,9.5em) -- (3.9cm,-6.9em) node[pos=0.5,above] {$\ 1/3$};

\end{tikzpicture}

\caption{\small Alternative realization of polychromatic molasses-like forces in a real molecular level structure using two parity-doublet $A^2\Pi$ electronic states with different total angular momentum projections $\Omega$.} \label{fig:sc2}
\end{figure}
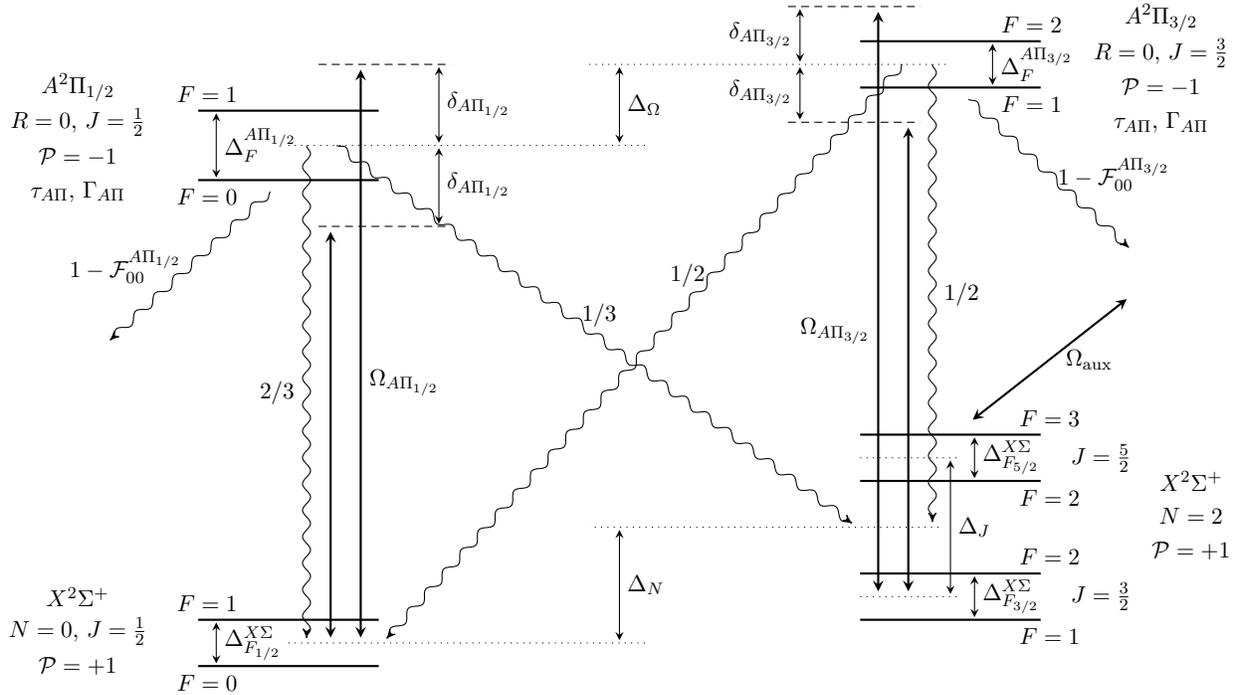

 If the spin-rotation splitting in the $B^2\Sigma^+$ state is not sufficiently large, a different scheme can be utilized, shown in Fig. \ref{fig:sc2}. Here, excited states $\lvert E1\rangle$ and $\lvert E2\rangle$ are two $A^2\Pi_{\Omega}$ states with a different total angular momentum projection of $\Omega=1/2$ and $\Omega=3/2$. In fact, in some molecules (like BaH) such scheme might work better, due to more beneficial FCF's compared to the $B^2\Sigma^+$ state \cite{Moore2019}. In this configuration, we exploit the fact that the excited states are parity doublets and given positive parity of the ground states, we can choose excited states of negative parity to create our light-coupled 2-level systems. In all other aspects this scheme is analogical to the one described before, and so the same constraints and criteria apply, with the spin-rotation splitting in the excited state being replaced by $\Omega$-splitting $\Delta_{\Omega}$.

\subsection{Analysis of Bichromatic Forces in Barium Monohydride \label{sec:BaH-BCF-slowing}}

\par Motivated by the prospects of ultracold hydrogen production via molecular laser cooling followed by photo-dissociation \cite{lane2015ultracold}, recently there has been an increased experimental \cite{mcnally2020,tarallo2016bah} and theoretical \cite{Moore2019,moore2018quantitative,gao2014laser} interest in direct laser cooling and trapping of alkaline-earth-metal monohydrides. High mass imbalance, low Doppler cooling limit, and small photon recoil velocity make BaH an extremely attractive candidate for producing ultracold atomic hydrogen via zero-energy photo-fragmentation \cite{lane2015ultracold}. Unfortunately, the very same inherent molecular characteristics make laser cooing of BaH experimentally challenging \cite{mcnally2020}. However, fine and hyperfine structure of BaH in the rotational states involved in the optical cycling process \cite{Iwata2017} together with technically accessible transition wavelengths make it an ideal candidate for stimulated slowing and cooling using polychromatic optical forces.   
\par In the electronic ground state $X^2 \Sigma^{+}$, the $\nu''=0$  $N''=1$ rovibrational state has an exceptionally large spin-orbit splitting of 8.6 GHz, while the hyperfine splitting is unresolvable in the $J'=1/2$ state \cite{Iwata2017}. This allows the transition from $X^2 \Sigma^{+}$ $J''=1/2$ to $A^2\Pi_{1/2}$ $J' = 1/2$, where the hyperfine splitting is similarly unresolvable, to be addressed simultaneously on all transitions with equal detuning between transition and carrier frequency, while leaving the $J''=3/2$ states unperturbed even when the bichromatic detuning is significantly larger than the decay lifetime ($\Delta_J / \Gamma_{A\Pi}   = 1180$, leaving plenty of range over which $\Gamma_{A\Pi} \ll \delta \ll \Delta_J$ holds), and thus enabling realization of the stimulated force significantly larger than any possible radiative force.
\par When this transition is driven with $\pi$-polarized light, a set of four separate, radiatively cross-coupled two-level transitions are all driven at equal strength and equal transition frequency. As a result, applying a set of BCF optical fields to this transition results in a force which (when neglecting any off-transition decays from the A state) is nearly identical to that achieved in simple two-level BCF. This includes that there is no formation of dark states, avoiding the requirement of dark-state destabilization as has been needed in prior application of BCF to molecules \cite{Aldridge2016}. This was verified by numerical simulation of the BCF force on the four ground state, four excited state system. Our numerical simulations solved the Liouville-von Neumann equations for density matrix evolution in the rotating wave, fixed-velocity approximations, as in previous simulations of the BCF on molecules \cite{aldridgethesis}.
\par An obvious weakness of this model is that, with the $J''=3/2$ states unaddressed, the off-diagonal decays from the A state are far from negligible. Even considering only decays to $X^2\Sigma^+$ $v''=0$ $N''=1$, one third of spontaneous decays of the excited states should end in $J''=3/2$. If this is allowed to continue undisturbed, the system will quickly go dark and stop feeling force after all population is pumped out of $J''=1/2$ into $J''=3/2$ and a small fraction into other rovibrational states.
\par In the context of creating a sustained force, this can be remedied by using additional optical fields to drive transitions from $J''=3/2$ in such a way that population eventually returns to the BCF-driven transition. In a simple BCF scheme, this can be done by addition of a CW optical field which drives $J''=3/2$ to $B {}^2\Sigma^{+}$ $N'=0$ $J'=1/2$. This state also decays primarily to $X^2\Sigma^+$ $v''=0$ $N''=1$, with one third of these decays ending in $J''=1/2$, back in the BCF cycle. This comprises an indirect repumping scheme for BCF as discussed in \cite{aldridgethesis} and previously implemented for SrOH BCF deflection \cite{Kozyryev2018}. This scheme is illustrated in Fig. \ref{fig:sc_BaH} but with weak repumping beams to address the $J''=3/2$ to $B {}^2\Sigma^{+}$ $N'=0$ $J'=1/2$ transition. Notably there are four sets of states, ground and excited states in both the BCF and repump transitions as depicted in Fig. \ref{fig:BCF_BaH_diagram}. These sets of states will be referred to as $\lvert G1 \rangle$, $\lvert E1 \rangle$, $\lvert G2 \rangle$, $\lvert E2 \rangle$ and the labels $g$, $e$, $g_R$ and $e_R$ will refer to the time-averaged ensemble population in each set, correspondingly.
\par With sufficiently strong optical fields, so that $\Omega \gg \Gamma$ in each case, population returning to the BCF or repump cycle from the opposite cycle can be taken as a small perturbation to the population dynamics, and overall population can be estimated accurately by assuming the relative populations in each cycle will be identical to that which they would have absent the other cycle and that relative population between cycles is determined by equilibrium rate equations. In particular, $e\Gamma_{12} = e_R\Gamma_{21}$, where $\Gamma_{12}$ is the decay rate from $\lvert E1 \rangle$ to $\lvert G2 \rangle$ and $\Gamma_{21}$ is the decay rate from $\lvert E2 \rangle$ to $\lvert G1 \rangle$ as shown in Fig. \ref{fig:BCF_BaH_diagram}.
\par Optical dark states will exist in the $J''=3/2$ to $B^2\Sigma^+$ transition, for any fixed choice of repump polarization. Assuming a remixing magnetic field and a saturated CW repump, the populations in the repump transition will equilibrate to have a proportionality equal to that of the number of states: $e_R/g_R = N_{\lvert E2 \rangle}/N_{\lvert G2 \rangle}$. In this case, with four excited states and eight ground states, 1/3 of the population in the repump cycle will be in the excited $B^2\Sigma^+$ state at any given time.
\par On the BCF transition manifold $\lvert G1 \rangle \rightarrow \lvert E1 \rangle$, the proportion of population between ground and excited states depends on the BCF (or polychromatic force) driving. This will produce a characteristic time-average excited state population $P_e = e/(e+g)$. The $J''=1/2$ to $A^2\Pi_{1/2}$ transition in BaH behaves nearly identically to a two-level system in response to BCF, as discussed above. The optimal BCF or 4-color PCF optical fields in a two-level system are known from previous works, along with the time-average excited state populations that results \cite{Galica2013}. After the decay of any transient behavior, the equilibrium populations between the two cycles will be reached when cross-decay occurs at equal rates. Taken together, these considerations are sufficient to determine the “participating fraction,” i.e. the fraction of molecular population which is in the BCF cycle at any given time:
\begin{displaymath}
e+g = \left(1 + P_e \frac{\Gamma_{12}}{\Gamma_{21}} \frac{N_{g_R}}{N_{e_R}}\right)^{-1}
\end{displaymath}

\par The effective time-averaged force at a given velocity can then be taken to be equal to the time-averaged force that would be achieved if the BCF cycle were closed, multiplied by this participating fraction which is a function of the time-averaged excited state fraction in the closed cycle at that velocity. This can be compared to the maximum radiative force that would result were both the $X^2\Sigma^+$  $J''=1/2 \leftrightarrow A^2\Pi_{1/2}$ and $X^2\Sigma^+$ $J''=3/2 \leftrightarrow B^2\Sigma^+$ transition manifolds driven with resonant, saturated CW optical fields. In that case, $P_e$ would equal one half, and the expected radiative force due to both transitions would be, where $k$ and $\Gamma$ are the wavenumber of and decay rate along the transition in question,
\begin{displaymath}
F_{\mathrm{rad}}=\hbar\left(k_{A\Pi}\Gamma_{A\Pi}\, e + k_{B\Sigma}\Gamma_{B\Sigma}\, e_R\right) \approx 1.98\,\mathrm{\,zN}.
\end{displaymath}

\par
\begin{figure}[!h]
\centering
\includegraphics[width=0.7\textwidth]{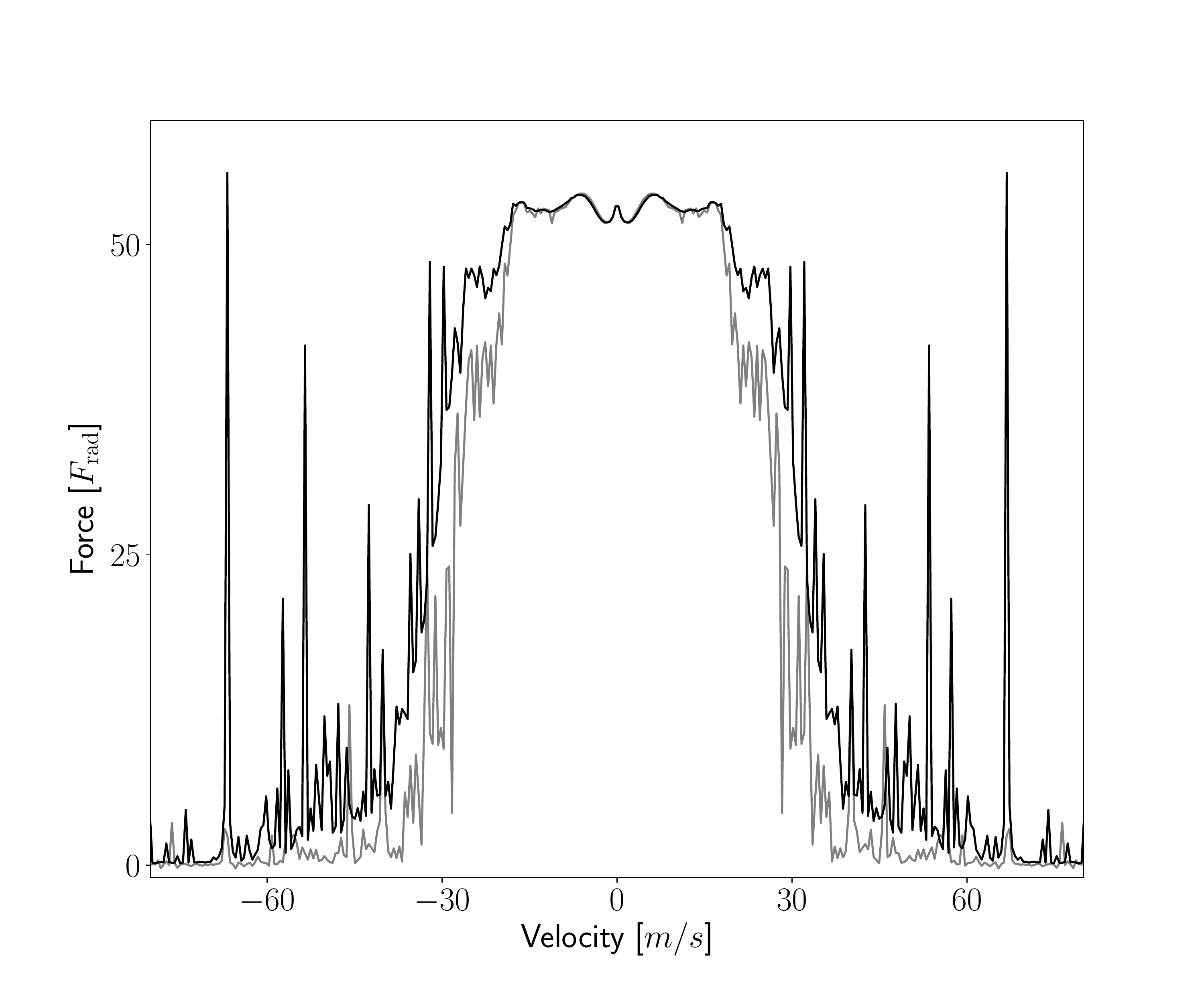}
\caption{\small Numerically simulated effective force profiles for $\delta = 189$ MHz bichromatic force driving of BaH on the $X {}^2 \Sigma^{+}$ $J''=1/2$ to $A {}^2\Pi_{1/2}$ $J' = 1/2$ transition with indirect repumping of $X {}^2 \Sigma^{+}$ $J''=3/2$ through $B {}^2\Sigma^{+}$ $N'=0,\,J'=1/2$. The black curve is from a simulation of all hyperfine projection states in the BCF cycle, and the grey curve is from a simulation of a single two-level transition in the BCF cycle. The two are indistinguishable as expected at sufficiently small molecular speeds but show some deviation as speed increases.}
\label{fig:BaH_BCF_sim}
\end{figure}

 \noindent This force can significantly exceeded the optimal radiative force with experimentally achievable BCF irradiances. Figure \ref{fig:BaH_BCF_sim} shows simulated effective force profiles for a bichromatic detuning of 189 MHz, which would require a per-beam irradiance of 22.5 W/cm$^2$, or in other words a total summed irradiance of 90 W/cm$^2$ across all four BCF beams, to have the optimal BCF Rabi frequency at this detuning. Given the experimentally realized laser power at 1060 nm \cite{galica2018polychromatic}, which is the wavelength for the $X-A$ optical cycling transition in BaH, our calculations indicate strong feasibility of using the described BCF-driving scheme to achieve rapid slowing of cryogenic BaH molecular beam in a short ($\sim$ few cm) distance. Using realistic experimental parameters (5 W and 1 mm radius beam), we anticipate that achieving a total summed irradiance of 500 W/cm$^2$ is feasible, leading to potential for even larger force enhancements. 

\subsection{SupER Molasses in BaH \label{sec:BaH-SupER-molasses}}
\par
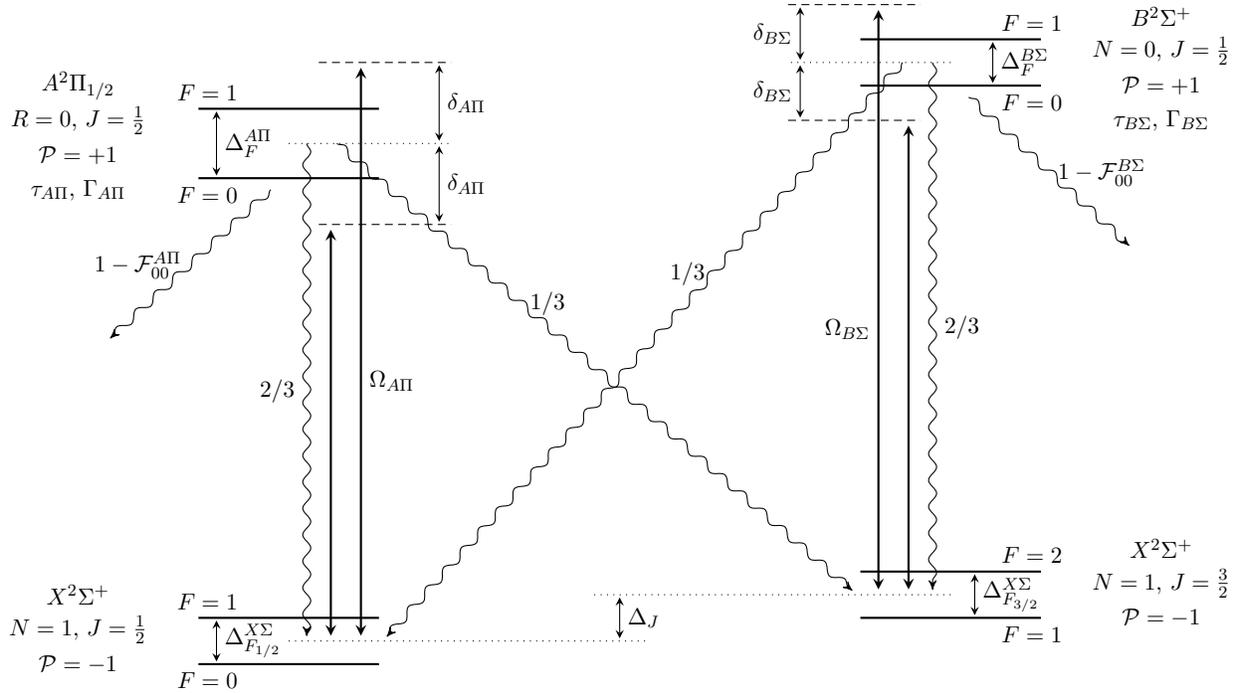
\begin{figure}[!h]
\centering

\begin{tikzpicture}[
scale=0.8, every node/.style={transform shape},
level/.style={thick},
energy/.style={thin,<->,shorten >=1pt,shorten <=1pt,>=stealth},
virtual/.style={densely dashed},
elong/.style={thin, dotted},
radiative/.style={black,->,>=stealth',shorten >=1pt,decorate,decoration={snake,amplitude=1.5}},
trans/.style={thick,<->,shorten >=2pt,shorten <=2pt,>=stealth},
]

\node at (-9cm,-10em) {$X^2\Sigma^+$};
\node at (-9cm,-11.5em) {$N=1,\,J=\frac{1}{2}$};
\node at (-9cm,-13em) {$\mathcal{P}=-1$};

\draw[level] (-4cm,-11em) -- (-7cm,-11em) node[above,pos=0.95] {$F=1$};
\draw[level] (-4cm,-13em) -- (-7cm,-13em) node[below,pos=0.95] {$F=0$};

\draw[elong] (-5.5cm,-12em) -- (0.5cm,-12em);

\draw[energy] (-6.7cm,-13em) -- (-6.7cm,-11em) node[midway,right] {$\Delta^{X\Sigma}_{F_{1/2}}$};

\node at (-9cm,12em) {$A^2\Pi_{1/2}$};
\node at (-9cm,10.5em) {$R=0,\,J=\frac{1}{2}$};
\node at (-9cm,9em) {$\mathcal{P}=+1$};
\node at (-9cm,7.5em) {$\tau_{A\Pi},\,\Gamma_{A\Pi}$};

\draw[level] (-4cm,11em) -- (-7cm,11em) node[above,pos=0.95] {$F=1$};
\draw[level] (-4cm,8em) -- (-7cm,8em) node[below,pos=0.95] {$F=0$};

\draw[virtual] (-5cm,6em) -- (-2.8cm,6em);
\draw[virtual] (-5cm,13em) -- (-2.8cm,13em);

\draw[elong] (-5.5cm,9.5em) -- (-2.8cm,9.5em);

\draw[energy] (-6.7cm,8em) -- (-6.7cm,11em) node[midway,right] {$\Delta^{A\Pi}_{F}$};

\draw[radiative] (-5.8cm,7.5em) -- (-8.5cm,1em) node[midway,left] {$1-\mathcal{F}^{A\Pi}_{00}$};
\draw[radiative] (-5.2cm,9.5em) -- (-5.2cm,-11.9em) node[midway,left] {$2/3\,$};

\draw[trans] (-4.8cm,-12em) -- (-4.8cm,6em);
\draw[trans] (-4.3cm,-12em) -- (-4.3cm,13em) node[pos=0.45,right] {$\Omega_{A\Pi}$};

\draw[energy] (-3cm,6em) -- (-3cm,9.5em) node[midway,right] {$\delta_{A\Pi}$};
\draw[energy] (-3cm,9.5em) -- (-3cm,13em) node[midway,right] {$\delta_{A\Pi}$};

\node at (9cm,-8em) {$X^2\Sigma^+$};
\node at (9cm,-9.5em) {$N=1,\,J=\frac{3}{2}$};
\node at (9cm,-11em) {$\mathcal{P}=-1$};

\draw[level] (4cm,-9em) -- (7cm,-9em) node[above,pos=0.95] {$F=2$};
\draw[level] (4cm,-11em) -- (7cm,-11em) node[below,pos=0.95] {$F=1$};

\draw[elong] (5.5cm,-10em) -- (-0.5cm,-10em);

\draw[energy] (5.9cm,-11em) -- (5.9cm,-9em) node[midway,right] {$\Delta^{X\Sigma}_{F_{3/2}}$};
\draw[energy] (0cm,-12em) -- (0cm,-10em) node[midway,right] {$\Delta_J$};

\node at (9cm,15em) {$B^2\Sigma^+$};
\node at (9cm,13.5em) {$N=0,\,J=\frac{1}{2}$};
\node at (9cm,12em) {$\mathcal{P}=+1$};
\node at (9cm,10.5em) {$\tau_{B\Sigma},\,\Gamma_{B\Sigma}$};

\draw[level] (4cm,14em) -- (7cm,14em) node[above,pos=0.95] {$F=1$};
\draw[level] (4cm,12em) -- (7cm,12em) node[below,pos=0.95] {$F=0$};

\draw[virtual] (2.8cm,10.5em) -- (5cm,10.5em);
\draw[virtual] (2.8cm,15.5em) -- (5cm,15.5em);

\draw[energy] (6.2cm,12em) -- (6.2cm,14em) node[midway,right] {$\Delta^{B\Sigma}_{F}$};

\draw[elong] (5.5cm,13em) -- (2.8cm,13em);

\draw[radiative] (5.8cm,11.5em) -- (8.5cm,5em) node[midway,right] {$1-\mathcal{F}^{B\Sigma}_{00}$};
\draw[radiative] (5.2cm,13em) -- (5.2cm,-9.9em) node[midway,right] {$\,2/3$};

\draw[trans] (4.8cm,-10em) -- (4.8cm,10.5em);
\draw[trans] (4.3cm,-10em) -- (4.3cm,15.5em) node[pos=0.45,left] {$\Omega_{B\Sigma}$};

\draw[energy] (3cm,10.5em) -- (3cm,13em) node[midway,left] {$\delta_{B\Sigma}$};
\draw[energy] (3cm,13em) -- (3cm,15.5em) node[midway,left] {$\delta_{B\Sigma}$};

\draw[radiative] (4.7cm,13em) -- (-3.9cm,-11.9em) node[pos=0.4,above] {$1/3 \ \ $};
\draw[radiative] (-4.7cm,9.5em) -- (3.9cm,-9.9em) node[pos=0.4,above] {$\ 1/3$};

\end{tikzpicture}

\caption{\small Specific realization of the 4-level toy model in barium monohydride possible due to its abnormally large spin-rotation splitting $\Delta_J$ in the $N=1$ rotational state of the $X^2\Sigma^+$ ground electronic state.} \label{fig:sc_BaH}
\end{figure}

 Given a high potential for an effective realization of BCF in BaH on the $X^2\Sigma^+\leftrightarrow  A^2\Pi_{1/2}$ transition as described in Sec.\ref{sec:BaH-BCF-slowing}, a different optical scheme can be used to realize the SupER molasses cooling configuration. As presented in Fig. \ref{fig:sc_BaH}, instead of using two states with different rotational quantum numbers $N$, we choose to use two $J$ states in $N''=1$ rotational state, which are separated by about $8.6$ GHz \cite{Iwata2017}. As the excited states we use two different electronic states: $A^2\Pi_{1/2}$ with positive parity and $B^2\Sigma^+$ in its ground rotational level $N'=0$.
\par In this system, many of the criteria listed before are fulfilled - both FCFs are greater than 0.95 \cite{Moore2019} and both transitions are far apart in the frequency space ($\lambda_{A\Pi}\approx 1060.7868$ nm and $\lambda_{B\Sigma}\approx 905.3197$ nm).  As was detailed in Sec. \ref{sec:BaH-BCF-slowing}, polychromatic forces can be created using the excitation to the $A^2\Pi_{1/2}$ state. The hyperfine splitting $\Delta_F^{A\Pi}$ in the $A^2\Pi_{1/2}$ electronic state and $\Delta^{X\Sigma}_{F_{1/2}}$ in $J''=1/2$ manifold of the $X^2\Sigma^+$ state are both very small (less than $4\,\Gamma_{A\Pi} \sim 2\pi\times 4$ MHz \cite{Iwata2017}), and so by using a $\pi$-polarized light fields with $\delta_{A\Pi}$ detuning having any reasonable value much larger than $\Gamma_{A\Pi}$ these forces will be created.

\par Transition to the $B^2\Sigma^+$ excited state is trickier to address. Hyperfine splitting in the $J''=3/2$ level of the $X^2\Sigma^+$ state is larger -- $\Delta^{X\Sigma}_{F_{3/2}}\approx 32\, \Gamma_{B\Sigma}\approx 2\pi \times 39$ MHz. So is the splitting in the $B^2\Sigma^+$ state -- $\Delta^{B\Sigma}_{F_{1/2}}\approx -43\, \Gamma_{B\Sigma}\approx -2\pi\times 52$ MHz \cite{Iwata2017}. Together, they lead to a $\sim 75\,\Gamma_{B\Sigma}$ frequency difference between $F''=2\leftrightarrow F'=1$ and $F''=1\leftrightarrow F'=0$ transitions. To observe polychromatic forces through this electronic transition, larger values of $\delta_{B\Sigma}$ detuning will be needed.

\par The $X^2\Sigma^+ \leftrightarrow B^2\Sigma^+$ transition considered will also create dark states in the $J''=3/2$ manifold. Fortunately, the g-factors are large enough \cite{Iwata2017} to provide efficient remixing with the help of a magnetic field of modest strength. Given that we already need high detunings $\delta_{B\Sigma}$, the Zeeman splittings should not cause any additional problems. Finally, while the branching ratios in this case are symmetric, the decay rates are different for both electronic states ($\Gamma_{A\Pi}\approx 2\pi\times 1.15$ MHz and $\Gamma_{B\Sigma}\approx 2\pi\times 1.21$ MHz); although, as was mentioned before, this asymmetry can be easily adjusted for by an appropriate choice of detunings and Rabi rates.

\par To obtain force profile in a realistic BaH level system, we have again solved the Liouville-von Neumann equations for density matrix evolution for a system that included Zeeman sublevels in states depicted in Fig. \ref{fig:sc_BaH}. For simplicity, we have solved it by assuming that $\mathcal{F}^{A\Pi}_{00}=\mathcal{F}^{B\Sigma}_{00}=1$. We have chosen to center the $X^2\Sigma^+\leftrightarrow A^2\Pi_{1/2}$ transition on the $F''=1\leftrightarrow F'=0$ frequency, while also assuming that $\Delta_F^{A\Pi}=2\pi \times 2$ MHz and $\Delta^{X\Sigma}_{F_{1/2}}=-2\pi \times 2$ MHz. The $X^2\Sigma^+\leftrightarrow B^2\Sigma^+$ transition was centered at frequency placed symmetrically between $F''=2\leftrightarrow F'=1$ and $F''=1\leftrightarrow F'=1$ transition frequencies.
\par For the detunings we have first chosen $\delta_{B\Sigma}=200\,\Gamma_{B\Sigma}$ and from there we obtained $\delta_{A\Pi}$ using Eq.(\ref{eqn:delta_criterion}), which resulted in $\delta_{A\Pi}\approx 234.35\,\Gamma_{A\Pi}\approx 2\pi\times 270$ MHz. Due to imperfect dark state remixing, we have then slightly increased value of the $\delta_{B\Sigma}$ to properly balance forces at $v=0$. In the end we have used $\delta_{B\Sigma}=208\,\Gamma_{B\Sigma}\approx 2\pi\times 252$ MHz. 
\par
\begin{figure}[!h]
\centering
\includegraphics[width=0.7\textwidth]{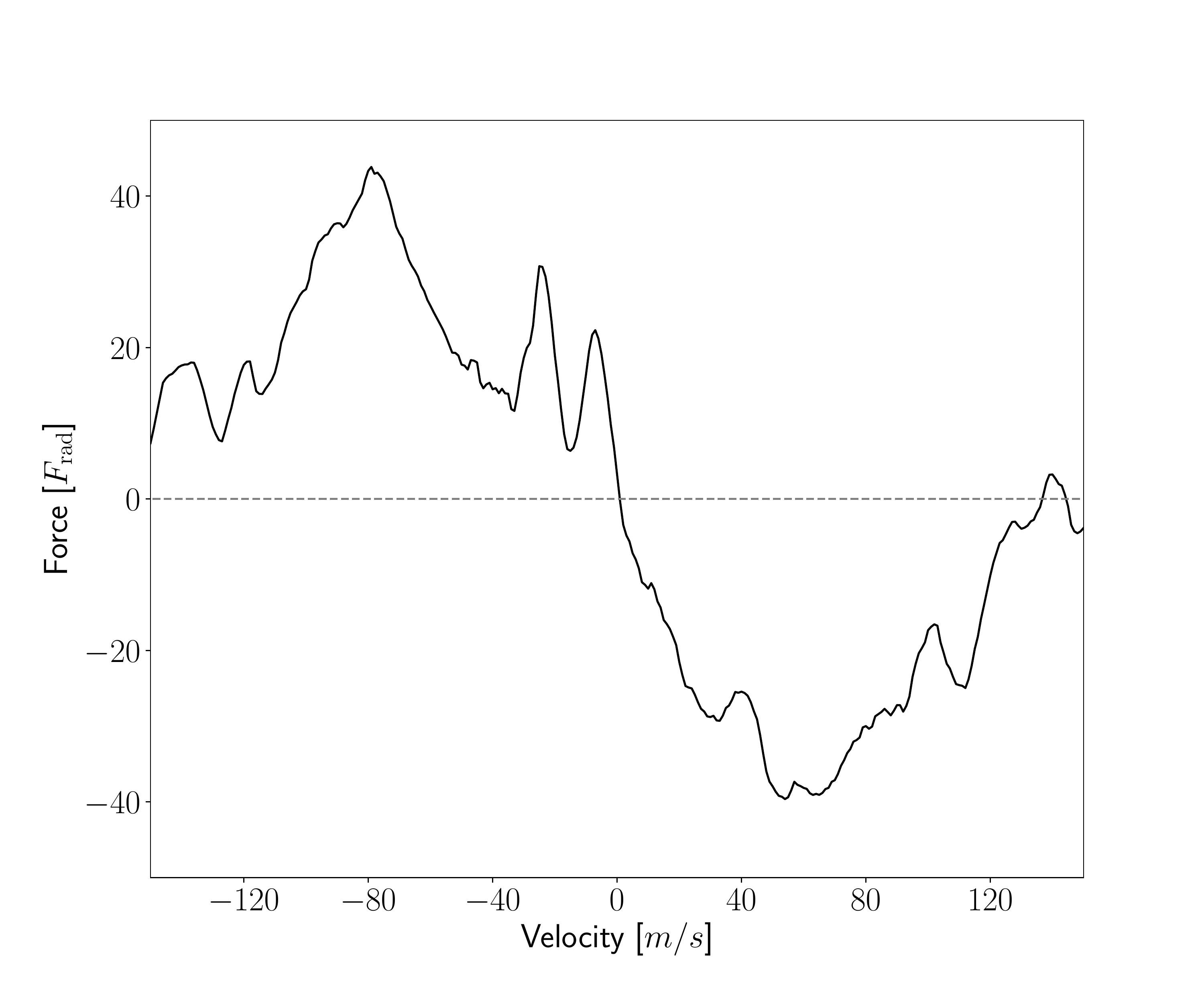}
\caption{\small Smoothed force profile of bichromatic SupER molasses in a real BaH level system obtained for $\delta_{A\Pi}\approx 234.35\,\Gamma_{A\Pi}\approx 2\pi\times 270$ MHz and $\delta_{B\Sigma}=208\,\Gamma_{B\Sigma}\approx 2\pi\times 252$ MHz in presence of 12 G ambient magnetic field, where the rest of the parameters were chosen according to principles discussed in previous sections. Large detunings are necessary due to large ($\sim 40-50$ MHz) hyperfine splittings in $J''=3/2$ rotational state in the ground $X^2\Sigma^+$ state and in the excited $B^2\Sigma^+$ state. The force profile is quite symmetric around $v=0$ and shows gigantic capture velocities consistent with $\delta/2k$ predictions. The force is provided in units of $F_{\mathrm{rad}}=1.98$ zN and the damping coefficient can be estimated to be $\beta\approx 3.24\, F_{\mathrm{rad}}s/m \approx 3.47\, \hbar k^2_{A\Pi}/2$.}
\label{fig:BaH_full_molasses}
\end{figure}

 The Rabi rates were those of optimal bichromatic fields, i.e.  $\Omega_{A\Pi}=\sqrt{3/2}\,\delta_{A\Pi}$ and $\Omega_{B\Sigma}=\sqrt{3/2}\,\delta_{B\Sigma}$. Assuming 3 mm diameter uniform beams, such Rabi rates would require about 2.2 W power per frequency component in the case of $\Omega_{A\Pi}$, and 1.9 W for $\Omega_{B\Sigma}$. We have assumed a presence of 12 G ambient magnetic field that defined the quantization axis. Zeeman splitting was obtained using experimentally obtained effective linear g-factors \cite{Iwata2017}. The $X^2\Sigma^+\leftrightarrow A^2\Pi_{1/2}$ transition light field was polarized along the quantization axis, while the other light field was perpendicular to it. Finally, we have Doppler-shifted the frequencies by appropriate amounts, that is $\Delta_1=-\delta_{A\Pi}/4\approx -2\pi\times 67.5$ MHz and $\Delta_2=\delta_{B\Sigma}/4\approx 2\pi\times 63.1$ MHz, and chose $\chi_1=-\chi_2=45^{\circ}$. The SupER molasses force profile obtained is shown in Fig. \ref{fig:BaH_full_molasses}.
\par The obtained force profile is quite symmetric, despite the fact that both bichromatic fields act on quite different level structures. It is also linear around zero velocity. We can estimate the slope to be $\beta\approx 3.24\, F_{\mathrm{rad}}/(\rm{m/s}) \approx 3.47\, \hbar k^2_{A\Pi}/2$, and while we do not know the exact value of $F_0$, we can place an upper bound equal to the maximum force seen in the profile, i.e. $F_0\lesssim 45\,F_{\mathrm{rad}}=39.45\, \hbar k_{A\Pi}\Gamma_{A\Pi}/2$. Using Eq.(\ref{eqn:limit_temp4}) with $\sigma^2(\varepsilon)\approx 21$ (asymmetric 4-level system) we can place an upper bound on the temperature to be $T_L\lesssim 193.8$ mK. We also see that the capture velocity gets as high as 120 m/s, which is consistent with the $\delta/2k$ estimate provided earlier. The temperature equivalent to capture velocity in these molasses is $T_{\mathrm{cap}}=M\delta^2/4k_Bk^2$, and for $v_{\mathrm{cap}}=120$ m/s in BaH is equal to $T_{\mathrm{cap}}\approx 60$ K, providing further evidence that the SupER molasses method does indeed realize a way to cool and confine molecules in kelvin-deep optical potentials as thought after for more than thirty years following initial speculations by Kazantzev \cite{kazantsev1987rectification} and Voitsekhovich \cite{voitsekhovich1989observation}.


\par The unusual structure at small negative velocities is due to summation of a close-to-regular bichromatic force profile obtained via the $X^2\Sigma^+\leftrightarrow A^2\Pi_{1/2}$ transition, and the off-center resonant (Doppleron) peak always appearing at $k(v-v_0)=\pm \delta/3$ in all bichromatic force profiles, and here created via the $X^2\Sigma^+\leftrightarrow B^2\Sigma^+$ transition. The latter transition's profile is centered around $v_0=\delta_{B\Sigma}/4 k_{B\Sigma}\approx 57.1\,$ m/s, so we would expect its off-center peak to appear at $v=v_0-\delta_{B\Sigma}/3k_{B\Sigma}\approx -12.5\,$ m/s, which is where we observe the dip.

\section{Conclusions and Future Prospects}

We have presented a novel experimentally viable method for achieving large optical molasses-like cooling forces for molecules using polychromatic optical fields driving coherent dynamics in a four-level system. Using direct numerical solutions of the time-dependent density matrix as well as Monte Carlo simulations of the cooling dynamics, we provide evidence that achieving rapid damping of a wide velocity capture range towards zero velocity should be possible for diatomic and polyatomic molecules with various constituents and geometries. Proposed Suppressed Emission Rate (SupER) molasses method relies on spontaneous emission coupling between two coherently-driven two-level systems and should be realizable with many complex nonlinear molecules for which scattering $\sim 100-1,000$ photons has been previously proposed \cite{augenbraun2020laser,klos2020prospects} or already experimentally demonstrated \cite{baum2020CaOH10k,mitra2020direct}. We anticipate that large velocity damping coefficients together with a broad velocity capture range will enable extension of laser-based cooling and coherent quantum control to novel molecular species with complex internal structures, weak optical transitions and abundant vibrational decay channels providing a fruitful experimental platform for realizing many exciting applications in fundamental physics and applied quantum technologies. 

\par Specifically, strontium methyl (SrCH\textsubscript{3}) has a number of advantageous characteristics not only for achieving ultracold temperatures via SupER molasses cooling but also for realizing diverse applications with such samples. 
An unpaired valence electron residing on the strontium atom allows for strong visible electronic transitions that can be used for laser manipulation of the internal molecular states. The spectrum of SrCH\textsubscript{3} has been extensively studied
in the past \cite{Dick2007}, and previously Kozyryev and co-workers have outlined details of achieving multiple-photon cycling using either $\tilde{X}-\tilde{A}$ (732 nm) or $\tilde{X}-\tilde{B}$ transition (676 nm) in symmetric-top molecules \cite{kozyryev2016MOR}. Because of the high degree of overlap for the vibrational wavefunctions in different electronic states, scattering of $>20$ photons per molecules can be achieved with only a single-color laser. With an addition of one repumping laser for the Sr-C stretching vibrational mode, scattering of $>100$ photons per molecule is possible. Realization of large SupER molasses cooling profiles should be possible since intensities
of over $600\,{\rm W/cm^{2}}$ can be achieved using commercial cw Ti:Sapphire lasers, which is a factor of $\sim10^{4}$ above the saturation intensity of the $\tilde{X}-\tilde{A}$ transition. Rich internal structure of SrCH\textsubscript{3} symmetric-top molecules pinned in optical lattices or tweezers will allow realization of nonconventional quantum
magnetism models (including Heisenberg XYZ \cite{wall2015realizing}) without the need for quantum degeneracy, creating a unique quantum simulation platform to probe strongly correlated many-body systems inaccessible to ultracold atom and diatomic molecule experiments.


\section*{Acknowledgements}
Work at Columbia has been supported by the W. M. Keck Foundation. K. Wenz and R.L. McNally would like to acknowledge support from the NSF IGERT Grant No. DGE-1069240. I. Kozyryev was supported by the Simons Junior Fellow Award. 

\appendix

\section{\label{sec:Variance-estimation} Momentum variance and force estimation for PCF in a $\pi$-pulse model}

In this Appendix section we provide a new conceptual framework employing a continuous-time Markov chain (CTMC) probabilistic model for estimating the magnitude of the BCF force and momentum transfer variance in a $\pi$-pulse model. Normally, in the $\pi$-pulse model the force estimation is done by assuming that some atoms or molecules can be in either correct (i.e. experiencing force in the desired direction) or wrong (i.e. experiencing force in the opposite direction) cycle, and that they might switch the cycle, if a spontaneous decay event occurs. If we just look at one atom that starts in a correct cycle (though, as we will later see, the initial condition does not influence the final result), it will be deterministically pushed in one direction by a process of excitation and stimulated emission. However, because it spends a non-zero amount of time in the excited state, it has a finite probability of decaying back to the ground state before the stimulated emission occurs. Because in such situation the pulse that was supposed to stimulate the emission will cause excitation instead, the atom is effectively in the wrong cycle. The average force is non-zero if average times spent in the excited and ground states are not equal.

\par The average fraction of time an atom\footnote{Throughout the Appendix we use the term ``atom'' to refer to either atoms or molecules as same conceptual arguments will apply to either system.} spends in the excited state on the correct cycle can be associated with an average fraction of particles in the wrong cycle. In the $\pi$-pulse model, this fraction can be obtained from the optical pulse shape and interval between consecutive beatnote pulses, and it is also what determines the average excited state population in an ensemble, as well as the photon scattering rate. Assuming that the fraction of time the atom spends in the excited state in the correct cycle is $\varepsilon$ and the natural decay rate of the excited state is $\Gamma$, the scattering rate for these atoms is simply $\varepsilon\Gamma$, which is to say that on average $\varepsilon$ fraction of them undergoes a decay. On the opposite cycle, the atom spends $1-\varepsilon$ fraction of time in the excited state, so then the scattering rate is $(1-\varepsilon)\Gamma$.

\par The photon scattering process is a random process described by a Poisson distribution. Therefore, the process of changing cycles is a random process as well, and the waiting time between events can be modelled as an exponential distribution with rate $\lambda_1=\varepsilon\Gamma$ on the correct cycle, and $\lambda_2=(1-\varepsilon)\Gamma$ on the opposite one. Hence, we consider a CTMC with two different transition rates - the correct-cycle state $C$ transitions to the wrong-cycle state $W$ with rate $\lambda_1$, while the wrong-cycle state $W$ transitions to state $C$ with rate $\lambda_2$. Graph shown in Fig. \ref{fig:CTMC_BCF} depicts this configuration:

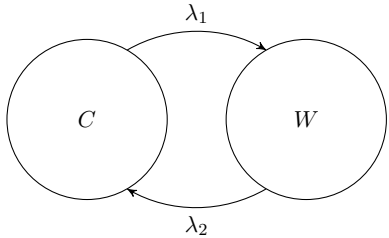
\begin{figure}[!h]
\centering

\begin{tikzpicture}[
scale=0.8, every node/.style={transform shape}
]
\draw[->,>=stealth'] (-3em,3em) arc[radius=6em, start angle=120, end angle=60] node[midway,above] {$\lambda_1$};
\draw[->,>=stealth'] (3em,-3em) arc[radius=6em, start angle=-60, end angle=-120] node[midway,below] {$\lambda_2$};

\draw (-4.732em,0) circle (3.46em) node{$C$};
\draw (4.732em,0) circle (3.46em) node{$W$};

\end{tikzpicture}
\caption{\small Simple schematic of BCF $\pi$-pulse model. The atom moves from state $C$ to $W$ with rate $\lambda_1$ and from $W$ to $C$ with rate $\lambda_2$.} \label{fig:CTMC_BCF}
\end{figure}

\par In this picture, to obtain both mean force and variance of momentum transfer we need to know probability $\mathbf{P}(t)$ of occupying one of the two states as a function of time. We can find it by using the Kolmogorov forward equation $\mathbf{P}'(t)=\mathbf{P}(t)\mathbf{Q}$, where $\mathbf{Q}$ is the generator matrix. In such CTMC the generator matrix is simply \cite{resnick1992adventures}:
\begin{displaymath}
\mathbf{Q}=\begin{bmatrix}
-\lambda_1&\lambda_1\\
\lambda_2&-\lambda_2
\end{bmatrix}
\end{displaymath}

\noindent Solution to this equation is:
\begin{displaymath}
\mathbf{P}(t)=\mathbf{P}(0)\exp(\mathbf{Q}t).
\end{displaymath}

\noindent By finding eigenvalues and eigenvectors of the generator matrix, one can calculate the exponent of the $\mathbf{Q}t$ matrix. In the end, by using $\mathbf{P}(0)=\mathit{I}$ we obtain: 
\begin{equation}
\mathbf{P}(t)=\begin{bmatrix}
\frac{\lambda_2}{\lambda_1+\lambda_2} +\frac{\lambda_1}{\lambda_1+\lambda_2}e^{-(\lambda_1+\lambda_2)t}&\frac{\lambda_1}{\lambda_1+\lambda_2} -\frac{\lambda_1}{\lambda_1+\lambda_2}e^{-(\lambda_1+\lambda_2)t}\\[1em]
\frac{\lambda_2}{\lambda_1+\lambda_2} -\frac{\lambda_2}{\lambda_1+\lambda_2}e^{-(\lambda_1+\lambda_2)t}&\frac{\lambda_1}{\lambda_1+\lambda_2} +\frac{\lambda_2}{\lambda_1+\lambda_2}e^{-(\lambda_1+\lambda_2)t}
\end{bmatrix}.
\label{eqn:state_prob}
\end{equation}

\noindent Having a certain initial state $\boldsymbol{\alpha}=[\alpha_1,\alpha_2]$ with $\alpha_1+\alpha_2=1$, the probability of occupying state $C$ or $W$ is simply $\boldsymbol{\alpha}\mathbf{P}(t)$. However, the steady-state probabilities at $t\gg \frac{1}{\lambda_1 + \lambda_2}$ are independent of the initial state:

\begin{displaymath}
\mathbf{P}(t\rightarrow\infty)=\begin{bmatrix}
\frac{\lambda_2}{\lambda_1+\lambda_2}&\frac{\lambda_1}{\lambda_1+\lambda_2}\\[1em]
\frac{\lambda_2}{\lambda_1+\lambda_2}&\frac{\lambda_1}{\lambda_1+\lambda_2}
\end{bmatrix}.
\end{displaymath}

\noindent We can now evaluate the expected momentum transfer in time $T$. Before a spontaneous emission occurs, momentum of $2\hbar k$ is exchanged between the light field and the atom with approximate constant frequency of $\delta/\pi$, where $\delta$ is the detuning of one of the laser components of the 2-color force. Therefore, if we define random variables:
\begin{eqnarray}
\Theta_C(t)&=\int_0^t{\mathbbm{1}_C dt'}\nonumber\\[0.3em]
\Theta_W(t)&=\int_0^t{\mathbbm{1}_W dt'},\label{eqn:occ_time}
\end{eqnarray}

\noindent where $\mathbbm{1}_{C/W}$ is an indicator variable for atom's state\footnote{The generic label $X$ to indicate the cycle in which atom resides during the dynamics under the influence of the coherent stimulated forces should not be confused with the ground electronic state for molecules, which is also customarily denoted as $X$.} $X$ being $C/W$ at time $t'$, and random variable $\Theta(t)$ is state's so-called occupancy time, we can find the expected value of momentum exchange in a given state given the initial condition $\boldsymbol{\alpha}$:

\begin{equation*}
E_{\alpha}p_X=2\hbar k \frac{\delta}{\pi}E_{\alpha}\Theta_X(t)=2\hbar k \frac{\delta}{\pi} \int_0^t{E_{\alpha}\mathbbm{1}_X dt'}=2\hbar k \frac{\delta}{\pi} \int_0^t{(\boldsymbol{\alpha}\mathbf{P})_X(t')dt'}.
\end{equation*}

\noindent The last part of that formula is the probability of occupying state $X$ at time $t'$ given initial state $\boldsymbol{\alpha}$. For example, for $\boldsymbol{\alpha}=[1,0]$ and $X=C$:
\begin{displaymath}
(\boldsymbol{\alpha}\mathbf{P})_C(t)=P_{11}(t)=\frac{\lambda_2}{\lambda_1+\lambda_2} +\frac{\lambda_1}{\lambda_1+\lambda_2}e^{-(\lambda_1+\lambda_2)t}.
\end{displaymath}

\par In our model, the atom moves in one direction in state $C$ and in the opposite direction in state $W$. Therefore, the expected total momentum transfer at time $T$ is: 
\begin{align}
E_{\alpha}p&=2\hbar k \frac{\delta}{\pi}E_{\alpha}(\Theta_C(T)-\Theta_W(T))\nonumber\\
&=2\hbar k \frac{\delta}{\pi} \int_0^T{[(\boldsymbol{\alpha}\mathbf{P})_{X=C}(t)-(\boldsymbol{\alpha}\mathbf{P})_{X=W}(t)]dt}. \label{eqn:exp_momentum}
\end{align}

\noindent The above equation can also be thought of as part of time-averaged expected value of the force:
\begin{displaymath}
\expval{F}=\frac{1}{T}\int_0^T f(t)dt=\frac{1}{T}\int_0^T \frac{dp}{dt}dt,
\end{displaymath}

\noindent with instantaneous force given by the integrand in Eq.(\ref{eqn:exp_momentum}). For previously chosen initial condition of $\boldsymbol{\alpha}=[1,0]$, Eq.(\ref{eqn:exp_momentum}) becomes:
\begin{displaymath}
E_{\alpha}p=2\hbar k \frac{\delta}{\pi}\int_0^T{\left(\frac{\lambda_2-\lambda_1}{\lambda_1+\lambda_2}+\frac{2\lambda_1}{\lambda_1+\lambda_2}e^{-(\lambda_1+\lambda_2)t}\right)dt}.
\end{displaymath}

\noindent Integration leads to:
\begin{displaymath}
E_{\alpha}p=2\hbar k \frac{\delta}{\pi}\left(\frac{\lambda_2-\lambda_1}{\lambda_1+\lambda_2}T+\frac{2\lambda_1}{(\lambda_1+\lambda_2)^2}\left(1-e^{-(\lambda_1+\lambda_2)T}\right)\right),
\end{displaymath}

\noindent which for $T\gg \left(\lambda_1 + \lambda_2\right)^{-1}$ simplifies to: 
\begin{displaymath}
E_{\alpha}p=2\hbar k \frac{\delta}{\pi}\frac{\lambda_2-\lambda_1}{\lambda_1+\lambda_2}T.
\end{displaymath}

\noindent Average force is then:
\begin{displaymath}
\expval{F}=2\hbar k \frac{\delta}{\pi}\left(\frac{\lambda_2-\lambda_1}{\lambda_1+\lambda_2}+\frac{2\lambda_1}{(\lambda_1+\lambda_2)^2}\frac{1-e^{-(\lambda_1+\lambda_2)T}}{T}\right),
\end{displaymath}

\noindent and it also simplifies to:
\begin{displaymath}
\expval{F}=2\hbar k \frac{\delta}{\pi}\frac{\lambda_2-\lambda_1}{\lambda_1+\lambda_2},
\end{displaymath}

\noindent where the last term can be thought of as proportion of time the atom or molecules spends in state $C$ minus proportion of time it spends in state $W$.

\par Notice that the result above is independent of the initial condition we have chosen, validating our initial assumption. That independence, of course, is related to the independence of the stationary state of CTMC on the initial conditions. The result for  large times $T$ can be also obtained simply from $\mathbf{P}(t\rightarrow \infty)$ - one can show that a time-averaged function of the states (here, state occupancy) is simply the expected value of the function with respect to the stationary distribution. Other way of deriving the average force in this model, would be to calculate average reward (force) per cycle.
\par If we plug in values for rates $\lambda_1$ and $\lambda_2$, we obtain:
\begin{equation}
\expval{F}=2\hbar k \frac{\delta}{\pi}(1-2\varepsilon), \label{eqn:exp_force}
\end{equation}

\noindent which for $\varepsilon=1/4$ is simply:
\begin{displaymath}
\expval{F}=\frac{\hbar k\delta}{\pi},
\end{displaymath}

\noindent agreeing with values estimated using other methods \cite{Soding1997,Yatsenko2004}. We can also re-write Eq.(\ref{eqn:exp_force}):
\begin{equation}
F_{\mathrm{II}}\equiv\expval{F}=\frac{\hbar k\delta}{\pi}\mu_{\mathrm{II}} (\varepsilon), \label{eqn:exp_force_mu}
\end{equation}

\noindent where $\mu_{\mathrm{II}}(\varepsilon)=2(1-2\varepsilon)$, and the Roman numeral II associates the quantity with a 2-state system.
\par To estimate variance of the momentum transfer we first note a few general things about the occupancy time random variables. First, it should be obvious that $\Theta_C(T)+\Theta_W(T)=T$, and thus:
\begin{displaymath}
\mathrm{Var}\,\Theta_C(T)=\mathrm{Var}\,(T-\Theta_W(T))=\mathrm{Var}\,\Theta_W(T).
\end{displaymath}

\noindent We also see that: 
\begin{align*}
0=\mathrm{Var}\,T&=\mathrm{Var}\,(\Theta_C(T)+\Theta_W(T))\\
&=\mathrm{Var}\,\Theta_C(T)+\mathrm{Var}(\Theta_W(T))+2\mathrm{Cov}\,(\Theta_C(T),\Theta_W(T))\\
&=2\mathrm{Var}\,\Theta_C(T)+2\mathrm{Cov}\,(\Theta_C(T),\Theta_W(T)),
\end{align*}

\noindent which shows us that $\mathrm{Cov}\,(\Theta_C(T),\Theta_W(T))=-\mathrm{Var}\,\Theta_C(T)$. Because the momentum transfer depends on the difference between occupancy times, we need to find:
\begin{displaymath}
\mathrm{Var}\,(\Theta_C(T)-\Theta_W(T))=4\,\mathrm{Var}\,\Theta_C(T).
\end{displaymath}

\noindent We therefore need to only find the variance in occupancy time of one of the CTMC states. 
\par To calculate variance in momentum transfer for a state up to time $T$ we start with writing the definition of variance: 
\begin{displaymath}
\mathrm{Var}\,p=(E_{\alpha}p^2)-(E_{\alpha}p)^2,
\end{displaymath}

\noindent where the last term was part of the previous calculation. We concentrate on the first term:
\begin{equation*}
E_{\alpha}p^2=4\hbar^2 k^2 \frac{\delta^2}{\pi^2}E_{\alpha}(\Theta_{C}(T)\Theta_{C}(T))=4\hbar^2 k^2 \frac{\delta^2}{\pi^2}\int_0^TE_{\alpha}(\mathbbm{1}_C(s)\mathbbm{1}_C(t))dsdt.
\end{equation*}

\noindent The expectation value of indicator random variables has to be calculated with care. We can write it in the following way assuming $s<t$: 
\begin{align*}
E_{\alpha}(\mathbbm{1}_C(s)\mathbbm{1}_C(t))&=P_{\alpha}(X(s)=C,X(t)=C)\\
&=P(X(t)=C|X(s)=C)P_{\alpha}(X(s)=C).
\end{align*}

\noindent Given previously chosen initial conditions and remembering that Markovian process is memoryless:
\begin{align*}
P_{\alpha}(X(s)=C)&=P_{11}(s)=\frac{\lambda_2}{\lambda_1+\lambda_2} +\frac{\lambda_1}{\lambda_1+\lambda_2}e^{-(\lambda_1+\lambda_2)s}\\
P(X(t)=C|X(s)=C)&=[1,0]\,\mathbf{P}_{X=C}(t-s)=P_{11}(t-s)\\
&=\frac{\lambda_2}{\lambda_1+\lambda_2}+\frac{\lambda_1}{\lambda_1+\lambda_2}e^{-(\lambda_1+\lambda_2)(t-s)}.
\end{align*}

\noindent After multiplication of above terms and re-defining $s\equiv\min{(s,t)}$ and $t\equiv\max{(s,t)}$, we obtain (for $\boldsymbol{\alpha}=[1,0]$):
\begin{flalign}
P_{\alpha}(X(s)=C,X(t)=C)&=\frac{1}{(\lambda_1+\lambda_2)^2}\left(\lambda_2^2+\lambda_1^2e^{-(\lambda_1+\lambda_2)\max{(s,t)}}+\right.\nonumber\\
&\left.+\lambda_1\lambda_2e^{-(\lambda_1+\lambda_2)\min{(s,t)}}+\lambda_1\lambda_2e^{-(\lambda_1+\lambda_2)|s-t|}\right). \label{eqn:P_var}
\end{flalign}

\noindent Integrating Eq.(\ref{eqn:P_var}) and using previously found expectation value of momentum transfer, one finds variance of the occupancy time: 
\begin{align}
\mathrm{Var}\,\Theta_C(T)=\frac{\lambda_1(\lambda_1-4\lambda_2)}{(\lambda_1+\lambda_2)^4}&+\frac{2\lambda_1\lambda_2}{(\lambda_1+\lambda_2)^3}T+\frac{4\lambda_1\lambda_2}{(\lambda_1+\lambda_2)^4}e^{-(\lambda_1+\lambda_2)T}+\nonumber\\[0.3em]
&+\frac{2\lambda_1(\lambda_2-\lambda_1)}{(\lambda_1+\lambda_2)^3}Te^{-(\lambda_1+\lambda_2)T}-\frac{\lambda_1^2}{(\lambda_1+\lambda_2)^4}e^{-2(\lambda_1+\lambda_2)T}. \label{eqn:var_occupancy}
\end{align}

\noindent The above result has a constant term, term linear in time, exponentially decaying terms and a mixed term. In general, it can be found that variance of a reward in CTMC can only have specific terms\footnote{The variance $\mathbf{V}(t)=\bm{\gamma} t+\mathbf{h}+\mathbf{c}(t)+\bm{\varepsilon}(t)$, where $\bm{\gamma}$ is a constant ``growth rate'' vector, $\mathbf{h}$ is a constant vector, vector $||\mathbf{c}(t)||\leq C$ is bounded for all $t$, and $\bm{\varepsilon}(t)$ is a vector function exponentially converging to 0.} \cite{VanDijk2006}, and terms in Eq.(\ref{eqn:var_occupancy}) fall into that category. For large times $T\rightarrow\infty$ only one term survives, and so the overall variance in momentum transfer can be found to be:
\begin{equation}
\mathrm{Var}\,p=4\hbar^2 k^2 \frac{\delta^2}{\pi^2}\times4\frac{2\lambda_1\lambda_2}{(\lambda_1+\lambda_2)^3}T=32\hbar^2 k^2 \frac{\delta^2}{\pi^2\Gamma}\varepsilon(1-\varepsilon)T. \label{eqn:var_momentum}
\end{equation}

\noindent This also allows us to find the diffusion coefficient $D$:
\begin{equation}
D\equiv\frac{1}{2}\frac{\partial}{\partial t}\mathrm{Var}\,p=16\,\hbar^2 k^2 \frac{\delta^2}{\pi^2\Gamma}\varepsilon(1-\varepsilon) \label{eqn:diff_coeff}
\end{equation}

\noindent Like before, we can define a new variable, $\sigma^2_{\mathrm{II}}(\varepsilon)=16\,\varepsilon(1-\varepsilon)$ and use the average force we found previously in Eq.(\ref{eqn:exp_force_mu}) to obtain:
\begin{equation}
D=F^2_{\mathrm{II}}\frac{\sigma^2_{\mathrm{II}}(\varepsilon)}{\mu^2_{\mathrm{II}}(\varepsilon)} \frac{1}{\Gamma}\label{eqn:diff_coeff_musigma}
\end{equation}

\par
\begin{figure}[!h]
\centering
\subfloat[\small $T=500\,\Gamma^{-1}$.]{{
\includegraphics[width=0.49\linewidth]{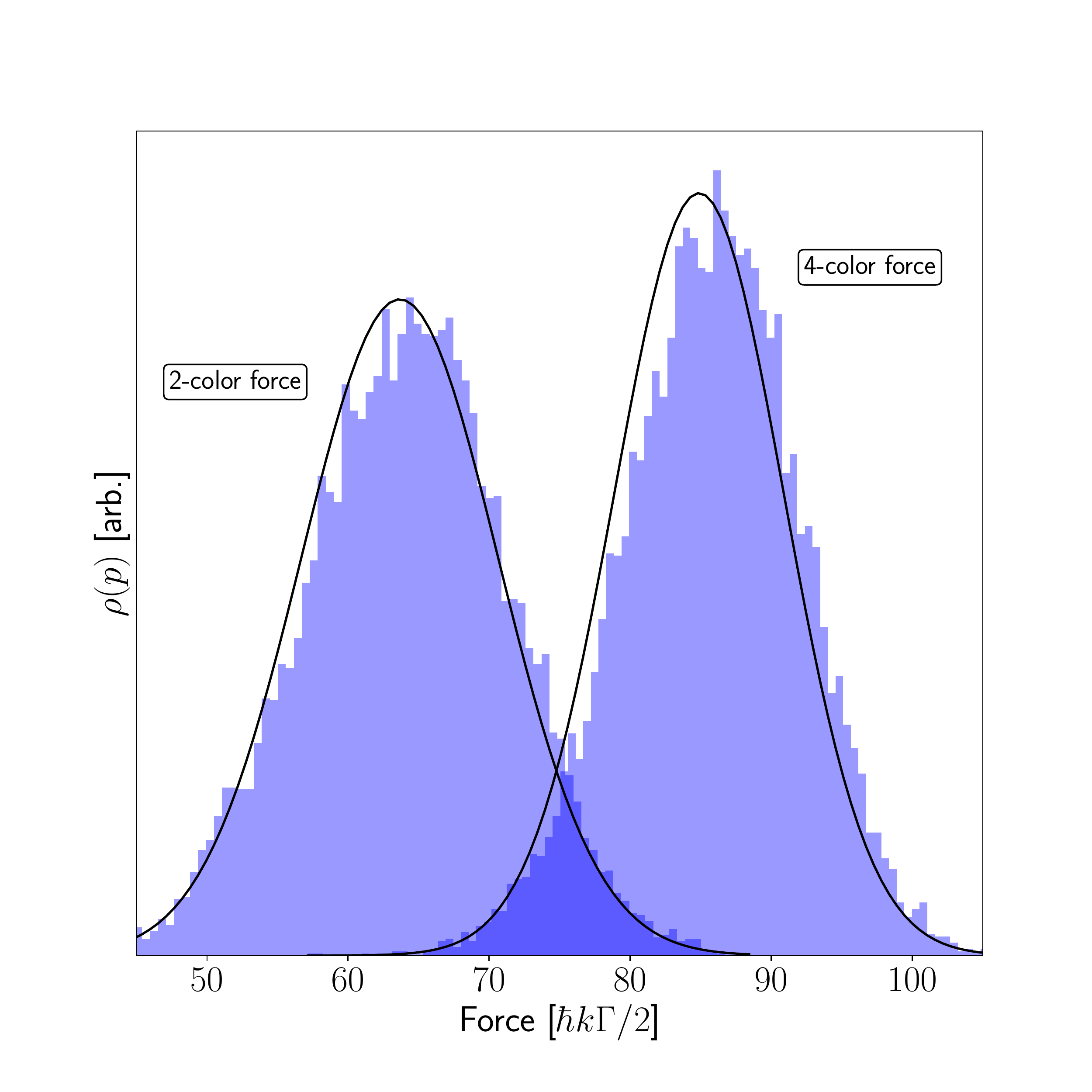}
}}
\subfloat[\small $T=5\times 10^4\,\Gamma^{-1}$.]{{
\includegraphics[width=0.49\linewidth]{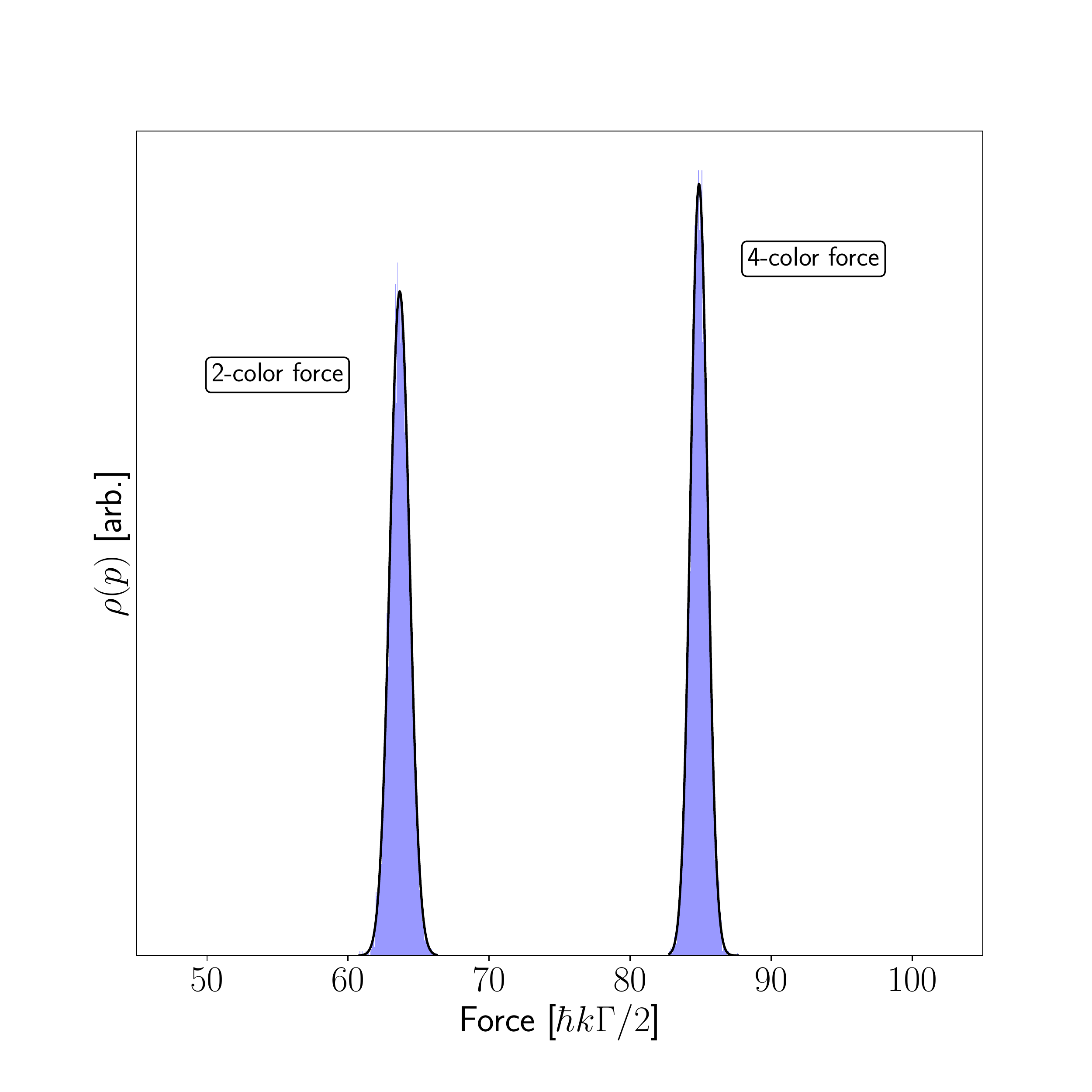}
}}
\caption{\small Histograms showing simulated force profiles for 2- and 4-color forces for $T = 500\,\Gamma^{-1}$ (a) and $T = 5\times 10^4\,\Gamma^{-1}$ (b) interaction time. Black lines represent Gaussian distributions with mean and variance estimated using Eq.(\ref{eqn:exp_force}) and Eq.(\ref{eqn:var_momentum}), not the fitted distribution to the histograms. The distribution is narrower for 4-color force (smaller $\varepsilon$) and for longer interaction times. Note that for shorter interaction times, while the distributions are already close to being Gaussian, the estimated mean and variance did not yet converge to predicted values.}
\label{fig:force_histograms}
\end{figure}

 There are several interesting aspects we can notice about the expression for calculated variance in momentum transfer presented in Eq.(\ref{eqn:var_momentum}). Firstly, it grows linearly with time, which is the same as for the radiative force. Secondly, it follows $\propto \delta^2$ proportionality, which for polychromatic forces having detunings of the order of $100\,\Gamma$ means that the variance will be quite substantial. However, if we consider ratio of the standard deviation to the mean:
\begin{displaymath}
\frac{\sqrt{\expval{p^2}-\expval{p}^2}}{\expval{p}}=2\sqrt{2}\frac{\sqrt{\varepsilon(1-\varepsilon)}}{1-2\varepsilon}\frac{1}{\sqrt{\Gamma\,T}},
\end{displaymath}

\noindent we realize that the distribution becomes narrower the longer the process. In Fig. \ref{fig:force_histograms} we show histograms of simulated polychromatic forces obtained from our model. The distribution of the force follows the distribution of the occupancy times. Occupancy time in single state in our model is a sum of independent exponential random variables and is therefore distributed with an Erlang distribution (special case of Gamma distribution), with mean and variance determined by the CTMC. Force, being proportional to difference in occupancy times, is distributed as difference of two Gamma distributions and closed form of its moment-generating function can be found \cite{Mathai1993}. Fortunately, for longer interaction times, due to the Central Limit Theorem, the distributions approach a Gaussian distribution, which we included in our figure. Its mean is given by Eq.(\ref{eqn:exp_force}), while its variance is related to Eq.(\ref{eqn:var_momentum}) - if $\sigma^2_p$ is variance in the momentum transfer distribution, $\sigma^2_f=\sigma_p^2/T^2$ will be the variance in the force distribution.

\par The value of the aforementioned ratio grows for $\varepsilon$ approaching $1/2$, but it, as well as the variance itself, can be brought arbitrarily close to zero by making $\varepsilon$ small. For instance,  adding additional colors reduces value of this parameter. At optimum force, $\varepsilon\approx\chi/\pi$, where $\chi$ is phase difference between counter-propagating beams in the polychromatic force (refer to Fig. \ref{fig:bcf_schematic}). For example, while for a 2-color force $\chi=\pi/4$, for a 4-color force $\chi\approx\pi/6$, and so $\varepsilon\approx 0.167$. Effects of decreasing $\varepsilon$ can be seen in Fig. \ref{fig:force_histograms}.
\par It is also worth noting that the velocity diffusion (known as beam ``pluming'') that could be observed when performing slowing or deflection using bichromatic forces should be pretty small comparing to the overall effect observed due to what was just mentioned. 
\par Finally, for all polychromatic forces, the Rabi rate $\Omega$ is of the order of the detuning $\delta$, so the variance $\mathrm{Var}\, p \propto \Omega^2$. Already Cohen-Tannoudji divided optical forces into two categories depending on their origin: dissipative and reactive \cite{CHT}. Polychromatic forces are reactive according to that definition, just like dipole forces. He showed that for such forces the momentum dissipation tensor $D$, which is associated with variance in momentum transfer and was shown in Eq.(\ref{eqn:diff_coeff_musigma}), should scale as square of the Rabi rate, which is consistent with our result. It is also consistent with value obtained in Refs. \cite{Partlow2004,Dalibard1985}.
\par Additionally, we can connect the average excited state population in an ensemble $\rho_{ee}$ with the average time $\varepsilon$ a single atom spends in the excited state, which simultaneously is the fraction of atoms currently in the wrong cycle. These can be tied together in a very simple way: $\varepsilon$ fraction of atoms spends $1-\varepsilon$ time on average in the excited state, while $1-\varepsilon$ of them spends $\varepsilon$ fraction of time in the excited state. We can then write: 
\begin{displaymath}
\rho_{ee}=2\,\varepsilon(1-\varepsilon).
\end{displaymath}

\noindent Because $0\leq\varepsilon\leq 0.5$ we obtain:
\begin{displaymath}
\varepsilon=\frac{1-\sqrt{1-2\rho_{ee}}}{2}.
\end{displaymath}

\noindent Finally, we can re-write the expected value of force and time-averaged variance of momentum transfer in terms of $\rho_{ee}$:
\begin{align}
\expval{F}&=2\sqrt{1-2\rho_{ee}}\frac{\hbar k\delta}{\pi}\label{eqn:force_bcf}\\[0.3em]
\frac{1}{T}\mathrm{Var}\,p&=16\rho_{ee}\left(\frac{\hbar k \delta}{\pi}\right)^2\frac{1}{\Gamma}.\label{eqn:pvar_bcf}
\end{align}

\section{Variance Estimation in PCF Molasses \label{sec:App-B}}

\par
\begin{figure}[!h]
\centering

\begin{tikzpicture}[
scale=0.9, every node/.style={transform shape}
]
\draw[->,>=stealth'] (-3.5cm,-2.134cm) arc[radius=9cm, start angle=193.72, end angle=166.28] node[midway,left] {$r_1\lambda_{C_1}$};
\draw[->,>=stealth'] (-2.5cm,2.134cm) arc[radius=9cm, start angle=13.72, end angle=-13.72] node[midway,left] {$r_1\lambda_{W_1}$};

\draw[->,>=stealth'] (2.5cm,-2.134cm) arc[radius=9cm, start angle=193.72, end angle=166.28] node[midway,right] {$r_2\lambda_{C_2}$};
\draw[->,>=stealth'] (3.5cm,2.134cm) arc[radius=9cm, start angle=13.72, end angle=-13.72] node[midway,right] {$r_2\lambda_{W_2}$};

\draw[->,>=stealth'] (-2.134cm,3.5cm) arc[radius=9cm, start angle=103.72, end angle=76.28] node[midway,above] {$(1-q_2)(1-r_1)\lambda_{W_1}$};
\draw[->,>=stealth'] (2.134cm,2.5cm) arc[radius=9cm, start angle=-76.28, end angle=-103.72] node[midway,above] {$(1-q_1)(1-r_2)\lambda_{W_2}$};

\draw[->,>=stealth'] (-2.134cm,-2.5cm) arc[radius=9cm, start angle=103.72, end angle=76.28] node[midway,below] {$q_2(1-r_1)\lambda_{C_1}$};
\draw[->,>=stealth'] (2.134cm,-3.5cm) arc[radius=9cm, start angle=-76.28, end angle=-103.72] node[midway,below] {$q_1(1-r_2)\lambda_{C_2}$};

\draw[->,>=stealth'] (2.2627cm,-2.3244cm) arc[radius=15cm, start angle=-122.51, end angle=-147.49] node[pos=0.32,below,sloped] {$(1-q_1)(1-r_2)\lambda_{C_2}$};
\draw[->,>=stealth'] (-2.2627cm,2.3244cm) arc[radius=15cm, start angle=57.49, end angle=32.51] node[pos=0.25,above,sloped] {$q_2(1-r_1)\lambda_{W_1}$};

\draw[->,>=stealth'] (-2.3244cm,-2.2627cm) arc[radius=15cm, start angle=147.49, end angle=122.51] node[pos=0.32,above,sloped] {$(1-q_2)(1-r_1)\lambda_{C_1}$};
\draw[->,>=stealth'] (2.3244cm,2.2627cm) arc[radius=15cm, start angle=-32.51, end angle=-57.49] node[pos=0.27,above,sloped] {$q_1(1-r_1)\lambda_{W_2}$};

\draw (-3cm,3cm) circle (1cm) node{$W_1$};
\draw (3cm,3cm) circle (1cm) node{$W_2$};
\draw (-3cm,-3cm) circle (1cm) node{$C_1$};
\draw (3cm,-3cm) circle (1cm) node{$C_2$};

\end{tikzpicture}
\caption{\small Schematic for a $\pi$-pulse model PCF molasses.} \label{fig:CTMC_BCF_mol}
\end{figure}
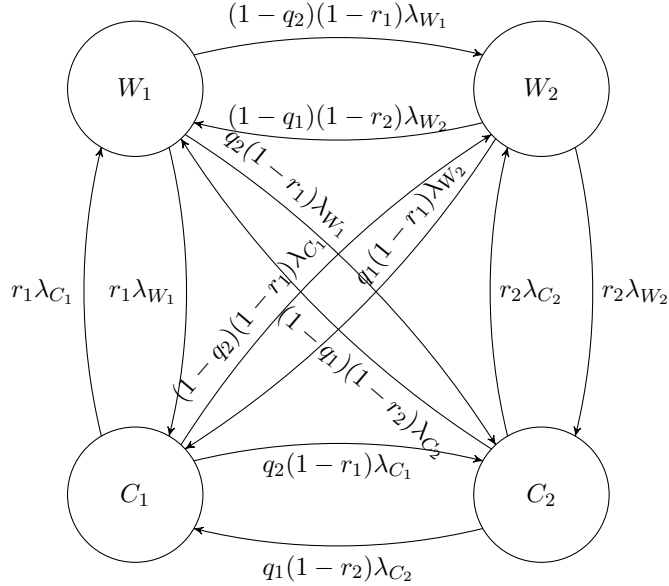

 To estimate variance in momentum transfer and, from there, the limiting temperature in SupER molasses, we will use the CTMC model introduced in App. \ref{sec:Variance-estimation}. In all generality, the diagram of our system is depicted in Fig. \ref{fig:CTMC_BCF_mol}. We consider 4 states: $C_1$, $W_1$, $C_2$ and $W_2$. The first two correspond to states an atom can be in when it is feeling PCF acting on one of two 2-level systems (Fig. \ref{fig:BCF_BaH_diagram}). Accordingly, states $C_2$ and $W_2$ correspond to the other 2-level system. States marked with letter $C$ are states, where, like in the simple PCF model in the previous section, the atom spends most of its time (``correct'' cycle). These states in PCF molasses will create force in opposing directions. Similarly, cycles marked with letter $W$ are the ones, where an atom spends less time (``wrong'' cycles). In those cycles the momentum is transferred in direction opposite to the direction in their respective $C$ states.
\par We assign the average time spent by atoms in states $C$ in the excited states $\ket{E1}$ and $\ket{E2}$ (Fig. \ref{fig:BCF_BaH_diagram}) with respect to the total time spent in one 2-level system as $\varepsilon_1$, $\varepsilon_2$ and $1-\varepsilon_1$, $1-\varepsilon_2$ for average proportion of time spent in states $\ket{G1}$ and $\ket{G2}$ respectively. Having defined these variables we can find rates for all of our states:
\begin{align*}
\lambda_{C_1}&=\varepsilon_1\Gamma_1\ &\lambda_{C_2}&=\varepsilon_2\Gamma_2\\[0.3em]
\lambda_{W_1}&=(1-\varepsilon_1)\Gamma_1\ &\lambda_{W_2}&=(1-\varepsilon_2)\Gamma_2.
\end{align*}

\noindent These rates are set up in a similar fashion as in the simple PCF model - we assume that when an atom is in one of the 2-level systems the situation is just like the model analyzed in App. \ref{sec:Variance-estimation}. Then, the rate at which the atom leaves the state due to spontaneous emission is just the natural decay rate $\Gamma$ times the proportion of time it spends in that excited state considering only the 2-level system in which the cycle of excitation and stimulated emission occurs for this atom or molecule, which is simply $\varepsilon$ or $1-\varepsilon$.

\par In our 4-state model the decay can always go to all three other states. A spontaneous decay that doesn't cause an atom or molecule to change a 2-level system (so when it switches between states $C_1$ and $W_1$ or $C_2$ and $W_2$) is assumed to have a branching ratio $r_1$ and $r_2$ for the first and second two level system respectively. Switching 2-level system occurs with branching ratios of $1-r_1$ and $1-r_2$. Comparing that to Fig. \ref{fig:BCF_BaH_diagram} gives:
\begin{align*}
\Gamma_{11}&=r_1\Gamma_1\ &\Gamma_{21}&=(1-r_2)\Gamma_2\\
\Gamma_{12}&=(1-r_1)\Gamma_1\ &\Gamma_{22}&=r_2\Gamma_2.
\end{align*}

\noindent When such a switch of a 2-level system happens, the decay can lead to either cycle $C$ or $W$, but the probabilities are not equal. In Fig. \ref{fig:CTMC_BCF_mol} probability of entering state $C_i$ is labeled $q_i$. However, in our situation they are simply the proportions of time an atom spends in respective states within each 2-level system, so $q_i=1-\varepsilon_i$\footnote{Exact probability should depend on phase difference between both 2-level subsystems' pulse trains. Here, we assume no phase coherence and, therefore, non-zero probability of ending in either of the 4 states.}. We now can write the generator matrix for this CTMC:
\[
\hspace*{-2cm}
Q\,=\,\bordermatrix{
&C_1&W_1&C_2&W_2\cr
C_1&-\varepsilon_1\Gamma_1&r_1\varepsilon_1\Gamma_1&(1-\varepsilon_2)(1-r_1)\varepsilon_1\Gamma_1&\varepsilon_2(1-r_1)\varepsilon_1\Gamma_1\cr
W_1&r_1(1-\varepsilon_1)\Gamma_1&-(1-\varepsilon_1)\Gamma_1&(1-\varepsilon_2)(1-r_1)(1-\varepsilon_1)\Gamma_1&\varepsilon_2(1-r_1)(1-\varepsilon_1)\Gamma_1\cr
C_2&(1-\varepsilon_1)(1-r_2)\varepsilon_2\Gamma_2&\varepsilon_1(1-r_2)\varepsilon_2\Gamma_2&-\varepsilon_2\Gamma_2&r_2\varepsilon_2\Gamma_2\cr
W_2&(1-\varepsilon_1)(1-r_2)(1-\varepsilon_2)\Gamma_2&\varepsilon_1(1-r_2)(1-\varepsilon_2)\Gamma_2&r_2(1-\varepsilon_2)\Gamma_2&-(1-\varepsilon_2)\Gamma_2
}.
\]

\par Such system is, however, difficult to solve analytically. Even finding the expected occupancy time or the stationary state $\eta$ of this CTMC, which corresponds to a left eigenvector associated with the zero eigenvalue, is very challenging. Fortunately, the symmetrized version of the system simplifies the situation. Before we move forward, we note a few characteristics of the expected value of the force in this more general system that emulates the population dynamics during the SupER molasses cooling process.  
\par If both 2-level systems are kept at optimum, for example with optimal parameters of $\Omega_{1,2}=\sqrt{3/2}\,\delta_{1,2}$ and $|\chi|=45^{\circ}$ for BCF, in the stationary state the proportion of time spent in state $C_1$ or $W_1$ with respect to total time spent in the first 2-level system (i.e. in either of those states) is the same as corresponding proportions in the second 2-level system:
\begin{displaymath}
\frac{\eta_{C_1}}{\eta_{C_1}+\eta_{W_1}}=\frac{\eta_{C_2}}{\eta_{C_2}+\eta_{W_2}}\qquad\frac{\eta_{W_1}}{\eta_{C_1}+\eta_{W_1}}=\frac{\eta_{W_2}}{\eta_{C_2}+\eta_{W_2}},
\end{displaymath}

\noindent where we assume that stationary state ratios are not necessarily the average times spent in energy levels of the system ($\varepsilon_i$ and $1-\varepsilon_i$). In the end, the expected time-averaged value of the force for both 2-level systems should be:
\begin{align*}
F_1&=2\hbar k_1\frac{\delta_1}{\pi}(\eta_{C_1}-\eta_{W_1})\\
F_2&=-2\hbar k_2\frac{\delta_2}{\pi}(\eta_{C_2}-\eta_{W_2}).
\end{align*}

\noindent In the formulas above we already assumed that both 2-level systems would generate opposing forces (due to $\chi_1=-\chi_2$). In molasses we'd like $F_1=F_2$, so that there's no net force at zero velocity. Taking that condition and by multiplying both sides by 1 we obtain:
\begin{displaymath}
k_1\delta_1\frac{\eta_{C_1}-\eta_{W_1}}{\eta_{C_1}+\eta_{W_1}}(\eta_{C_1}+\eta_{W_1})=k_2\delta_2\frac{\eta_{C_2}-\eta_{W_2}}{\eta_{C_2}+\eta_{W_2}}(\eta_{C_2}+\eta_{W_2}).
\end{displaymath}

\noindent Because the ratios are the same for both 2-level systems, these terms will drop out:
\begin{displaymath}
k_1\delta_1(\eta_{C_1}+\eta_{W_1})=k_2\delta_2(\eta_{C_2}+\eta_{W_2}).
\end{displaymath}

\noindent We're now left with total proportions of time spent in the first and second 2-level systems. Atom decays from the first 2-level system to the second with a total rate $(1-r_1)\Gamma_1$ and from the second back to the first with a rate $(1-r_2)\Gamma_2$. This creates its own two-state CTMC, which we have already solved, so:
\begin{align*}
\eta_{C_1}+\eta_{W_1}&=\frac{(1-r_2)\Gamma_2}{(1-r_1)\Gamma_1+(1-r_2)\Gamma_2}\\
\eta_{C_2}+\eta_{W_2}&=\frac{(1-r_1)\Gamma_1}{(1-r_1)\Gamma_1+(1-r_2)\Gamma_2}.
\end{align*}

\noindent Using the above, we finally arrive at a criterion for detunings $\delta_i$ that has to be met to properly balance power in an asymmetric system (like in the BaH molecule considered in Sec. \ref{sec:BaH-BCF-slowing} and \ref{sec:BaH-SupER-molasses}):
\begin{equation}
\frac{\delta_1}{\delta_2}=\frac{\hbar\omega_2}{\hbar\omega_1}\frac{\Gamma_1}{\Gamma_2}\frac{1-r_1}{1-r_2}. \label{eqn:delta_criterion}
\end{equation}

\noindent Criterion shown in Eq.(\ref{eqn:delta_criterion}) can be understood intuitively: detuning, which determines rate of the cycle of spontaneous and stimulated emission, has to be higher, if energy of the scattered photon is smaller or if spontaneous emission causing switching of the 2-level system occurs more often, which is determined be either the decay rate or the branching ratio.

\par
\begin{figure}[!h]
\centering

\begin{tikzpicture}[
scale=0.9, every node/.style={transform shape}
]
\draw[->,>=stealth'] (-3.5cm,-2.134cm) arc[radius=9cm, start angle=193.72, end angle=166.28] node[midway,left] {$\frac{2}{3}\varepsilon\Gamma\ $};
\draw[->,>=stealth'] (-2.5cm,2.134cm) arc[radius=9cm, start angle=13.72, end angle=-13.72] node[midway,left] {$\frac{2}{3}(1-\varepsilon)\Gamma\!\!\!$};

\draw[->,>=stealth'] (2.5cm,-2.134cm) arc[radius=9cm, start angle=193.72, end angle=166.28] node[midway,right] {$\frac{2}{3}\varepsilon\Gamma$};
\draw[->,>=stealth'] (3.5cm,2.134cm) arc[radius=9cm, start angle=13.72, end angle=-13.72] node[midway,right] {$\frac{2}{3}(1-\varepsilon)\Gamma$};

\draw[->,>=stealth'] (-2.134cm,3.5cm) arc[radius=9cm, start angle=103.72, end angle=76.28] node[midway,above] {$\frac{1}{3}\varepsilon(1-\varepsilon)\Gamma$};
\draw[->,>=stealth'] (2.134cm,2.5cm) arc[radius=9cm, start angle=-76.28, end angle=-103.72] node[midway,above] {$\frac{1}{3}\varepsilon(1-\varepsilon)\Gamma$};

\draw[->,>=stealth'] (-2.134cm,-2.5cm) arc[radius=9cm, start angle=103.72, end angle=76.28] node[midway,below] {$\frac{1}{3}(1-\varepsilon)\varepsilon\Gamma$};
\draw[->,>=stealth'] (2.134cm,-3.5cm) arc[radius=9cm, start angle=-76.28, end angle=-103.72] node[midway,below] {$\frac{1}{3}(1-\varepsilon)\varepsilon\Gamma$};

\draw[->,>=stealth'] (2.2627cm,-2.3244cm) arc[radius=15cm, start angle=-122.51, end angle=-147.49] node[pos=0.32,below,sloped] {$\frac{1}{3}\varepsilon^2\Gamma$};
\draw[->,>=stealth'] (-2.2627cm,2.3244cm) arc[radius=15cm, start angle=57.49, end angle=32.51] node[pos=0.25,above,sloped] {$\frac{1}{3}(1-\varepsilon)^2\Gamma$};

\draw[->,>=stealth'] (-2.3244cm,-2.2627cm) arc[radius=15cm, start angle=147.49, end angle=122.51] node[pos=0.32,above,sloped] {$\frac{1}{3}\varepsilon^2\Gamma$};
\draw[->,>=stealth'] (2.3244cm,2.2627cm) arc[radius=15cm, start angle=-32.51, end angle=-57.49] node[pos=0.27,above,sloped] {$\frac{1}{3}(1-\varepsilon)^2\Gamma$};

\draw (-3cm,3cm) circle (1cm) node{$W_1$};
\draw (3cm,3cm) circle (1cm) node{$W_2$};
\draw (-3cm,-3cm) circle (1cm) node{$C_1$};
\draw (3cm,-3cm) circle (1cm) node{$C_2$};

\end{tikzpicture}
\caption{\small Schematic for a symmetrized and simplified $\pi$-pulse model PCF molasses.} \label{fig:CTMC_BCF_mol_simp}
\end{figure}
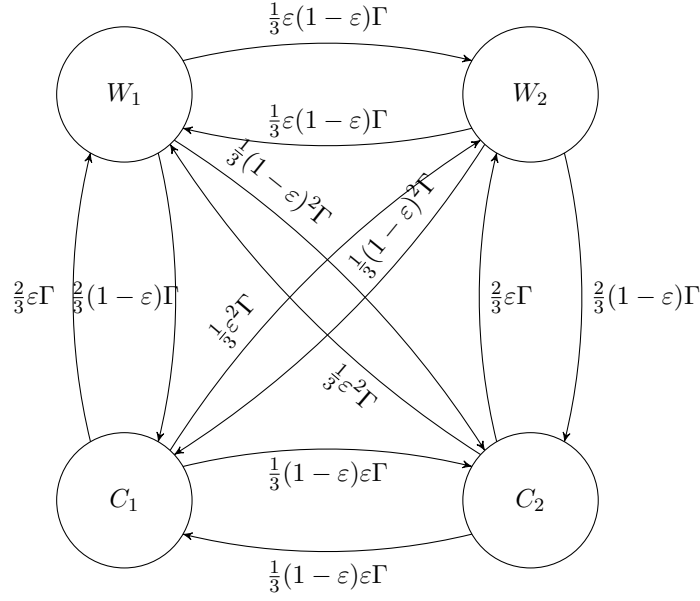

 Now, we can move forward with simplification of the model to obtain analytical estimates. As in the main section, we symmetrize the CTMC and assume that all relevant parameters are identical in both 2-level systems: $\Gamma_1=\Gamma_2\equiv\Gamma$, $\varepsilon_1=\varepsilon_2\equiv\varepsilon$, $\delta_1=\delta_2\equiv\delta$, $k_1=k_2\equiv k$. The simplified state graph for the CTMC is shown in Fig. \ref{fig:CTMC_BCF_mol_simp}. We also used values for branching ratios in BaH: $r_1=r_2=2/3$. This model was used in Monte Carlo simulations, results for which are shown in Sec. \ref{sec:MC-simulations}.

\newpage

\noindent This system has a generator matrix: 
\[
Q\,=\,\begin{pmatrix}
-\varepsilon\Gamma&\frac{2}{3}\varepsilon\Gamma&\frac{1}{3}(1-\varepsilon)\varepsilon\Gamma&\frac{1}{3}\varepsilon^2\Gamma\\[0.3em]
\frac{2}{3}(1-\varepsilon)\Gamma&-(1-\varepsilon)\Gamma&\frac{1}{3}(1-\varepsilon)^2\Gamma&\frac{1}{3}\varepsilon(1-\varepsilon)\Gamma\\[0.3em]
\frac{1}{3}(1-\varepsilon)\varepsilon\Gamma&\frac{1}{3}\varepsilon^2\Gamma&-\varepsilon\Gamma&\frac{2}{3}\varepsilon\Gamma\\[0.3em]
\frac{1}{3}(1-\varepsilon)^2\Gamma&\frac{1}{3}\varepsilon(1-\varepsilon)\Gamma&\frac{2}{3}(1-\varepsilon)\Gamma&-(1-\varepsilon)\Gamma
\end{pmatrix}
\]

\noindent For such simplified system we can identify the stationary distribution:
\begin{align*}
\eta_{C_1}=\eta_{C_2}=\frac{\varepsilon^2-4\varepsilon+3}{4\varepsilon^2-4\varepsilon+6}\\[0.3em]
\eta_{W_1}=\eta_{W_2}=\frac{\varepsilon(\varepsilon+2)}{4\varepsilon^2-4\varepsilon+6}
\end{align*}

\noindent As expected, in symmetrized system proportion of time spent in both 2-level systems is the same, that is $\eta_{C_1}+\eta_{W_1}=\eta_{C_2}+\eta_{W_2}=1/2$. And so the time-averaged expected value of the force created by both 2-level systems is:
\begin{align*}
\expval{F_1}=-\expval{F_2}=2\hbar k\frac{\delta}{\pi}(\eta_{C_1}-\eta_{W_1})&=2\hbar k\frac{\delta}{\pi}\frac{3-6\varepsilon}{4\varepsilon^2-4\varepsilon+6}\\
&=3\hbar k\frac{\delta}{\pi}\frac{1-2\varepsilon}{2\varepsilon^2-2\varepsilon+3}.
\end{align*}

\noindent Just like in the case of a 2-level system (see Eq. \ref{eqn:exp_force_mu}), we can write the force exerted through one of the subsystems as:
\begin{equation}
    F_{\mathrm{IV}}=\frac{\hbar k\delta}{\pi}\mu_{\mathrm{IV}}(\varepsilon), \label{eqn:avg_force_4lvl}
\end{equation}

\noindent with:
\begin{displaymath}
\mu_{\mathrm{IV}}(\varepsilon)=\frac{1-2\varepsilon}{\frac{2}{3}\varepsilon^2-\frac{2}{3}\varepsilon+1}.
\end{displaymath}

\par Calculating variance of the occupancy time, and therefore the momentum transfer, is more challenging. We first note that by using similar tricks as in the previous section, we can show that variance of the momentum transfer, which is proportional to $\Theta_{C_1}(T)-\Theta_{W_1}(T)-\Theta_{C_2}(T)+\Theta_{W_2}(T)$, is:
\begin{align*}
\mathrm{Var}\,p&=4\hbar^2 k^2\frac{\delta^2}{\pi^2}\mathrm{Var}\,\left(\Theta_{C_1}(T)-\Theta_{W_1}(T)-\Theta_{C_2}(T)+\Theta_{W_2}(T)\right)\\
&=4\hbar^2 k^2\frac{\delta^2}{\pi^2}\times 4\mathrm{Var}\,(\Theta_{C_1}(T)+\Theta_{W_2}(T))\\
&=16\hbar^2 k^2\frac{\delta^2}{\pi^2}\mathrm{Var}\,(\Theta_{C_1}(T)+\Theta_{W_2}(T))
\end{align*}

\noindent where $\mathrm{Var}\,\Theta(T)$ is variance of any of the occupancy times. To calculate it, we first take a step back and look at general solutions of the Kolmogorov forward equation. Because the generator matrix of CTMC is negative semi-definite, its eigenvalues are non-positive. The zero eigenvalue is related to the stationary distribution, while others add exponentially decaying parts to the matrix $\mathbf{P}$. In all generality, we can write: 
\begin{equation}
P_{ij}(t)=\eta_{j}+\sum_{k=1}^{n-1}{u^k_{ij}e^{\nu_k t}}, \label{eqn:P_matrix_gen}
\end{equation}

\noindent where $P_{ij}(t)$ describes probability of being in state $j$ at time $t$ given the system in state $i$ at time $0$. Therefore, $\eta_j$, the component of the stationary distribution for state $j$, is the same for any initial state $i$. Here, $\nu_k$ is the $k$-th eigenvalue ($\nu_0=0$) and $u^k_{ij}$ is a function of eigenvectors multiplying the exponential part. For an initial state $\boldsymbol{\alpha}$ and an $n$-state system we have: 
\begin{align*}
E_{\alpha}\Theta_j(T)&=\int_0^T{\boldsymbol{\alpha}\mathbf{P}(t)dt}=\int_0^T{\sum_{i=1}^n{\alpha_i P_{ij}}dt}\\[0.3em]
&=\int_0^T{\left[\sum_{i=1}^n{\alpha_i\left(\eta_{j}+\sum_{k=1}^{n-1}{u^k_{ij}e^{\nu_k t}}\right)}\right]dt}\\[0.3em]
&=\eta_j T\sum_{i=1}^n{\alpha_i}+\sum_{i=1}^n{\alpha_i}\sum_{k=1}^{n-1}{\frac{u^k_{ij}}{\nu_k}e^{\nu_k T}}-\sum_{i=0}^n{\alpha_i}\sum_{k=1}^{n-1}{\frac{u^k_{ij}}{\nu_k}}\\[0.3em]
&=\eta_j T+\sum_{i=1}^n\sum_{k=1}^{n-1}{\frac{\alpha_i u^k_{ij}}{\nu_k}e^{\nu_k T}}-\sum_{i=1}^n\sum_{k=1}^{n-1}{\frac{\alpha_i u^k_{ij}}{\nu_k}},
\end{align*}

\noindent where at the end we used the fact that $\sum_i\alpha_i=1$. From the above and the fact for all $k>0$ eigenvalues $\nu_k<0$, we easily see that at $T\rightarrow\infty$ only the $\eta_j$ term survives. In variance calculations we need the square of the expectation value:
\begin{align*}
\left[E_{\alpha}\Theta_j(T)\right]^2&=\eta^2_j T^2-2\eta_j\sum_{i=1}^n\sum_{k=1}^{n-1}{\frac{\alpha_i u^k_{ij}}{\nu_k}}T+\left(\sum_{i=1}^n\sum_{k=1}^{n-1}{\frac{\alpha_i u^k_{ij}}{\nu_k}}\right)^2+(\mathrm{exponentially\ decaying\ terms})\\[0.3em]
&\stackrel{T\rightarrow\infty}{=}\eta^2_j T^2-2\eta_j\sum_{i=1}^n\sum_{k=1}^{n-1}{\frac{\alpha_i u^k_{ij}}{\nu_k}}T.
\end{align*}

\par To obtain variance, we also require the expectation value of the square of the occupancy time. We first find that (for $t>s$): 
\begin{align*}
E_{\alpha}\Theta^2_j(T)&=\int_0^T\int_0^T\boldsymbol{\alpha}P(X(t)=j,X(s)=j)dtds\\[0.3em]
&=\int_0^T\int_0^T\boldsymbol{\alpha}P(X(s)=j)P(X(t)=j|X(s)=j)dtds\\[0.3em]
&=\int_0^T\int_0^T\sum_{i=1}^n\alpha_iP_{ij}(s)P_{jj}(t-s)dtds.
\end{align*}

\noindent Plugging in appropriate values for the probabilities and, as before, by re-defining $s\equiv\min{(s,t)}$ and $t\equiv\max{(s,t)}$, we get:
\begin{align*}
E_{\alpha}\Theta^2_j(T)&=\int_0^T\int_0^T{\left(\eta_j^2+\eta_j\sum_{k=1}^{n-1}u^k_{jj}e^{\nu_k|t-s|}+\eta_j\sum_{i=1}^n\sum_{k=1}^{n-1}\alpha_iu^k_{ij}e^{\nu_k\min{(s,t)}}+\right.}\\[0.3em]
&+\left.\sum^n_{i=1}\sum^{n-1}_{k=1}\sum^{n-1}_{l=1}\alpha_iu^l_{ij}u^k_{jj}e^{\nu_k \max{(s,t)}}e^{(\nu_k-\nu_l)\min{(s,t)}}\right)dtds\\[0.3em]
&=\eta_j^2 T^2-2\eta_j T\sum_{k=1}^{n-1}\frac{u^k_{jj}}{\nu_k}-2\eta_j T\sum_{i=1}^n\sum_{k=1}^{n-1}\frac{\alpha_i u^k_{ij}}{\nu_k}+(\mathrm{constant\ and\ exponentially\ decaying\ terms})\\[0.3em]
&\stackrel{T\rightarrow\infty}{=}\eta_j^2 T^2-2\eta_j T\sum_{k=1}^{n-1}\frac{u^k_{jj}}{\nu_k}-2\eta_j T\sum_{i=1}^n\sum_{k=1}^{n-1}\frac{\alpha_i u^k_{ij}}{\nu_k}.
\end{align*}

\noindent When calculating the variance for large $T$, both the quadratic term as well as the term that depends on the initial conditions will cancel out, leading to:
\begin{equation}
\mathrm{Var}\,\Theta_j(T)=-2\eta_j\sum_{k=1}^{n-1}\frac{u^k_{jj}}{\nu_k}\,T \label{eqn:var_p_gen}
\end{equation}

\noindent which is a result that is, as expected, linear in time and independent of the initial conditions. Analogically, one can show that for large $T$:
\begin{equation}
\mathrm{Cov}\,(\Theta_i(T),\Theta_j(T))=-\left(\eta_i\sum_{k=1}^{n-1}\frac{u^k_{ij}}{\nu_k}+\eta_j\sum_{k=1}^{n-1}\frac{u^k_{ji}}{\nu_k}\right)\,T \label{eqn:cov_p_gen}
\end{equation}

\par In our simplified system we can find the eigenvalues:
\begin{align*}
\nu_0&=0&\nu_2&=-\frac{2+\nu_1+r}{2}\\[0.3em]
\nu_1&=-\frac{2}{3}\varepsilon^2+\frac{2}{3}\varepsilon-1&\nu_3&=-\frac{2+\nu_1-r}{2},
\end{align*}

\noindent where 
\begin{displaymath}
r=\frac{1}{3}\sqrt{4\varepsilon^4-8\varepsilon^3+32\varepsilon^2-28\varepsilon+9}.
\end{displaymath}

\noindent First, we should note that $\nu_3$ eigenvalue becomes 0 at $\varepsilon=0$. In fact, at $\varepsilon=0$ the CTMC stops being recurrent and so there's no well-defined stationary distribution. Physically, an atom will be trapped in one of the $C$ states, thus moving continuously in one direction. In such situation we simply obtain a deterministic continuous momentum transfer with zero variance. 
\par In case of non-zero $\varepsilon$ we should expect that variance will not behave as for BCF in 2-level system. For small $\varepsilon$ atom or molecule will spend a lot of time in one 2-level system, before jumping to the other one, so the variance will be high. Indeed, using Eq.(\ref{eqn:var_p_gen}) and Eq.(\ref{eqn:cov_p_gen}) for our generator matrix $Q$, we obtain the variance, which can be written as:
\begin{displaymath}
\frac{1}{T}\mathrm{Var}\,p=16\,\hbar^2 k^2\frac{\delta^2}{\pi^2\Gamma}\frac{26\,\varepsilon(1-\varepsilon)-9}{20\,\varepsilon(1-\varepsilon)\nu_1},
\end{displaymath}

\noindent  where the numerator is non-zero at $\varepsilon=0$. Therefore, as expected, the variance diverges, when $\varepsilon$ becomes small, and is smallest at $\varepsilon=0.5$ reaching value of:
\begin{displaymath}
\frac{1}{T}\mathrm{Var}\,p=\frac{48}{5}\hbar^2 k^2\frac{\delta^2}{\pi^2\Gamma}.
\end{displaymath}

\noindent For experimentally achievable $\varepsilon$ (that is $\varepsilon\gtrsim 0.1$), variance stays very close to the given limiting value (changes by at most a factor of 2). At $\varepsilon=0.5$, we wouldn't generate any force at either of 2-level systems. For BCF $\varepsilon=0.25$ and the variance is approximately:
\begin{displaymath}
\frac{1}{T}\mathrm{Var}\,p_{BCF}\approx 20.11\,\hbar^2 k^2\frac{\delta^2}{\pi^2\Gamma}.
\end{displaymath}

\par Using the found eigenvalues, we can re-write the function $\mu_{\mathrm{IV}}$ found in Eq.(\ref{eqn:avg_force_4lvl}) as:
\begin{displaymath}
\mu_{\mathrm{IV}}=\frac{1-2\varepsilon}{-\nu_1},
\end{displaymath}

\noindent and similarly, like in Eq.(\ref{eqn:diff_coeff_musigma}), we can write the diffusion coefficient as:
\begin{displaymath}
D=F^2_{\mathrm{IV}}\frac{\sigma^2_{\mathrm{IV}}(\varepsilon)}{\mu^2_{\mathrm{IV}}(\varepsilon)} \frac{1}{\Gamma},
\end{displaymath}

\noindent with:
\begin{displaymath}
\sigma^2_{\mathrm{IV}}=\frac{2}{5}\frac{26\,\varepsilon(1-\varepsilon)-9}{\varepsilon(1-\varepsilon)\nu_1}.
\end{displaymath}

\noindent We can also notice a relationship between forces $F_{\mathrm{II}}$ in the 2-level system and $F_{\mathrm{IV}}$ in this model. Namely:
\begin{displaymath}
F_{\mathrm{IV}}=\frac{F_{\mathrm{II}}}{-2\nu_1}.
\end{displaymath}

\noindent This shows that we should expect forces acting on the two-level subsystem discussed here to be just re-scaled versions of the normal 2-level polychormatic forces with re-scaling factor of $-1/2\nu_1$, which is equal to exactly $4/7$ for the bichromatic fields. 
\par In a more general case we can simply use Eq.(\ref{eqn:var_p_gen}) and Eq.(\ref{eqn:cov_p_gen}) directly on whatever combination of occupancy times is appropriate in the system. In general we can write:
\begin{displaymath}
\frac{1}{T}\mathrm{Var}\,p=\mathrm{Var}\left(\sum_ia_i\Theta_i\right)=\sum_ia^2_i\mathrm{Var}\,\Theta_i+2\sum_{i<j}a_ia_j\mathrm{Cov}\,(\Theta_i,\Theta_j),
\end{displaymath}

\noindent where sum is over all the states in the model. For example, in a more realistic BCF molasses model in BaH described at the beginning of this section with $\Gamma_1\neq\Gamma_2$, $k_1\neq k_2$ and $\delta_1\neq\delta_2$, but with $\varepsilon\equiv\varepsilon_1=\varepsilon_2$, we would obtain:
\begin{align*}
\frac{1}{T}\mathrm{Var}\,p&=4\frac{\hbar^2}{\pi^2}\mathrm{Var}\left(k_1\delta_1(\Theta_1-\Theta_2)-k_2\delta_2(\Theta_3-\Theta_4)\right)\\
&=4\frac{\hbar^2k^2_1\delta^2_1}{\pi^2}\mathrm{Var}\left(\Theta_1-\Theta_2-\gamma(\Theta_3-\Theta_4)\right),
\end{align*}

\noindent with $\gamma\equiv k_2\delta_2/k_1\delta_1$. Omitting the term preceding the variance of occupancy times, we have $a_1=1$, $a_2=-1$, $a_3=-\gamma$ and $a_4=\gamma$. Because the forces in asymmetric systems have to be balanced according to Eq.(\ref{eqn:delta_criterion}) to create molasses centered at zero velocity, we know that $\gamma=\Gamma_2(1-r_2)/\Gamma_1(1-r_1)$.
\par Evaluating Eq.(\ref{eqn:var_p_gen}) and Eq.(\ref{eqn:cov_p_gen}) algorithmically can be done with ease as long as we are able to find eigenvectors and diagonalize generator matrix $Q$. In general, assuming the eigenvectors of $Q$ are columns in a matrix $V$ and eigenvalues are diagonal elements of $D$, we have $Q=VDV^{-1}$ and so:
\begin{displaymath}
P=e^{Qt}=Ve^{Dt}V^{-1},
\end{displaymath}

\noindent where exponential of eigenvalue matrix simply has exponents of eigenvalues on its diagonal. After matrix multiplication one obtains values in cells of $P$ as given in Eq.(\ref{eqn:P_matrix_gen}). To easily get values for $\eta_j$ and $u^k_{ij}$ we can instead create a matrix $U^k=VD^k V^{-1}$, where $D^k$ is defined as:
\begin{displaymath}
D^k_{ij}=\begin{cases}
1&\mathrm{if}\ k=i=j\\
0&\mathrm{otherwise}
\end{cases}.
\end{displaymath}

\noindent Then, we simply obtain $u^k_{ij}$ that we need in Eq.(\ref{eqn:var_p_gen}) and Eq.(\ref{eqn:cov_p_gen}) as the $ij$-th cell of matrix $U^k$, i.e. $U^k_{ij}=u^k_{ij}$. The same matrix gives us $\eta_j=U^0_{ij}$, where the equality holds for any index $0\leq i\leq n-1$ in a system with $n$ states. In summary, evaluating variance and covariance of occupancy times, and therefore variance of momentum transfer in $\pi$-pulse models for PCF, boils down to finding eigenvalues and eigenvectors of the generator matrix of the appropriate CTMC. Using this method we can numerically find that in a more realistic, asymmetric and balanced BaH system:
\begin{displaymath}
\frac{1}{T}\mathrm{Var}\,p_{BCF}\approx 20.79\,\hbar^2 k_1^2\frac{\delta_1^2}{\pi^2\Gamma_1}.
\end{displaymath}

\par Finally, we should note that applicability of this model and all the formulas to the actual polychromatic forces is limited to situations when the interaction can be actually approximated by $\pi$-pulses. This, for example, occurs at $\chi=\pi/4$ for bichromatic fields and $\chi=\pi/6$ for 4-color fields. These parameters, however, don't need to yield maximum attainable force. While they do in the case of 2-color forces, already in the case of 4-color fields such choice of $\chi$ provides strong force over a very wide range of velocities, but not the maximum at small velocities, which appears at lower values of $\chi$.

\bibliography{refs}

\end{document}